\documentclass[prd,amsmath,aps,floats,amssymb, floatfix,superscriptaddress,nofootinbib,twocolumn,preprintnumbers]{revtex4-2}

\setcitestyle{authoryear,round}

\usepackage[T1]{fontenc}
\usepackage{aecompl}

\usepackage{newtxtext,newtxmath}
\usepackage{mdframed,bm}
\usepackage{multirow}
\usepackage[T1]{fontenc}
\usepackage{ae,aecompl}
\usepackage{graphicx}	% Including figure files
\usepackage{amsmath}	% Advanced maths commands
\usepackage{amssymb}	% Extra maths symbols
\usepackage{booktabs}
\usepackage{color}
\usepackage{enumitem}
\usepackage[colorlinks=true]{hyperref}
\hypersetup{
    urlcolor=blue,
    citecolor=blue,
    linkcolor=blue,
  }

\makeatletter
  \let\NAT@sort\z@
  \let\NAT@cmprs\z@
\makeatother

\usepackage{graphicx}% Include figure files
\usepackage{dcolumn}% Align table columns on decimal point
\usepackage{bm}% bold math

\preprint{APS/123-QED}

\usepackage{amsmath}
\usepackage{fontawesome5}
\usepackage{color}
\usepackage{xcolor}
\usepackage{ulem}

\usepackage{graphicx}	% Including figure files
\usepackage{amsmath}	% Advanced maths commands

\usepackage{fontawesome5}
\usepackage{color}
\usepackage{xcolor}
\usepackage{xspace}

\usepackage{enumitem}
\setlist{nosep}

\newcommand{\wrp}{w_{p}}
\newcommand{\rp}{r_{p}}
\newcommand{\pimax}{\pi_{\rm max}}
\newcommand{\kpar}{k_\parallel}
\newcommand{\kperp}{k_\perp}
\newcommand{\ktot}{k_{\rm t}}

\newcommand{\mpc}{\, {\rm Mpc}}

\newcommand{\Rtwohc}{R_{\rm 200c}}
\newcommand{\Mtwohc}{M_{\rm 200c}}

\newcommand{\ctwohc}{c_{\rm 200c}}

\newcommand{\Asz}{A_{\rm SZ}}
\newcommand{\Bsz}{B_{\rm SZ}}
\newcommand{\Csz}{C_{\rm SZ}}
\newcommand{\xisz}{\xi_{\rm SZ}}

\newcommand{\mass}{\mbox{$M_{\mbox{\scriptsize 500c}}$}}
\newcommand{\mmass}{\mbox{$M^{\mbox{\scriptsize med}}_{\mbox{\scriptsize 500c}}$}}
\newcommand{\medianmass}{$1.65 \times 10^{14}$ \mbox{$M_\odot/h_{70}$}} 
\newcommand{\mlow}{7.9 \times 10^{13}} 
\newcommand{\mhigh}{16.2 \times 10^{14}}

\newcommand{\RA}{\textrm{R.A.} }
\newcommand{\DEC}{\textrm{Dec.} }

\defcitealias{Bleem:2026}{B26}

\begin{document}

\title{Measuring the Clustering of tSZ{-Selected} Galaxy Clusters with SPT-3G}% Force line breaks with \\

\author{E.~S.~Martsen}
\thanks{emartsen@uchicago.edu}
\affiliation{Department of Astronomy and Astrophysics, University of Chicago, 5640 South Ellis Avenue, Chicago, IL, 60637, USA}
\affiliation{Kavli Institute for Cosmological Physics, University of Chicago, 5640 South Ellis Avenue, Chicago, IL, 60637, USA}

\author{D.~Anbajagane}
\affiliation{Department of Astronomy and Astrophysics, University of Chicago, 5640 South Ellis Avenue, Chicago, IL, 60637, USA}
\affiliation{Kavli Institute for Cosmological Physics, University of Chicago, 5640 South Ellis Avenue, Chicago, IL, 60637, USA}
\affiliation{Department of Physics, Stanford University, 382 Via Pueblo Mall, Stanford, CA, 94305, USA}
\affiliation{Kavli Institute for Particle Astrophysics and Cosmology, Stanford University, 452 Lomita Mall, Stanford, CA, 94305, USA}

\author{L.~E.~Bleem}
\affiliation{High-Energy Physics Division, Argonne National Laboratory, 9700 South Cass Avenue, Lemont, IL, 60439, USA}
\affiliation{Kavli Institute for Cosmological Physics, University of Chicago, 5640 South Ellis Avenue, Chicago, IL, 60637, USA}
\affiliation{Department of Astronomy and Astrophysics, University of Chicago, 5640 South Ellis Avenue, Chicago, IL, 60637, USA}
\author{B.~A.~Benson}
\affiliation{Fermi National Accelerator Laboratory, MS209, P.O. Box 500, Batavia, IL, 60510, USA}
\affiliation{Kavli Institute for Cosmological Physics, University of Chicago, 5640 South Ellis Avenue, Chicago, IL, 60637, USA}
\affiliation{Department of Astronomy and Astrophysics, University of Chicago, 5640 South Ellis Avenue, Chicago, IL, 60637, USA}
\author{E.~Anderes}
\affiliation{Department of Statistics, University of California, One Shields Avenue, Davis, CA 95616, USA}
\author{A.~J.~Anderson}
\affiliation{Fermi National Accelerator Laboratory, MS209, P.O. Box 500, Batavia, IL, 60510, USA}
\affiliation{Kavli Institute for Cosmological Physics, University of Chicago, 5640 South Ellis Avenue, Chicago, IL, 60637, USA}
\affiliation{Department of Astronomy and Astrophysics, University of Chicago, 5640 South Ellis Avenue, Chicago, IL, 60637, USA}
\author{B.~Ansarinejad}
\affiliation{School of Physics, University of Melbourne, Parkville, VIC 3010, Australia}
\author{M.~Archipley}
\affiliation{Department of Astronomy and Astrophysics, University of Chicago, 5640 South Ellis Avenue, Chicago, IL, 60637, USA}
\affiliation{Kavli Institute for Cosmological Physics, University of Chicago, 5640 South Ellis Avenue, Chicago, IL, 60637, USA}
\author{L.~Balkenhol}
\affiliation{Sorbonne Universit\'e, CNRS, UMR 7095, Institut d'Astrophysique de Paris, 98 bis bd Arago, 75014 Paris, France}
\author{D.~R.~Barron}
\affiliation{Department of Physics and Astronomy, University of New Mexico, Albuquerque, NM, 87131, USA}
\author{P.~S.~Barry}
\affiliation{School of Physics and Astronomy, Cardiff University, Cardiff, CF24 3AA, UK}
\author{K.~Benabed}
\affiliation{Sorbonne Universit\'e, CNRS, UMR 7095, Institut d'Astrophysique de Paris, 98 bis bd Arago, 75014 Paris, France}
\author{A.~N.~Bender}
\affiliation{High-Energy Physics Division, Argonne National Laboratory, 9700 South Cass Avenue, Lemont, IL, 60439, USA}
\affiliation{Kavli Institute for Cosmological Physics, University of Chicago, 5640 South Ellis Avenue, Chicago, IL, 60637, USA}
\affiliation{Department of Astronomy and Astrophysics, University of Chicago, 5640 South Ellis Avenue, Chicago, IL, 60637, USA}
\author{F.~Bianchini}
\affiliation{Kavli Institute for Particle Astrophysics and Cosmology, Stanford University, 452 Lomita Mall, Stanford, CA, 94305, USA}
\affiliation{Department of Physics, Stanford University, 382 Via Pueblo Mall, Stanford, CA, 94305, USA}
\affiliation{SLAC National Accelerator Laboratory, 2575 Sand Hill Road, Menlo Park, CA, 94025, USA}
\author{S.~Bocquet}
\affiliation{University Observatory, Faculty of Physics, LMU Munich, Scheinerstr.~1, 81679 Munich, Germany}
\author{F.~R.~Bouchet}
\affiliation{Sorbonne Universit\'e, CNRS, UMR 7095, Institut d'Astrophysique de Paris, 98 bis bd Arago, 75014 Paris, France}
\author{E.~Camphuis}
\affiliation{Sorbonne Universit\'e, CNRS, UMR 7095, Institut d'Astrophysique de Paris, 98 bis bd Arago, 75014 Paris, France}
\author{M.~G.~Campitiello}
\affiliation{High-Energy Physics Division, Argonne National Laboratory, 9700 South Cass Avenue, Lemont, IL, 60439, USA}
\author{J.~E.~Carlstrom}
\affiliation{Kavli Institute for Cosmological Physics, University of Chicago, 5640 South Ellis Avenue, Chicago, IL, 60637, USA}
\affiliation{Enrico Fermi Institute, University of Chicago, 5640 South Ellis Avenue, Chicago, IL, 60637, USA}
\affiliation{Department of Physics, University of Chicago, 5640 South Ellis Avenue, Chicago, IL, 60637, USA}
\affiliation{High-Energy Physics Division, Argonne National Laboratory, 9700 South Cass Avenue, Lemont, IL, 60439, USA}
\affiliation{Department of Astronomy and Astrophysics, University of Chicago, 5640 South Ellis Avenue, Chicago, IL, 60637, USA}
\author{J.~Carron}
\affiliation{Istituto ricerche solari Aldo e Cele Dacc\`o (IRSOL), Faculty of Informatics, Universit\`a della Svizzera italiana, 6605 Locarno, Switzerland}
\affiliation{Universit\'e de Gen\`eve, D\'epartement de Physique Th\'eorique, 24 Quai Ansermet, CH-1211 Gen\`eve 4, Switzerland}
\author{C.~L.~Chang}
\affiliation{High-Energy Physics Division, Argonne National Laboratory, 9700 South Cass Avenue, Lemont, IL, 60439, USA}
\affiliation{Kavli Institute for Cosmological Physics, University of Chicago, 5640 South Ellis Avenue, Chicago, IL, 60637, USA}
\affiliation{Department of Astronomy and Astrophysics, University of Chicago, 5640 South Ellis Avenue, Chicago, IL, 60637, USA}
\author{P.~M.~Chichura}
\affiliation{Department of Physics, University of Chicago, 5640 South Ellis Avenue, Chicago, IL, 60637, USA}
\affiliation{Kavli Institute for Cosmological Physics, University of Chicago, 5640 South Ellis Avenue, Chicago, IL, 60637, USA}
\author{A.~Chokshi}
\affiliation{Department of Astronomy and Astrophysics, University of Chicago, 5640 South Ellis Avenue, Chicago, IL, 60637, USA}
\author{T.-L.~Chou}
\affiliation{Department of Astronomy and Astrophysics, University of Chicago, 5640 South Ellis Avenue, Chicago, IL, 60637, USA}
\affiliation{Kavli Institute for Cosmological Physics, University of Chicago, 5640 South Ellis Avenue, Chicago, IL, 60637, USA}
\affiliation{National Taiwan University, No. 1, Sec. 4, Roosevelt Road, Taipei 106319, Taiwan}
\author{A.~Coerver}
\affiliation{Department of Physics, University of California, Berkeley, CA, 94720, USA}
\author{T.~M.~Crawford}
\affiliation{Department of Astronomy and Astrophysics, University of Chicago, 5640 South Ellis Avenue, Chicago, IL, 60637, USA}
\affiliation{Kavli Institute for Cosmological Physics, University of Chicago, 5640 South Ellis Avenue, Chicago, IL, 60637, USA}
\author{C.~Daley}
\affiliation{Universit\'e Paris-Saclay, Universit\'e Paris Cit\'e, CEA, CNRS, AIM, 91191, Gif-sur-Yvette, France}
\affiliation{Department of Astronomy, University of Illinois Urbana-Champaign, 1002 West Green Street, Urbana, IL, 61801, USA}
\author{T.~de~Haan}
\affiliation{High Energy Accelerator Research Organization (KEK), Tsukuba, Ibaraki 305-0801, Japan}
\author{K.~R.~Dibert}
\affiliation{Department of Astronomy and Astrophysics, University of Chicago, 5640 South Ellis Avenue, Chicago, IL, 60637, USA}
\affiliation{Kavli Institute for Cosmological Physics, University of Chicago, 5640 South Ellis Avenue, Chicago, IL, 60637, USA}
\author{M.~A.~Dobbs}
\affiliation{Department of Physics and McGill Space Institute, McGill University, 3600 Rue University, Montreal, Quebec H3A 2T8, Canada}
\affiliation{Canadian Institute for Advanced Research, CIFAR Program in Gravity and the Extreme Universe, Toronto, ON, M5G 1Z8, Canada}
\author{M.~Doohan}
\affiliation{School of Physics, University of Melbourne, Parkville, VIC 3010, Australia}
\author{D.~Dutcher}
\affiliation{Joseph Henry Laboratories of Physics, Jadwin Hall, Princeton University, Princeton, NJ 08544, USA}
\author{C.~Feng}
\affiliation{Department of Astronomy, University of Science and Technology of China, Hefei 230026, China}
\affiliation{School of Astronomy and Space Science, University of Science and Technology of China, Hefei 230026}
\affiliation{Department of Physics, University of Illinois Urbana-Champaign, 1110 West Green Street, Urbana, IL, 61801, USA}
\author{K.~R.~Ferguson}
\affiliation{Department of Physics and Astronomy, University of California, Los Angeles, CA, 90095, USA}
\affiliation{Department of Physics and Astronomy, Michigan State University, East Lansing, MI 48824, USA}
\author{N.~C.~Ferree}
\affiliation{California Institute of Technology, 1200 East California Boulevard., Pasadena, CA, 91125, USA}
\affiliation{Kavli Institute for Particle Astrophysics and Cosmology, Stanford University, 452 Lomita Mall, Stanford, CA, 94305, USA}
\affiliation{Department of Physics, Stanford University, 382 Via Pueblo Mall, Stanford, CA, 94305, USA}
\author{K.~Fichman}
\affiliation{Department of Physics, University of Chicago, 5640 South Ellis Avenue, Chicago, IL, 60637, USA}
\affiliation{Kavli Institute for Cosmological Physics, University of Chicago, 5640 South Ellis Avenue, Chicago, IL, 60637, USA}
\author{A.~Foster}
\affiliation{Joseph Henry Laboratories of Physics, Jadwin Hall, Princeton University, Princeton, NJ 08544, USA}
\author{N.~Frontiere}
\affiliation{Computational Science Division, Argonne National Laboratory, 9700 South Cass Avenue, Lemont, IL, 60439, USA}
\author{S.~Galli}
\affiliation{Sorbonne Universit\'e, CNRS, UMR 7095, Institut d'Astrophysique de Paris, 98 bis bd Arago, 75014 Paris, France}
\author{A.~E.~Gambrel}
\affiliation{Kavli Institute for Cosmological Physics, University of Chicago, 5640 South Ellis Avenue, Chicago, IL, 60637, USA}
\author{A.~K.~Gao}
\affiliation{Department of Physics, University of Illinois Urbana-Champaign, 1110 West Green Street, Urbana, IL, 61801, USA}
\author{F.~Ge}
\affiliation{California Institute of Technology, 1200 East California Boulevard., Pasadena, CA, 91125, USA}
\affiliation{Kavli Institute for Particle Astrophysics and Cosmology, Stanford University, 452 Lomita Mall, Stanford, CA, 94305, USA}
\affiliation{Department of Physics, Stanford University, 382 Via Pueblo Mall, Stanford, CA, 94305, USA}
\affiliation{Department of Physics \& Astronomy, University of California, One Shields Avenue, Davis, CA 95616, USA}
\author{S.~Gradis}
\affiliation{Universit"at Innsbruck, Institut f"ur Astro- und Teilchenphysik, Technikerstr. 25/8, 6020 Innsbruck, Austria}
\author{F.~Guidi}
\affiliation{Department of Physics \& Astronomy, University of California, One Shields Avenue, Davis, CA 95616, USA}
\affiliation{Sorbonne Universit\'e, CNRS, UMR 7095, Institut d'Astrophysique de Paris, 98 bis bd Arago, 75014 Paris, France}
\author{S.~Guns}
\affiliation{Department of Physics, University of California, Berkeley, CA, 94720, USA}
\author{N.~W.~Halverson}
\affiliation{CASA, Department of Astrophysical and Planetary Sciences, University of Colorado, Boulder, CO, 80309, USA }
\affiliation{Department of Physics, University of Colorado, Boulder, CO, 80309, USA}
\author{E.~Hivon}
\affiliation{Sorbonne Universit\'e, CNRS, UMR 7095, Institut d'Astrophysique de Paris, 98 bis bd Arago, 75014 Paris, France}
\author{G.~P.~Holder}
\affiliation{Department of Physics, University of Illinois Urbana-Champaign, 1110 West Green Street, Urbana, IL, 61801, USA}
\author{W.~L.~Holzapfel}
\affiliation{Department of Physics, University of California, Berkeley, CA, 94720, USA}
\author{J.~C.~Hood}
\affiliation{Kavli Institute for Cosmological Physics, University of Chicago, 5640 South Ellis Avenue, Chicago, IL, 60637, USA}
\author{A.~Hryciuk}
\affiliation{Department of Physics, University of Chicago, 5640 South Ellis Avenue, Chicago, IL, 60637, USA}
\affiliation{Kavli Institute for Cosmological Physics, University of Chicago, 5640 South Ellis Avenue, Chicago, IL, 60637, USA}
\author{N.~Huang}
\affiliation{Department of Physics, University of California, Berkeley, CA, 94720, USA}
\author{T.~Jhaveri}
\affiliation{Department of Astronomy and Astrophysics, University of Chicago, 5640 South Ellis Avenue, Chicago, IL, 60637, USA}
\affiliation{Kavli Institute for Cosmological Physics, University of Chicago, 5640 South Ellis Avenue, Chicago, IL, 60637, USA}
\author{F.~K\'eruzor\'e}
\affiliation{High-Energy Physics Division, Argonne National Laboratory, 9700 South Cass Avenue, Lemont, IL, 60439, USA}
\author{A.~R.~Khalife}
\affiliation{Sorbonne Universit\'e, CNRS, UMR 7095, Institut d'Astrophysique de Paris, 98 bis bd Arago, 75014 Paris, France}
\author{M.~Klein}
\affiliation{University Observatory, Faculty of Physics, LMU Munich, Scheinerstr.~1, 81679 Munich, Germany}
\author{L.~Knox}
\affiliation{Department of Physics \& Astronomy, University of California, One Shields Avenue, Davis, CA 95616, USA}
\author{K.~Kornoelje}
\affiliation{Department of Astronomy and Astrophysics, University of Chicago, 5640 South Ellis Avenue, Chicago, IL, 60637, USA}
\affiliation{Kavli Institute for Cosmological Physics, University of Chicago, 5640 South Ellis Avenue, Chicago, IL, 60637, USA}
\affiliation{High-Energy Physics Division, Argonne National Laboratory, 9700 South Cass Avenue, Lemont, IL, 60439, USA}
\author{C.-L.~Kuo}
\affiliation{Kavli Institute for Particle Astrophysics and Cosmology, Stanford University, 452 Lomita Mall, Stanford, CA, 94305, USA}
\affiliation{Department of Physics, Stanford University, 382 Via Pueblo Mall, Stanford, CA, 94305, USA}
\affiliation{SLAC National Accelerator Laboratory, 2575 Sand Hill Road, Menlo Park, CA, 94025, USA}
\author{P.~Larsen}
\affiliation{Computational Science Division, Argonne National Laboratory, 9700 South Cass Avenue, Lemont, IL, 60439, USA}
\author{K.~Levy}
\affiliation{School of Physics, University of Melbourne, Parkville, VIC 3010, Australia}
\author{Y.~Li}
\affiliation{Kavli Institute for Cosmological Physics, University of Chicago, 5640 South Ellis Avenue, Chicago, IL, 60637, USA}
\author{A.~E.~Lowitz}
\affiliation{Kavli Institute for Cosmological Physics, University of Chicago, 5640 South Ellis Avenue, Chicago, IL, 60637, USA}
\author{C.~Lu}
\affiliation{Department of Physics, University of Illinois Urbana-Champaign, 1110 West Green Street, Urbana, IL, 61801, USA}
\author{G.~P.~Lynch}
\affiliation{Department of Physics \& Astronomy, University of California, One Shields Avenue, Davis, CA 95616, USA}
\author{T.~J.~Maccarone}
\affiliation{Department of Physics \& Astronomy, Box 41051, Texas Tech University, Lubbock TX 79409-1051, USA}
\author{A.~S.~Maniyar}
\affiliation{Kavli Institute for Particle Astrophysics and Cosmology, Stanford University, 452 Lomita Mall, Stanford, CA, 94305, USA}
\affiliation{Department of Physics, Stanford University, 382 Via Pueblo Mall, Stanford, CA, 94305, USA}
\affiliation{SLAC National Accelerator Laboratory, 2575 Sand Hill Road, Menlo Park, CA, 94025, USA}
\author{F.~Menanteau}
\affiliation{Department of Astronomy, University of Illinois Urbana-Champaign, 1002 West Green Street, Urbana, IL, 61801, USA}
\affiliation{Center for AstroPhysical Surveys, National Center for Supercomputing Applications, Urbana, IL, 61801, USA}
\author{M.~Millea}
\affiliation{Department of Physics, University of California, Berkeley, CA, 94720, USA}
\author{J.~Montgomery}
\affiliation{Department of Physics and McGill Space Institute, McGill University, 3600 Rue University, Montreal, Quebec H3A 2T8, Canada}
\author{Y.~Nakato}
\affiliation{Department of Physics, Stanford University, 382 Via Pueblo Mall, Stanford, CA, 94305, USA}
\author{T.~Natoli}
\affiliation{Kavli Institute for Cosmological Physics, University of Chicago, 5640 South Ellis Avenue, Chicago, IL, 60637, USA}
\author{Y.~Omori}
\affiliation{Department of Astronomy and Astrophysics, University of Chicago, 5640 South Ellis Avenue, Chicago, IL, 60637, USA}
\affiliation{Kavli Institute for Cosmological Physics, University of Chicago, 5640 South Ellis Avenue, Chicago, IL, 60637, USA}
\affiliation{NSF-Simons AI Institute for the Sky (SkAI), 172 E. Chestnut St., Chicago, IL 60611, USA}
\author{A.~Ouellette}
\affiliation{Department of Physics, University of Illinois Urbana-Champaign, 1110 West Green Street, Urbana, IL, 61801, USA}
\author{Z.~Pan}
\affiliation{High-Energy Physics Division, Argonne National Laboratory, 9700 South Cass Avenue, Lemont, IL, 60439, USA}
\affiliation{Kavli Institute for Cosmological Physics, University of Chicago, 5640 South Ellis Avenue, Chicago, IL, 60637, USA}
\affiliation{Department of Physics, University of Chicago, 5640 South Ellis Avenue, Chicago, IL, 60637, USA}
\author{P.~Paschos}
\affiliation{Enrico Fermi Institute, University of Chicago, 5640 South Ellis Avenue, Chicago, IL, 60637, USA}
\author{K.~A.~Phadke}
\affiliation{Department of Astronomy, University of Illinois Urbana-Champaign, 1002 West Green Street, Urbana, IL, 61801, USA}
\affiliation{Center for AstroPhysical Surveys, National Center for Supercomputing Applications, Urbana, IL, 61801, USA}
\affiliation{NSF-Simons AI Institute for the Sky (SkAI), 172 E. Chestnut St., Chicago, IL 60611, USA}
\author{K.~Prabhu}
\affiliation{Department of Physics \& Astronomy, University of California, One Shields Avenue, Davis, CA 95616, USA}
\author{W.~Quan}
\affiliation{High-Energy Physics Division, Argonne National Laboratory, 9700 South Cass Avenue, Lemont, IL, 60439, USA}
\affiliation{Department of Physics, University of Chicago, 5640 South Ellis Avenue, Chicago, IL, 60637, USA}
\affiliation{Kavli Institute for Cosmological Physics, University of Chicago, 5640 South Ellis Avenue, Chicago, IL, 60637, USA}
%\author{S.~Raghunathan}
%\affiliation{Department of Physics \& Astronomy, University of California, One Shields Avenue, Davis, CA 95616, USA}
%\affiliation{Center for AstroPhysical Surveys, National Center for Supercomputing Applications, Urbana, IL, 61801, USA}
\author{M.~Rahimi}
\affiliation{School of Physics, University of Melbourne, Parkville, VIC 3010, Australia}
\author{A.~Rahlin}
\affiliation{Department of Astronomy and Astrophysics, University of Chicago, 5640 South Ellis Avenue, Chicago, IL, 60637, USA}
\affiliation{Kavli Institute for Cosmological Physics, University of Chicago, 5640 South Ellis Avenue, Chicago, IL, 60637, USA}
\author{C.~L.~Reichardt}
\affiliation{School of Physics, University of Melbourne, Parkville, VIC 3010, Australia}
\author{M.~Rouble}
\affiliation{Department of Physics and McGill Space Institute, McGill University, 3600 Rue University, Montreal, Quebec H3A 2T8, Canada}
\author{J.~E.~Ruhl}
\affiliation{Department of Physics, Case Western Reserve University, Cleveland, OH, 44106, USA}
\author{A.~C.~Silva~Oliveira}
\affiliation{California Institute of Technology, 1200 East California Boulevard., Pasadena, CA, 91125, USA}
\affiliation{Kavli Institute for Particle Astrophysics and Cosmology, Stanford University, 452 Lomita Mall, Stanford, CA, 94305, USA}
\affiliation{Department of Physics, Stanford University, 382 Via Pueblo Mall, Stanford, CA, 94305, USA}
\author{A.~Simpson}
\affiliation{Department of Astronomy and Astrophysics, University of Chicago, 5640 South Ellis Avenue, Chicago, IL, 60637, USA}
\affiliation{Kavli Institute for Cosmological Physics, University of Chicago, 5640 South Ellis Avenue, Chicago, IL, 60637, USA}
\author{J.~A.~Sobrin}
\affiliation{Department of Physics, Villanova University, 800 E Lancaster Ave, Villanova, PA 19085, USA}
\affiliation{Fermi National Accelerator Laboratory, MS209, P.O. Box 500, Batavia, IL, 60510, USA}
\author{A.~A.~Stark}
\affiliation{Center for Astrophysics \textbar{} Harvard \& Smithsonian, 60 Garden Street, Cambridge, MA, 02138, USA}
\author{J.~Stephen}
\affiliation{Enrico Fermi Institute, University of Chicago, 5640 South Ellis Avenue, Chicago, IL, 60637, USA}
\author{C.~Tandoi}
\affiliation{Department of Astronomy, University of Illinois Urbana-Champaign, 1002 West Green Street, Urbana, IL, 61801, USA}
\author{C.~Trendafilova}
\affiliation{CERCA/ISO, Department of Physics, Case Western Reserve University, Cleveland, OH 44106, USA}
\affiliation{Center for AstroPhysical Surveys, National Center for Supercomputing Applications, Urbana, IL, 61801, USA}
\author{J.~D.~Vieira}
\affiliation{Department of Astronomy, University of Illinois Urbana-Champaign, 1002 West Green Street, Urbana, IL, 61801, USA}
\affiliation{Department of Physics, University of Illinois Urbana-Champaign, 1110 West Green Street, Urbana, IL, 61801, USA}
\affiliation{Center for AstroPhysical Surveys, National Center for Supercomputing Applications, Urbana, IL, 61801, USA}
\author{A.~G.~Vieregg}
\affiliation{Kavli Institute for Cosmological Physics, University of Chicago, 5640 South Ellis Avenue, Chicago, IL, 60637, USA}
\affiliation{Department of Astronomy and Astrophysics, University of Chicago, 5640 South Ellis Avenue, Chicago, IL, 60637, USA}
\affiliation{Enrico Fermi Institute, University of Chicago, 5640 South Ellis Avenue, Chicago, IL, 60637, USA}
\affiliation{Department of Physics, University of Chicago, 5640 South Ellis Avenue, Chicago, IL, 60637, USA}
\author{A.~Vitrier}
\affiliation{Sorbonne Universit\'e, CNRS, UMR 7095, Institut d'Astrophysique de Paris, 98 bis bd Arago, 75014 Paris, France}
\author{Y.~Wan}
\affiliation{Department of Astronomy, University of Illinois Urbana-Champaign, 1002 West Green Street, Urbana, IL, 61801, USA}
\affiliation{Center for AstroPhysical Surveys, National Center for Supercomputing Applications, Urbana, IL, 61801, USA}
\author{N.~Whitehorn}
\affiliation{Department of Physics and Astronomy, Michigan State University, East Lansing, MI 48824, USA}
\author{W.~L.~K.~Wu}
\affiliation{California Institute of Technology, 1200 East California Boulevard., Pasadena, CA, 91125, USA}
\affiliation{Kavli Institute for Particle Astrophysics and Cosmology, Stanford University, 452 Lomita Mall, Stanford, CA, 94305, USA}
\affiliation{SLAC National Accelerator Laboratory, 2575 Sand Hill Road, Menlo Park, CA, 94025, USA}
\author{M.~R.~Young}
\affiliation{Fermi National Accelerator Laboratory, MS209, P.O. Box 500, Batavia, IL, 60510, USA}
\affiliation{Kavli Institute for Cosmological Physics, University of Chicago, 5640 South Ellis Avenue, Chicago, IL, 60637, USA}
\author{J.~A.~Zebrowski}
\affiliation{Kavli Institute for Cosmological Physics, University of Chicago, 5640 South Ellis Avenue, Chicago, IL, 60637, USA}
\affiliation{Department of Astronomy and Astrophysics, University of Chicago, 5640 South Ellis Avenue, Chicago, IL, 60637, USA}
\affiliation{Fermi National Accelerator Laboratory, MS209, P.O. Box 500, Batavia, IL, 60510, USA}
\collaboration{SPT-3G Collaboration}
\noaffiliation
%

%\author{Dhayaa Anbajagane}
%\affiliation{%
% Department of Astronomy and Astrophysics\\ 
% The University of Chicago
%}%

%\author{Lindsey Bleem}
%\affiliation{%
% High Energy Physics Division\\ 
% Argonne National Laboratory
%}%

%\author{Bradford Benson}
%\affiliation{%
% Department of Astronomy and Astrophysics\\ 
% The University of Chicago
%}%
%\affiliation{%
% Fermilab National Laboratory
%}%

\date{\today \\}% It is always \today, today,

\begin{abstract}

We present a measurement of the spatial clustering of galaxy clusters selected via the thermal Sunyaev-Zel’dovich (tSZ) signal. {Using a cluster catalog derived from five years of SPT-3G Main field survey data}, we define a sample spanning $0.25 < z < 1.1$, as well as two subsamples spanning $0.25 < z < 0.67$ and $0.67 < z < 1.1$. We detect cluster clustering at $12.8\sigma$, the highest {significance} detection in a {tSZ-selected sample} to date. {We detail our analytic modeling and theoretical covariance models, including validations performed on synthetic data created from numerical simulations. The measurements and inference pass a variety of null tests, and show minimal sensitivity to common spatially varying systematics that impact optical clustering measurements, such as depth, stellar density, and extinction. We use our pipeline to place constraints on the halo bias of our sample, $b_{\textrm{h}} = {5.30\pm{0.24}}$, and the normalization of {our} mass-observable relation, $\ln\Asz = {0.43^{+0.17}_{-0.21}}$.} {We also demonstrate the {useful} synergies between cluster clustering and abundance measurements, which manifest {in} the breaking of high-dimensional parameter degeneracies.} Samples of clusters selected using the tSZ effect serve as a unique probe of clustering {that is insensitive to common systematics}, relative to conventional galaxy datasets, and the {detection} significance {of this measurement} will increase {substantially} with {existing data from the SPT-3G { Ext-10k survey} and upcoming data from SPT-3G+ and the Simons Observatory.}
\end{abstract}


%\tableofcontents

% \preprint{FERMILAB-PUB-25-0644-CSAID-PPD}
\maketitle

\section{Introduction}

The formation and evolution of the matter distribution in our Universe, referred to as large-scale structure (LSS), offers a powerful probe into the nature of dark matter and dark energy \citep{weinberg_observational_2013}. Structure forms in a hierarchical fashion, with small density fluctuations collapsing (due to gravity) into a cosmic web of dark matter halos. These halos are then populated with galaxies and galaxy clusters that we observe today \citep[e.g.,][]{lacey_merger_1993, kravtsov_formation_2012}. Galaxy clusters lie within the most massive dark matter halos, found at the peaks of the matter density field, and their abundances and spatial distributions are sensitive to the underlying cosmology driving the expansion of the universe and the growth of structure \citep{allen_cosmological_2011,weinberg_observational_2013}. As such, they deliver unique insights into our Universe, and their statistics are powerful tools for constraining cosmological parameters. 

In the past two decades, the primary way of extracting cosmological constraints from galaxy cluster samples has been through their abundances as a function of mass and redshift \citep{vikhlinin_chandra_2009, mantz_weighing_2015, bocquet_cluster_2024b, ghirardini_2024_erosita, DES:2025:6x2pt_Y6}. As the most massive gravitationally {collapsed} objects, clusters lie on the extreme tail of the halo mass function. This makes them particularly sensitive probes of the total matter in the Universe ($\Omega_{\rm m}$) and of the amplitude of density fluctuations ($\sigma_8$){, as well as of their evolution with redshift}. However, the reliance on a halo mass function also means the constraints are particularly sensitive to uncertainties in the mass-observable scaling relations, which is used to assign masses to the clusters and determine their subsequent abundances \citep{allen_cosmological_2011, bocquet_cluster_2024b, Grandis:2024:eROSITAMass, Kleinebreil:2025:eROSITAMass, DES:2025:6x2pt_Y6}. Thus, as demonstrated in the above works, the ability to precisely and accurately constrain the masses of clusters is one of the main limitations in obtaining cosmological parameter constraints from cluster counts.

As cluster samples reach greater number densities, {it becomes possible to use their spatial clustering} --- quantified through their 2-point correlation function --- as a probe of cosmology and as a way of self-calibrating the cluster masses \citep{majumdar_self-calibration_2004, lima_self-calibration_2005, cromer_2019}. Clustering studies have been extensively performed on populations of galaxies --- using large galaxy catalogs provided by surveys like the Sloan Digital Sky Survey \citep[SDSS;][]{sdss_collaboration_angular_2002}, the Dark Energy Survey \citep[DES;][]{rodriguez-monroy_dark_2022}, the Dark Energy Spectroscopic Instrument \citep[DESI;][]{Adame_2025} and the Kilo-Degree Survey \citep[KiDS;][]{dvornik_kids-1000_2023} --- while their application to galaxy clusters has been less frequent but still extends over the last two decades. {This spatial clustering,  often referred to as ``cluster clustering,''} has been measured previously for optical and X-ray selected cluster samples \citep{baxter_constraining_2016, marulli_xxl_2018, lindholm_clustering_2021, Fumagalli_2024, seppi_srgerosita_2024}, with recent work using the {tSZ-selected} {sample} from \textit{Planck} \citep{Lesci:2023:PlanckClustering}. The amplitude of the tSZ effect --- which is the inverse-Compton scattering of Cosmic Microwave Background (CMB) photons off the hot gas in clusters --- is independent {of} redshift, giving the tSZ a unique advantage over optical and X-ray surveys in the detection of high-redshift galaxy clusters \citep{sunyaev_observations_1972}.

\begin{figure}
    \centering
    \includegraphics[width=\linewidth]{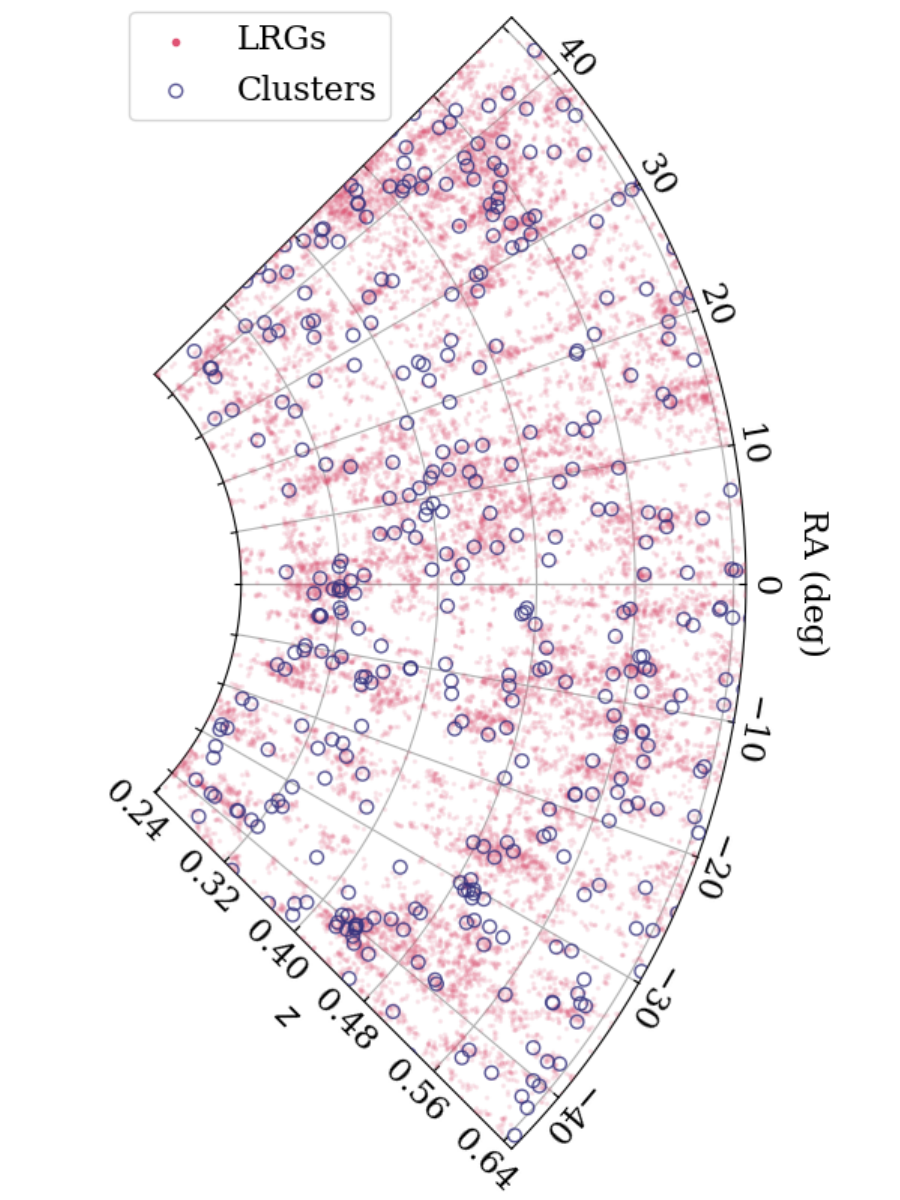}
    \caption{Illustrative example of the distribution of some SPT-3G {c}lusters (purple circles) tracing the LSS. Clusters are projected within a small declination slice of ${\rm \DEC} \pm 2^{\circ}$ centered at ${\rm \DEC} = -50^\circ$. Also plotted are {luminous red galaxies (LRGs)} from the DES redMaGiC sample \citep{rodriguez-monroy_dark_2022} (light pink dots), which have been filtered using k-nearest neighbors (for visual purposes) to emphasize the underlying peaks in the structure.}
    \label{fig:wedge}
\end{figure}

In this work, we make the first measurement of cluster clustering using a cluster sample produced from high-resolution millimeter-wave data from the South Pole Telescope (SPT). {In particular, we utilize a cluster catalog derived from five years of observations with the SPT third-generation camera (SPT-3G) of the SPT-3G Main Field in the
southern hemisphere} {\citep[][hereafter \citetalias{Bleem:2026}]{Bleem:2026}}. This catalog {has the highest cluster density} available thus far in the millimeter regime. We {compare} our measurement with our model for cluster clustering and quantify the statistical power, while also placing constraints on the mass-observable relation of the SPT-3G sample. {An illustrative example of clustering in a portion of our cluster catalog is shown in Figure \ref{fig:wedge}}.

We organize this paper as follows: in Section \ref{sec:data}, we describe the SPT-3G cluster sample. The methods used to estimate the correlation are given in Section \ref{sec:Corr Estimator}. The theoretical correlation model, the model of the covariance, and the methods for inferring the scaling relation parameters are described in Section \ref{sec:theory}. Simulation-based model validations are presented in Section \ref{sec:simvalidations} while data-based consistency checks are shown in Section \ref{sec:data_validations}. The results are presented in Section \ref{sec:results}.  Finally, we summarize this work and discuss future prospects in Section \ref{sec:conclusions}.

{For consistency with \citetalias{Bleem:2026},} we assume a fiducial $\Lambda$CDM cosmology with $\sigma_8=0.80$, $\Omega_b = 0.046$, $\Omega_m = 0.30$, $h = 0.7$, $n_s(k_s=0.002) = 0.96$, and $\Sigma m_\nu=0.06$ eV, except for when using the Last Journey Simulation data where we instead assume the cosmology given in \cite{Heitmann2021} for consistency. Cluster masses are reported in terms of  $M_{200c}$, which is defined as the mass enclosed within a radius, $r_{200c}$, at which the average enclosed density is 200$\times$ the critical density at the cluster redshift.

\section{SPT-3G Cluster Catalog} \label{sec:data}

This work uses data from
SPT-3G, the third generation survey camera on the SPT. Installed on the telescope in 2017, SPT-3G has a focal plane consisting of 16,000 tri-chroic TES detectors, with observing bands centered at 95, 150, and 220 GHz \citep{carlstrom_2010, benson_spt-3g_2014, Sobrin_2022}. In addition to covering the peak of the CMB spectrum, these wavelengths allow us to efficiently measure the tSZ effect --- we measure both the peak of the signal amplitude at {150} GHz and the null signal at 220 GHz. This makes SPT-3G an ideal instrument for detecting galaxy clusters out to high redshifts{, as demonstrated by the catalog of \citetalias{Bleem:2026}}.

SPT-3G has three {primary science} surveys: the Main survey, which is
the {smallest and} deepest SPT survey, the Summer survey, and the Wide survey, which is the shallowest
survey but covers the widest area of the sky \citep{prabhu_2024}. In this work we focus on the cluster sample produced from {1,604 deg$^{2}$ of data taken over five} years of observations of the SPT-3G Main survey \citepalias{Bleem:2026}. 
This sample consists of 8,892 tSZ cluster candidates detected above a significance, or signal-to-noise ratio (SNR) maximized over filter scales, of $\xi{\geq}4$. Using both archival and dedicated targeted optical and infrared (IR) observations from the DES \citep{abbott21}, the DECam Legacy
Survey \citep[DECaLs;][]{dey19}, the Wide Field Infrared Explorer \citep[WISE;][]{wright10}, the \textit{Spitzer} Space Telescope, and the 6.5~m Magellan Telescopes, around $\sim7,200$ candidates were confirmed as galaxy clusters via the presence of significant red-sequence and/or IR galaxy overdensities. The expected purity of the confirmed sample exceeds 96\%. Additionally, {the sample} spans $0.037<z\lesssim 2$ in redshift, $z$, with a median redshift of $z_{\rm med}=0.73$ and has masses ranging from $\mlow < \mass [M_\odot/h_{70}] < \mhigh$ with a median mass of \mmass=\medianmass. 

 The redshift estimation for the sample was done via the multi-component matched filter (MCMF) technique \citep{Klein_2018,Klein_2019,Klein_2024a}, which utilizes redshift- and magnitude-dependent color filters to search for overdensities of red-sequence or IR galaxies in the aforementioned optical and infrared observations. The uncertainty in the redshift estimation is assumed to be Gaussian with a width of $\sigma_z$ that is a function of redshift as $\sigma_z = \sigma_{i}\times(1+z)$, where $\sigma_i$ is the uncertainty coefficient for each halo derived from the MCMF technique. 
 
\begin{figure}
    \centering
    \includegraphics[width=\linewidth]{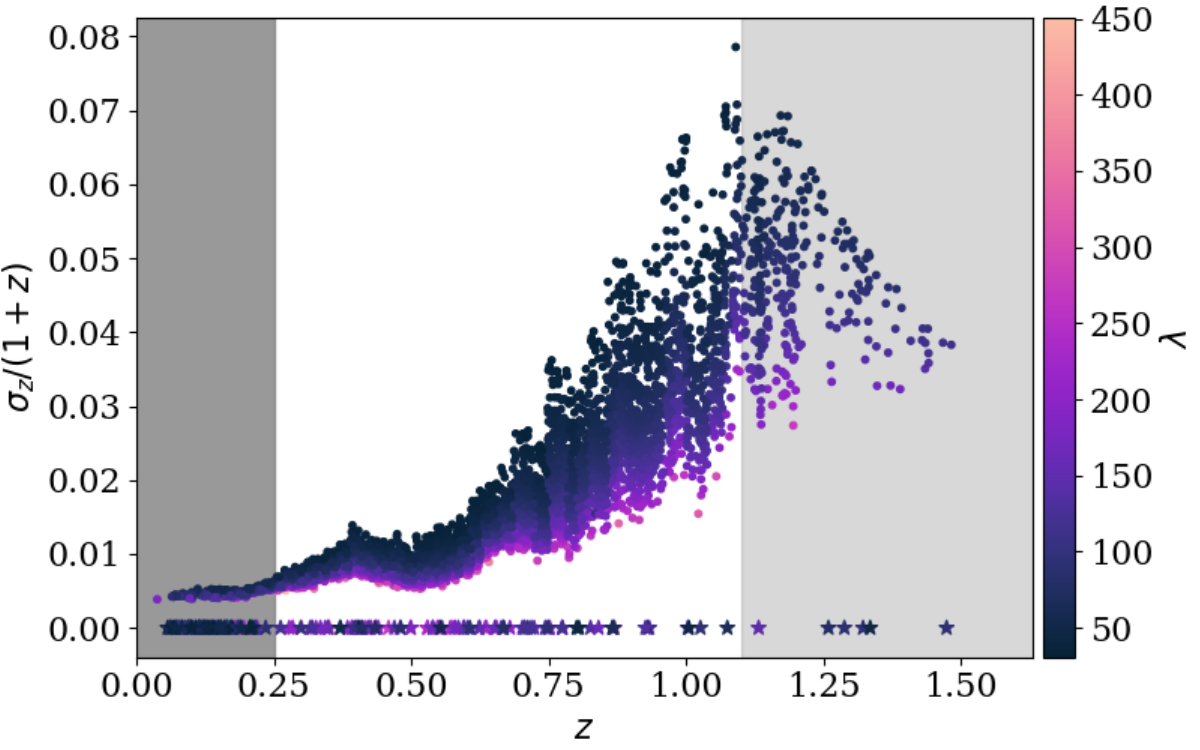}
    \caption{Distribution of $\sigma_z/(1+z)$ to $z$ of the SPT-3G cluster sample given from the MCMF redshift estimation. {Clusters with spectroscopic redshifts are denoted by stars.} The {cluster} richness {($\lambda$)} is given by the color {code}. {The dark grey shaded region indicate a portion of the sample that was not used in this analysis at all, while the light grey shaded} region indicates the portion of the sample omitted in our fiducial analysis and only used to demonstrate the potential statistical power of the ultra-high-$z$ subsample.}
    \label{fig:pz_v_z}
\end{figure}

 While the sample extends to $z\sim2$, we only use clusters with  $z < 1.1$ for the fiducial analysis.\footnote{{While our parameter inference is limited to $z < 1.1$, where clusters have reliable, well-calibrated redshift uncertainties,} we will briefly comment in Section \ref{sec:results} on the statistical power of clustering with the $1.1 < z < 1.5$ subsample, {and the prospects for} mass calibration of this high-redshift regime.} Below this redshift, the galaxy confirmation data is highly complete, and the cluster redshifts are derived primarily from optical data with small, well-controlled uncertainties (typically $\sigma_z/(1+z)\sim0.01{7}$). Above $z \sim 1.1$, cluster redshifts increasingly rely on IR data which have larger, more unreliable uncertainties ($\sigma_z/(1+z)\sim0.0{34}$). The evolution of $\sigma_z/(1+z)$ with $z$ is {illustrated} in Figure \ref{fig:pz_v_z}.

 We additionally limit this work to only clusters with $z > 0.25$ owing to the {effects of filtering during the map-making process on} the cluster selection below this redshift. We thus define three samples that we examine in this work{:} a full-$z$ sample that spans the range $0.25 < z < 1.1$ and two redshift subsamples {to test redshift based errors in our models or data and obtain halo bias measurements for multiple redshifts. Further splits are not useful at this point in time due to diluted statistical power and uninformative tests. These subsamples consist of} a low-$z$ subsample in the range of $0.25 < z < 0.67$ and a high-$z$ subsample in the range of $0.67 < z < 1.1$, with $z = 0.67$ being the median value in the full-$z$ sample. 
More details on the sample construction can be found in \citetalias{Bleem:2026}.

\section{Projected Correlation Measurement}\label{sec:Corr Estimator}

Our key observable is the two-point spatial correlation of galaxy clusters. Here this is measured using the Landy-Szalay (LS) estimator \citep{landy_szalay}{, which compares the counts of data object pairs to counts of random object pairs, both separated by a distance $r$},
\begin{equation}\label{eqn:LS_estimate}
    \xi_h(r) = \frac{DD(r) - 2DR(r) + RR(r)}{RR(r)},
\end{equation}
where $DD$, $DR$, and $RR$ are counts of data-data, data-random, and random-random pairs, respectively. {Here, the random components refers to an artificial catalog of points that occupy the same geometry as the data catalog but distributes the points in a uniformly random manner across the field. This is ensured by applying a pixel-level mask that replicates the overall footprint of the data catalog including holes due to point sources or bad data. Since the randoms catalog occupies the same mask as the data, the random-random and data-random pairs act as an explicit calibrator for mask geometry and make the LS estimator robust to spatial mask effects}. In practice, we evaluate Equation \eqref{eqn:LS_estimate} as a function of perpendicular and parallel components to the line of sight, denoted as $\rp$ and $r_\pi$, respectively. Redshift estimates from broad-band photometric datasets generally do not allow one to estimate line-of-sight distances as the redshift uncertainties translate to radial distance uncertainties of $\Delta r_\pi \gtrsim 200 \mpc/h$. However, photometric redshifts for clusters {are obtained by averaging the estimated redshifts of their satellite galaxies, and so these estimates are significantly more precise tha{n} those of the individual galaxies, and allow one to estimate line-of-sight separations between clusters. As seen in Figure \ref{fig:pz_v_z}, half our sample has a redshift uncertainty of $\sigma_{z}/(1+z) \leq 0.02$ which translates to an uncertainty in the line-of-sight position of $\sigma_{\chi} \sim 35 \mpc/h$, where $\chi$ represents comoving distance. Redshift estimates with larger uncertainties are still highly valuable, as they enable us to omit pairs with $\Delta r_\pi \gtrsim 500 \mpc/h$, which otherwise would significantly increase the noise in the clustering estimate.}

Rewriting Equation \eqref{eqn:LS_estimate} and accounting for different sample sizes between the data and randoms catalog, we get
\begin{equation}
    \xi_h(\rp, r_\pi) = 1 + \frac{N_{RR}}{N_{DD}}\frac{DD(\rp, r_\pi)}{RR(\rp, r_\pi)} - 2\frac{N_{RR}}{N_{DR}}\frac{DR(\rp, r_\pi)}{RR(\rp, r_\pi)}.
\end{equation}
The randoms catalog is generated by sampling uniformly on a sphere within the SPT-3G footprint. The uniform coverage of the SPT{-3G M}ain field means there is minimal angular variation in the completeness of the catalog. In Appendix \ref{appx:syscorr}, we explore how the clustering signal varies with several observational effects, including the SPT-3G noise levels, for which we find no statistically significant correlations. The randoms catalog is assigned redshifts by building a spline from the redshift distribution of the SPT-3G catalog and inverse sampling it to assign redshifts to the randoms catalog. This randoms catalog is generated with $>100\times$ the size of the data catalog so that the shot noise from the $RR$ and $DR$ terms is subdominant to the $DD$ one.

While we estimate the two-dimensional correlation (in $\rp$ and $r_\pi$), our final estimate is obtained by integrating this along the line of sight. This step reduces the effect of photometric redshift uncertainties as well as of redshift space distortions (RSD); the latter is the impact of line-of-sight bulk velocities of the galaxy clusters, which are correlated with large-scale structure, on the inferred redshifts \citep{Kaiser_1987_RSD,hamilton_1992, Hamilton_1998_RSD_Review}. Both components can complicate the measurement and modeling of three-dimensional correlation functions, whereas the projected correlation is far less sensitive to both effects (and therefore, to errors in the modeling of these effects). Our projected correlation is constructed as
\begin{equation}\label{eq:proj_corr}
    \wrp(\rp) = 2\int_{0}^{r_\pi,_{\rm max}}\xi_h(\rp, r_\pi)dr_{\pi}
\end{equation}
to a maximum distance of $\pimax$. We set $\pimax = 80 \mpc/h$, which we find maximizes our signal-to-noise in simulations
given our data quality ({Section} \ref{sec:sims:splits}{)}. This integration primarily helps discard pairs that are distant in $r_\pi${. Such pairs contribute} noise to the correlation estimate, since they cannot be cosmologically correlated.\footnote{In $\Lambda$CDM, the spatial correlation of the matter/galaxy field averages to zero after $r \gtrsim 200 \mpc/h$.} The choice of using a projected correlation over an angular correlation --- where the latter only considers transverse distances, not perpendicular ones --- boosts our signal-to-noise by approximately a factor of 2. We find choices of $\pimax = 100 \textrm{ and } 120 \mpc/h$ degrade our SNR by $\sim10\%$ (Section \ref{sec:data_validations}), and larger projections will only further degrade the SNR.

We measure Equation \eqref{eq:proj_corr} within 10 logarithmically spaced bins across $10\mpc/h < \rp < 100 \mpc/h$. The largest scale is set by scanning different choices and roughly identifying where our signal-to-noise plateaus\footnote{{\citet{Fumagalli_2026} finds increased constraining power by including scales up to $130\mpc/h$. However, the smaller area covered by the SPT-3G Main Survey results in increased cosmic variance at large separations and thus there is no improvement adding such scales for this work.}}, while the smallest is set by modeling criteria. Below $\rp < 10 \mpc/h$, the linear bias prescription we use is no longer viable {(}See Section \ref{sec:theory} for more details{)}. All correlations are computed using the \textsc{CorrFunc} package \citep{sinha_2020_corrfunc}. We cross-checked that we obtain the same results when using the \textsc{TreeCorr} pipeline \citep{Jarvis:2004:Treecorr} as well.

\section{Theoretical Modeling and Inference}\label{sec:theory}

We now detail our model for $\wrp$, the projected correlation function of our cluster sample. We make frequent use of Fourier-space observables and use the notation of $\kpar$ ($\kperp$) to refer to a Fourier-mode along (perpendicular to) the line-of-sight direction, and $\ktot = \sqrt{k_\parallel^2 + k_\perp^2}$ to denote the magnitude of the Fourier mode.

\subsection{Correlation from Matter Power Spectrum}\label{subsec:corr_from_mps}

The projected correlation function, $\wrp(\rp)$, is derived from the 3D halo correlation function as shown in Equation \eqref{eq:proj_corr}. This 3D correlation can be computed from the halo power spectrum as
\begin{multline}
    \xi_h(r_p, r_\pi) = \\[10pt]
    \,\, \int_0^{\infty} \frac{\kperp d\kperp}{2\pi} J_0(\kperp \rp)
    \int_{-\infty}^\infty \frac{d\kpar}{\pi} P_{h}(\ktot) \cos(\kpar r_\pi).\\
\end{multline}
Here, $J_0$ is the zero-order Bessel function, and {the halo power spectrum is defined as
\begin{align}\label{eqn:HaloPk}
    P_{h}(\kpar, \kperp) = \,\, P_{m}(\ktot) \bigg(b_h + f\frac{\kpar^2}{\ktot^2}\bigg)^2
\end{align}
where $f \equiv \ln D / \ln a$ is the logarithmic derivative of the (linear) growth rate. The second term accounts for the linear Kaiser RSD effect. All quantities are implicitly functions of redshift as well. For the scales we consider, $\rp > 10 \mpc/h$, the linear-bias prescription is adequate. This was noted in \citet{McClintock:2019:Aemulusbias} and is confirmed in our simulation validation tests (Section \ref{sec:simvalidations}). In Equation \eqref{eqn:HaloPk},} the term $P_{m}$ is the nonlinear matter power spectrum, evaluated using the HMCode interface within CAMB \citep{Lewis_Challinor_2011_CAMB, Mead_2021}.\footnote{We do not consider baryonic effects. We tested their impact using a variety of prescriptions \citep{Schneider2019Baryonification,Arico:2021:Bacco, Anbajagane:2024:Baryons} and found the effect, for the scales we consider, is less than 1\% and therefore completely negligible.} The above expression is often simplified in terms of a low-order multipole series, which alleviates the need to perform the entire line-of-sight integral. In our case, we retain the entire integral so that we can more robustly include anisotropic effects (which impact the line-of-sight clustering differently from the transverse one) such as redshift errors and RSDs.

We can directly perform the line-of-sight integration, used to convert $\xi_h(\rp, r_\pi)$ to $\wrp(\rp)$, in Fourier space as
\begin{align}
    \wrp(\rp) = &\,\, \int_0^{\infty} \frac{\kperp d\kperp}{2\pi} \bigg\{J_0(\kperp \rp) \times \nonumber\\
    & \hspace{10pt} 2\int_0^\infty \frac{d\kpar}{\pi} P_{h}(\kpar, \kperp) \frac{2\sin(\kpar \pi_{\rm max})}{\kpar} \bigg\}\\[15pt]
    = &\,\, 2\int_{0}^{\infty} \frac{\kperp d\kperp}{2\pi} J_0(\kperp \rp) \times P_\perp(\kperp).
\end{align}
where $\pimax$ sets the maximum scale in the line-of-sight integration, and the factors of 2 come from exploiting the parity-even nature of the projection integrals over $r_\pi$ and $\kpar$. {In the second line, we have relabeled the inner integral by the shorthand $P_\perp(\kperp)$ for simplicity}. We perform this inner line-of-sight integral from $0 \leq \kpar \leq 20/\pi_{\rm max}$ via Simpson integration, while the outer integral is done via FFTlog using the routines from the \texttt{Core Cosmology Library} \citep[CCL;][]{chisari_2019_ccl}.

While this projected correlation function is less sensitive to redshift uncertainties {than a fully 3D correlation}, it is still not completely immune. The photometric redshift errors from MCMF are well-approximated by a Gaussian distribution, and this distribution can therefore be characterized by its dispersion as a function of measured redshift. In practice, higher-redshift clusters have larger redshift uncertainties (due to both the absence of good calibration samples and noisier photometry). We include this redshift error in our modeling via a Gaussian smoothing in the line-of-sight correlation by modifying the power spectrum as
\begin{equation}
    P_{h}(\ktot, z) \rightarrow P_{h}(\ktot, z)\, e^{-\kpar^2\sigma_\chi^2(z)}
\end{equation}
where $\sigma_\chi(z)$ is the redshift error in units of comoving ${\rm Mpc}/h$ and as a function of redshift. 

\subsection{Modeling Halo Bias from the Mass-Observable Relation}\label{subsec:bias_model}

{The primary observable used in SPT cluster science is the tSZ detection significance, defined as the maximum SNR of a detected object across a series of filtered maps. Each filtered map is a linear combination of single-frequency SPT maps designed to maximize sensitivity to the tSZ frequency spectrum which is then spatially filtered to maximize SNR on objects with a particular angular profile \citep{Melin:2006:MMF, Bleem:2026}. The maximization of SNR over both filter scale and position in the map leads to a bias from the three degrees of freedom  \citep{Vanderlinde_2010}. Following standard convention within SPT, we adopt the notation of $\xi$ for the biased detection significance, and $\zeta$ for the unbiased one. }{The relation between the two is given as
\begin{equation} \label{eq:mass_obs2}
    P(\xisz|\zeta) = \mathcal{N}\!\left(\sqrt{\zeta^{2}+3},\,1\right).
\end{equation}
As $\xi$ also commonly refers to the two-point correlations, which are discussed at length in this work, we henceforth use $\xisz$ instead to refer to the detection significance.
 
 To forward model a halo mass into a detection significance, we utilize a set of halo-model tools \citep{Cooray:2002:HaloModel} alongside the tSZ mass-observable relation. For the latter, we use the one given in \citetalias{Bleem:2026},
\begin{align} \label{eq:mass_obs}
    \langle \ln\zeta \rangle = &\,\,\, \ln \Asz + B_{\rm SZ}\ln\left(\frac{M_{200c}}{1\times10^{14}M_{\odot} h^{-1}}\right) \nonumber\\
    & +C_{\rm SZ}\ln\left.(\frac{E(z)}{E(0.6)}\right).
\end{align}
The lognormal intrinsic scatter in $\zeta$, at fixed halo mass, has been measured to be $\sim20\%$ \citep{benson_cosmological_2013, bocquet_cluster_2024b}. 

The halo bias can then be translated from a function of mass to a function of ${\ln}\zeta$ using
\begin{equation}\label{eq: zeta_bias}
    b({\ln}\zeta,z) = \frac{1}{A}\int d M \, b(M,z)\frac{dn}{dM}P({\ln}\zeta| M,z)
\end{equation}
where $b(M,z)$ is a mass-bias model, $\frac{dn}{dM}$ is a halo-mass function, and $A = \int d M \, \frac{dn}{dM}P({\ln}\zeta| M,z)$ is the normalization coefficient. The distribution $P(\ln\zeta|M,z)$ encompasses the scaling relation
\begin{equation}
    P({\ln}\zeta|M,z) = \frac{1}{\sqrt{2\pi\sigma_{\ln \zeta}^{2}}}\exp\left[ -\frac{(\ln \zeta-\langle \ln \zeta|\ln M,z\rangle)^{2}}{2\sigma_{\ln \zeta}^{2}}\right]
\end{equation}
with $\langle \ln \zeta|\ln M,z\rangle$ given by Equation~\ref{eq:mass_obs}. 

Care must be taken when making the choice of model for $b(M,z)$. While many works in the literature utilize the simulation-calibrated model of \citet{tinker_2010}, this and other similar models perform worse for the high-mass, high-redshift regime that is required for tSZ modeling. In short, {tSZ-selected sample}s consist of the rarest objects in the Universe, and simulation-based models find it challenging to simulate large-enough volumes, at high-enough resolution, to resolve a statistically significant sample of such rare objects. We instead use the Aemulus bias emulator \citep{McClintock:2019:Aemulusbias}, which is trained on a large set of multi-resolution, large-box simulations, for our bias model. This emulator is accurate to better than 3\% across the entire mass-redshift range of {this analysis}. While we note above that simulations are often limited in resolving large samples of tSZ-detected clusters, the Last Journey simulation --- which we use extensively in our modeling and validation {and is discussed further in Section \ref{sec:simvalidations}}--- is uniquely suited for mock-tSZ analyses given its large cosmic volume while preserving high spatial resolution  \citep{Heitmann2021}.

We can now convert our estimate of $b(\zeta,z)$ to $b(\xi_{\rm SZ},z)$ through the expression
\begin{equation}
    b(\xisz,z) = \frac{1}{A} \int d\zeta \, b(\zeta,z)P(\xisz|\zeta, z)
\end{equation}
where once again $A = \int d\zeta \, P(\xisz|\zeta, z)$ is a normalization coefficient. The final quantity, $b(\xisz,z)$, is the average halo bias expected for clusters of a given detection significance and redshift, while accounting for selection effects, the mass-observable relation, and the shape of the halo mass function. This is the same formalism used in analyses of tSZ cluster abundance \citep{bocquet_cluster_2019, bocquet_cluster_2024a, bocquet_cluster_2024b}, as well as in cluster clustering in optical surveys \citep{To:2021:6x2pt, DES:2025:6x2pt_Y6}.

\subsection{Full $\wrp$ Model}\label{subsec: final theory model}

Putting the above pieces together, and now folding in the redshift distribution of the sample, our final expression for modeling the theoretical correlation function is given by

%\begin{align}
%    \wrp(\rp) = & \, \int_0^\infty \frac{d\kperp \kperp}{2\pi} J_0(\kperp \rp)  \int_0^2 dz \,\, \frac{n(z)^2}{dV/dz}\,\, \nonumber\\[10pt]
%    & \times \int_{-\infty}^{\infty} \frac{d\kpar}{2\pi}  \bigg[P(\kpar, \kperp, z)\, e^{-\kpar^2\sigma_\chi^2(z)} \frac{2\sin(\kpar \pimax)} {\kpar}\bigg] \nonumber\\[10pt]
%    & \times \bigg[\bigg(\frac{1}{\int_{0}^{200} d\xi\,\, {n(z, \xi)}}\int_{4}^{200}d\xi \,\, b(z, \xi) n(z, \xi)\bigg) \nonumber\\[10pt]
%    & + f(z)\frac{\kpar^2}{\kpar^2 + \kperp^2} \bigg]^2
%  \label{eq:final theory}
%\end{align}

 \begin{widetext}
 \begin{align}
     \wrp(\rp) = & \, \int_0^\infty \frac{d\kperp \kperp}{2\pi} J_0(\kperp \rp)  \int_0^2 dz \,\, \frac{n(z)^2}{dV/dz}\,\, \int_{-\infty}^{\infty} \frac{d\kpar}{2\pi}  \bigg\{P(\kpar, \kperp, z)\, e^{-\kpar^2\sigma_\chi^2(z)} \frac{2\sin(\kpar \pimax)} {\kpar} \nonumber\\[10pt]
     & \times \bigg[\bigg(\frac{1}{\int_{\xisz, \textrm{min}}^{\xisz, \textrm{max}} d\xisz\,\, {n(z, \xisz)}}\int_{\xisz, \textrm{min}}^{\xisz, \textrm{max}}d\xi \,\, b(z, \xisz) n(z, \xisz)\bigg) + f(z)\frac{\kpar^2}{\kpar^2 + \kperp^2}\bigg]^2 \bigg\}.
   \label{eq:final theory}
\end{align}
\end{widetext}
{where $dV/dz$ is the differential volume element.} We have now explicitly written down the redshift dependence and include the integration over the redshift kernel. The quantity $n(z, \xisz)$ represents the distribution of our tSZ cluster sample, and we use $n(z) = \int d\xisz\,\, n(z, \xisz)$ for simplicity as most calculations have no $\xisz$-dependence.

{As we make our measurements in bins of $\rp$, the correlation model must similarly be treated in terms of $\rp$ bins.} We compute the bin-averaged correlation as
\begin{equation}
    \wrp(\hat{r}_p) = \frac{1}{\pi r^2_{p, {\rm max}} - \pi r^2_{p, {\rm min}}} \int_{r_{p, {\rm min}}}^{r_{p, {\rm max}}} 2\pi \rp \wrp(\rp) d\rp.
\end{equation}
The averaging is done with annuli between the edges of a given bin. In practice, bin-averaging has minimal impact on our final predictions.

\subsection{Likelihoods and Sampling}\label{sec:theory:likelihood}

{We adopt a Gaussian likelihood following other analyses of cluster clustering \citep[e.g.,][]{lindholm_clustering_2021, To:2021:6x2pt, Lesci:2023:PlanckClustering, DES:2025:6x2pt_Y6}.} We therefore perform inference by maximizing the logarithmic likelihood: $\ln \mathcal{L} = -\frac{\chi^{2}}{2}$ where $\chi^{2}$ is given by
\begin{equation}\label{eqn:chi2}
    \chi^{2} = (D - M)^{T} \cdot C^{-1} \cdot (D - M)
\end{equation}
where $D$ are our measurements for the projected correlation function, $M$ is the theoretical model for this correlation, and $C$ is the covariance matrix discussed {below} in Section \ref{sec:theory:cov}. {We have explicitly checked the Gaussianity of our likelihood using full-sky Poisson realizations\footnote{{We use Poisson realizations, rather than a large-box simulation with cosmological clustering, as our covariance is dominated by shot noise rather than cosmic variance (Appendix \ref{appx:CovMatrix}), and because this approach allows us to generate as many Poisson realizations as we wish, whereas we only have $\mathcal{O}(10)$ cosmological realizations of the SPT-3G footprint when relying on the Last Journey simulation.}} and confirmed the $\chi^2$ of the {measurements} across 500 different {realizations} follow a $\chi^2$ distribution with 10 degrees of freedom, with a $p$-value of $p = 0.59$ as determined by the Kolmogorov-Smirnov test \citep{Kolmogorov1933, Smirnov1948}.}

The modeling pipeline, as described above, is implemented within the \textsc{Cosmosis} inference framework \citep{Zuntz_2015}. All sampling presented in this work is performed using the \textsc{Emcee} method \citep{Foreman-Mackey_2013}. Given the scope of our work, we fix all cosmology parameters during inference to values from \citetalias{Bleem:2026} except when working with simulations where we instead adopt those from \citet{Heitmann2021} which adopts the cosmology from \citet{planck2020}. Likewise, the scaling relation parameters are fixed to the best-fit values from \citetalias{Bleem:2026} unless noted otherwise. Both the cosmology and scaling relation parameters assumed for each case are listed in Table \ref{tab:cosmo}. The vast majority of our analyses will perform a one-parameter analysis of $\ln \Asz$, the amplitude of the mass--$\xisz$ observable relation, as this is a simple way to quantify the statistical precision of the measurement. Future analyses will be combined with other SPT datasets and sample over a far larger parameter space.
\begingroup
\setlength{\tabcolsep}{12pt}
\begin{table}[]
    \centering
    \begin{tabular}{c|c c}
    Parameter & SPT-3G  & Last Journey\\
    \hline\hline
    $\sigma_8$ & $0.80$ &  $0.8102$\\
    $\Omega_{\rm m}$ & $0.30$ & $0.310$\\
    $h$ & $0.7$ & $0.6766$\\
    $\Omega_{\rm b}$ & \multicolumn{2}{c}{0.046}\\
    $n_{\rm s}$ & \multicolumn{2}{c}{0.965} \\
    $\Sigma m_\nu$ & \multicolumn{2}{c}{0.06} \\
    $\ln\Asz$ & \multicolumn{2}{c}{$\mathcal{U}[\ln0.1,\ln 10]$} \\
    $\Bsz$ & \multicolumn{2}{c}{1.64} \\
    $\Csz$ & \multicolumn{2}{c}{0.91}\\
    \end{tabular}
    \caption{Adopted cosmological and scaling relation parameters. {The difference in the first three parameters originates from the choice to keep consistent with all generations of SPT cluster catalogs since \citet{bleem_galaxy_2015} when working with the data, while using the input cosmology from \citet{planck2020} when working with the Last Journey Simulations.}}
    \label{tab:cosmo}
\end{table}
\endgroup

\subsection{Covariance Matrix}\label{sec:theory:cov}

{As can be seen from  Equation \eqref{eqn:chi2}, we require the covariance matrix, C, of the projected correlation, $\wrp$.} {There are multiple approaches to estimating/modeling such a covariance, including analytic, simulation-based, and data-based jackknife/bootstrap methods.} All results presented in this work utilize an analytic covariance matrix, the model for which is described in detail in Appendix \ref{appx:CovMatrix}. In brief, we follow \citet{Fumagalli:2024:CovEuclid} and include a variety of contributions when calculating our covariance matrix. We construct these contributions in Fourier space and then translate them into $\wrp$ using the linear transformations summarized in Equation \eqref{eq:final theory}. The final covariance matrix model comprises three separate terms --- (i) the Gaussian shot noise term, (ii) the Gaussian cosmic variance term, and (iii) the leading-order non-Gaussian term. A detailed description of this model is provided in Appendix \ref{appx:CovMatrix}, alongside comparisons to simulations.

To {validate the analytic covariance}, we measure jackknife-based covariance from both the data and the simulated datasets discussed in the above sections. The leave-on-out jackknife-based matrix is estimated as
\begin{equation}\label{eq:cov}
    C_\textrm{jackknife}(\wrp^a,\wrp^b) = \frac{N_{\textrm {patch}}-1}{N_{\textrm {patch}}}\sum_{k=1}^{N_{\textrm {patch}}}(\wrp^{a,k}-\langle \wrp^{a} \rangle)(\wrp^{b,k}-\langle \wrp^{b} \rangle).
\end{equation}
where {the data are split into $N_{\textrm {patch}}$ subsets, $k$ is in the index over the patches, and }$a,b$ are indices {of a} given $\rp$-separation bin. While jackknife covariances {estimated from the data} are generally limited in their accuracy, this accuracy drops precipitously when the size of the {removed} patch is smaller than the largest correlation scale in the analysis. Our maximum separation scale is $\rp = 100 \mpc/h$ and thus we set $N_{\textrm {patch}}$ accordingly to ensure each patch is at least that scale. For a $1600 \textrm{ deg}^{2}$ field at our redshifts, this equates to $N_{\textrm {patch}} \sim 25$. As the small number of patches limits the size of our {measurement} datavector, we instead opt to use the full-sky area of the simulations for the covariance validation tests and rescale the covariance by

\begin{equation}
    C_{1600} = C_{\textrm{fullsky}} \times \frac{A_{\textrm{survey}} = 1600}{A_{\textrm{sky}} = \frac{360^{2}}{\pi}}
\end{equation}
to obtain the average jackknife covariance for a $1600 \textrm{ deg}^{2}$ patch. This allows us to use $N_{\textrm{patch}} = 100$, which is $10\times$ the length of our {measurement} datavector while comfortably remaining larger than our largest desired scale. We further discuss comparisons between our fiducial theoretical covariance and our simulation-based covariance in Appendix \ref{appx:CovMatrix:Valid}. 
{We have checked, by computing the covariance from 18 independent realizations, that we do not see any mask-based effects on our measurements that are ignored by pursuing this full-sky approach.}

\section{Simulation Validations}\label{sec:simvalidations}

{We validate the analysis and modeling pipeline through  simulations}, {the construction of which} is described in Section \ref{subsec:LJ_sims}. In the rest of this section, we describe the results from our validation tests. {The reader interested primarily in the data constraints can move to Section \ref{sec:results}.}

\subsection{Mock Catalogs from Last Journey}\label{subsec:LJ_sims}

We validate the analysis pipeline through simulations. To construct 3G-like mock catalogs, we use the lightcone halo catalog from the Last Journey simulation \citep{Heitmann2021}. The latter is a cosmological N-body simulation, performed using the Hybrid-Hardware Accelerated Cosmology Code \citep[HACC,][]{habib2016} on the Mira supercomputer, {containing} 1.24 trillion particles in a cube of side length 5.025~G{p}c, all under a \textit{Planck} best-fit cosmology \citep{planck2020}. 

Halo properties were stored at 101 time steps between redshift 0 and 10, and the continuous lightcone was built from these discrete snapshots via interpolation of the halo's most massive progenitor in the merger tree. The precise procedure is described in Section~4.3 of \cite{Heitmann2021}. The halos are identified using a friends-of-friends algorithm using linking length $b = 0.168$. In this work, we exclusively use the spherical overdensity measurements of these halos, including $\Rtwohc$ and $\Mtwohc$. The halo concentration, $\ctwohc$, is obtained by fitting measurements of density profiles with a Navarro–Frenk–White (NFW) profile \citep{NFW_1996, NFW_1997}. The full fitting procedure is described in \citet{Child_2018}.

Given the large volume of Last Journey, the lightcone covers the full sky up to $z \sim 2$.\footnote{Simulations with smaller volumes often have to replicate their boxes, by tiling them out, in order to construct a cosmological lightcone \citep{Kacprzak_2023_COSMOGRID, Omori_2024_Agora, Anbajagane_2024_ULAGAM}. Last Journey avoids any subtleties arising from such repetition.} To make use of the full sky while still retaining likeness to the SPT-3G catalog (which covers only 1600 $\deg^2$), we define 18 independent 1600 $\deg^2$ patches in the full sky and split the halo catalog accordingly. For simplicity, we do not apply the {full, }pixel-level mask from the SPT-3G data in our mock analyses {and instead use only plain rectangles for the shape of the footprint}. We confirmed this simplification makes no difference in our validations, as the estimators we use are robust to mask effects (Section \ref{sec:Corr Estimator}).

To construct mock catalogs comparable to the real catalog, we apply the redshift selection (as defined in Section \ref{sec:data}) to the input halo catalog. We assign each halo unbiased and biased SZ cluster detection significances, $\zeta$ and $\xisz$ respectively, using Equation~\ref{eq:mass_obs}  and Equation~\ref{eq:mass_obs2} alongside the true halo mass provided in the catalog. We use the best-fit values for the scaling relation parameters from \citetalias{Bleem:2026} as input to the above procedure. To achieve a similar $\xisz$ distribution, cluster density, and selection as the SPT-3G catalog, the mock catalogs are cut to only contain halos with $\xisz > 4$. 
A comparison of the $z$ and $\xisz$ distributions of the real and mock catalogs is shown in Figure \ref{fig:z_xi_dists}.

\begin{figure}
    \centering
    \includegraphics[width=\linewidth]{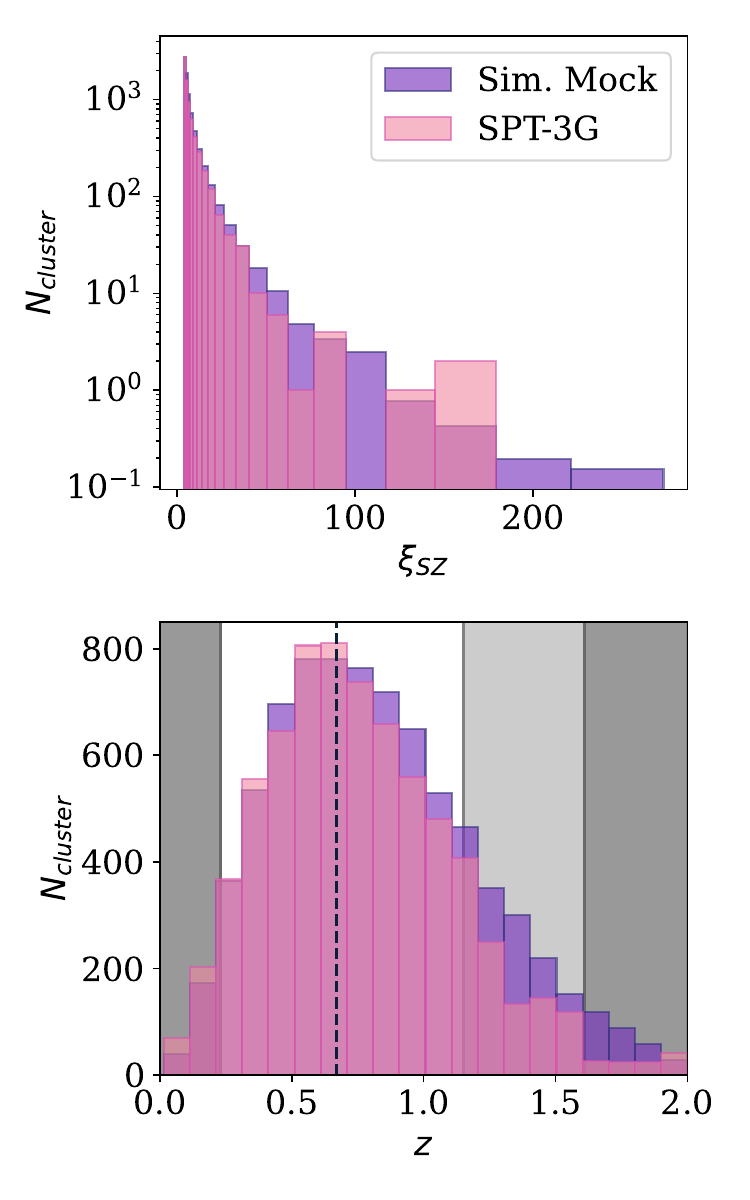}
    \caption{Comparison of $\xisz$ (top) and redshift (bottom) distributions between the mean of the simulated mock catalogs (purple) and the true SPT-3G Main field cluster catalog described in \citetalias{Bleem:2026} (pink). Dark grey shaded regions indicate portions of the samples that aren't used in this analysis while the light grey shaded region indicate the portion only used for the ultra-zigh-$z$ investigation described in {Section \ref{sec:results}}.}
    \label{fig:z_xi_dists}
\end{figure}

\subsection{Parameter Recovery in the Ideal Case}\label{subsec:No_photoz}

We first verify that the analysis pipeline can recover the input mass-scaling relation parameters in the idealized case of no redshift uncertainties. We measure the correlation functions in each of the 18 mock catalogs and use their mean as the datavector for this test. Averaging over the mocks suppresses statistical fluctuations in the measured correlation functions by a factor of $\sqrt{18} \approx 4$, enabling a more precise test for biases in parameter recovery. {Importantly, we do not correspondingly reduce the covariance used in the likelihood: the covariance matrix remains that of the fiducial analysis, corresponding to a single $1600\deg^2$ field. Thus, the averaging is used only to reduce realization noise in this validation test, rather than to emulate a survey with 18 times the fiducial area.}

The significance of the clustering signal can be quantified by evaluating Equation \eqref{eqn:chi2} but with the model M set to zero, i.e. {$\textrm{SNR} \equiv \sqrt{\chi^{2}} = \sqrt{d^T C^{-1} d}$}. We calculate this for each of the 18 mock measurements and find ${\rm SNR} = 13.3 \pm 1.2$ across the samples. Using the model described in Section \ref{sec:theory}, we also obtain a best-fit theoretical prediction for each of the 18 measurements, which gives an average probability-to-exceed (PTE) of $p = 0.431$. This both shows our model is a good fit to the data and that the covariance matrix accurately represents the noise in the measurement. We find $\ln \Asz = 0.60^{+0.16}_{-0.17}$. This gives a difference from the input value of $\ln \Asz = 0.62$ of $\Delta\ln \Asz = 0.02$, which is $0.14\sigma$. In the ideal case, we can recover the parameters adequately.

\subsection{Modeling of Photometric Redshift Errors}

As discussed in Section \ref{sec:data}, we use photometric redshift estimates with uncertainties of $\sigma_{\rm pz} \approx 0.013(1+z)$ on average. Errors of this level will affect the measured correlation by scattering the clusters in the line-of-sight direction which suppresses clustering on small scales and smears out the BAO peak, {boosting clustering on intermediate scales} \citep{Estrada:2009:Clustering}. While we minimize our sensitivity to redshift error modeling by using the projected correlation function (compared to the full 3D correlation), it is still necessary to model the effect. We validate our $\sigma_{z}$ correction described in Section \ref{sec:theory} by including redshift uncertainty into the mock catalogs as $z_{\rm eff} = z_{\rm true} + \mathcal{N}(0, \sigma_{\rm pz})$. In particular, we randomly assign each simulated halo a $\sigma_{\rm pz}$ corresponding to a random SPT-3G cluster at a similar redshift to the mock halo. The redshift slices used in this query are $z_{\rm halo} \pm 0.02$.

To test the robustness of this correction under different measurement choices, we measured the correlation for the modified catalogs for values of $\pimax = [60, 80, 100, 120] \mpc/h$. We find that our model remains a good fit to the data with {$p$-values in the range $0.1 < p < 0.6$} in all cases. The measured $\wrp$ for the fiducial $\ln \Asz$ and $\pimax = 80\mpc/h$ case is shown in Figure \ref{fig:sim_3_bin_wp}. We find $\ln \Asz = 0.61^{+0.22}_{-0.17}$ in this case, indicating a{n} increase of $17.3\%$ in $\sigma_{\ln \Asz}$ due to redshift uncertainties. We recover the input parameters within $0.07\sigma$, validating our model in the presence of realistic photometric errors. This finding remains true at all $\pimax$ values we tested, as well as with several alternate input values of $\ln \Asz$.

\subsection{Parameter Recovery under Catalog Splits and Model Variations}\label{sec:sims:splits}

We further test the robustness of our measurement and modeling pipeline by sub-dividing the mock catalogs into four different subsamples. The first two subsamples are low- and high-redshift halves split at the sample's median redshift of $z = 0.67$ while the last two samples are low- and high-$\xisz$ halves split at $\xisz = 5.5$. These splits check the robustness of our modeling across the properties of our sample, rather than for just the average sample. The results of these splits are shown in Figure \ref{fig:sim_validations}. All splits are within $0.3\sigma$ of the input values. The accurate recovery also indicates good accuracy in the chosen HMF and halo bias models. Such models often fail at higher redshifts, where the presence of massive halos exponentially diminishes. However, we can accurately recover our input parameter in all regimes spanned by our catalog.

We also explicitly check the impact of swapping between different prescriptions for the halo bias and the HMF. As a reminder, we relied on the Aemulus model from \citet{McClintock:2019:Aemulusbias} as they achieve better than 3\% accuracy across the unique high-mass/high-redshift range spanned by tSZ clusters. In Figure \ref{fig:sim_validations}, we perform parameter inferences on the simulated {measurement}, but using alternative choices for the halo bias or HMF. We find that different choices of the HMF cause no appreciable change in the recovered $\ln \Asz$. This is somewhat expected as any mass-independent, or nearly mass-independent, fractional errors in the HMF cancel in Equation \eqref{eq: zeta_bias} as the HMF appears in both numerator and denominator. Along those lines, we expect the choice of bias model to be significantly more relevant, and indeed, we see using the model from \citet[][henceforth T10]{tinker_2010} results in a shift of $1\sigma$ from the input {whereas using the emulator from M19 gives a shift of $<0.3\sigma$}. This is consistent with our expectations given the known accuracy of the T10 bias model --- the uncertainty at high mass is 10\% or higher in many cases, whereas our measurement constrains the bias to $5\%$ or better. The {difficulty} of the calibration of the linear halo bias model for galaxy clusters has similarly been noted by \cite{seppi_srgerosita_2024}, who opted to use a Halo Occupation Distribution (HOD) model, and \cite{Euclid_2024_Bias}, who calibrated their own semi-analytic model.  

\begin{figure*}
    \centering
    \includegraphics[width=1\linewidth]{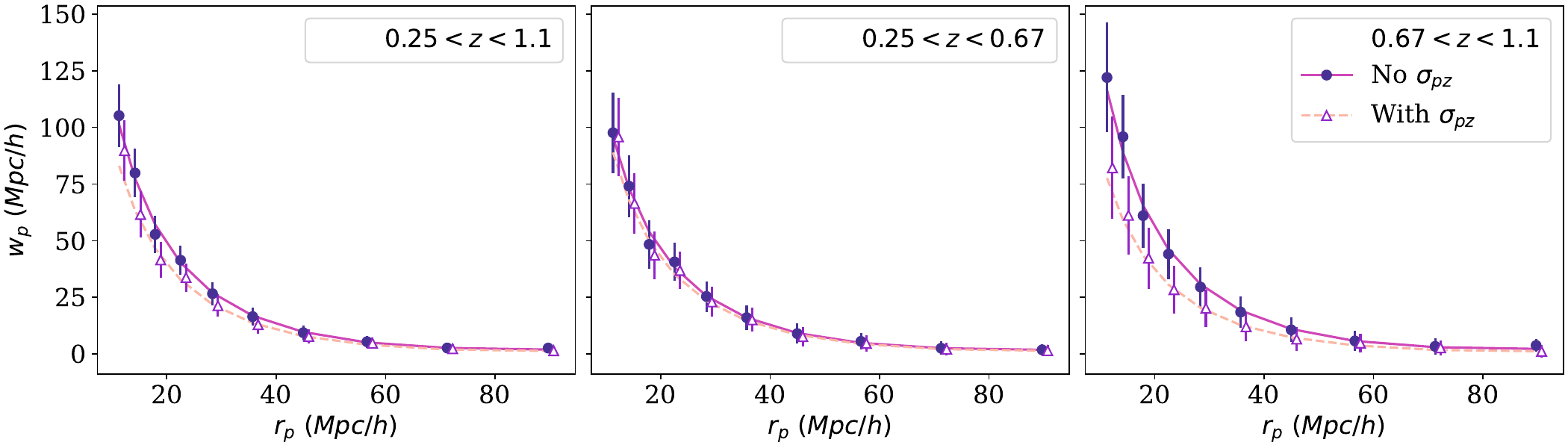}
    \caption{The average projected correlation across the 18 mock catalogs for the full-$z$ (left), low-$z$ (center), and high-$z$ (right) samples. Solid lines {and} circles show the fiducial {model and} measurements{, respectively,} without photometric redshift errors while dashed lines and triangles show the {model and} measurements in the presence of realistic 3G-like photometric redshift errors. There is a clear suppression of correlation strength caused by the redshift errors, and is more prominent at higher-$z$ where these errors have larger amplitudes. {Analogous measurements from the data are shown in Figure \ref{fig:wp_results}.}}
    \label{fig:sim_3_bin_wp}
\end{figure*}

\subsection{Impact of Cluster Concentration on Bias and Parameter Recovery}

Mass and redshift, while the most prominent cluster properties, are not the only ones that determine the halo bias of a sample. Halos can cluster differently depending on their local environment and the nature of their ``assembly.'' Such effects, deemed ``assembly bias,'' are known to alter the effective halo bias of a given cluster sample \citep{Wechsler_2006,  Wu_Rozo_Wechsler_2008, baxter_constraining_2016}. The halo concentration is a key property that can trace assembly bias. This is particularly relevant for a cluster catalog, where the key observable ($\xisz$ in our case) is {thought to be} correlated with halo concentration {due to its impact on the cluster potential \citep{Komatsu_Seljak_2001}}. {However, measurements of the concentration of clusters are notoriously difficult, especially for the high-redshift samples that dominate tSZ-selected catalogs. Currently, models for the SZ-concentration relation have only been tested on hydrodynamical simulations which have found that} a selection on an SZ observable, at fixed halo mass, inherently selects objects that are more concentrated at that mass \citep{Baxter:2024:Concentration}. Thus, the predicted halo bias (using just mass and redshift) may not be adequate to represent the selected sample. 

As described in Section \ref{subsec:LJ_sims}, the Last Journey simulation provides halo concentrations, $\ctwohc$, for all halos. This allows us to use our mock catalogs to measure the concentration dependence of the cluster sample's bias. To do so, we now assign $\zeta$ significances to our mock catalog using an updated model. For a halo, $i$, with mass, $M_i$, and at redshift, $z_i$, 
\begin{equation}\label{eqn:SZ_assembybias}
    \zeta_{\rm mod}(M_i, z_i) = \zeta_{\rm orig}(M_i, z_i)\times \bigg(\frac{c_i}{\langle c|M_i, z_i\rangle}\bigg)^\beta,
\end{equation}
where $c_i$ is the concentration estimate of halo $i$. In practice, Equation \eqref{eqn:SZ_assembybias} is ensuring our model retains its mean relation, $\langle \zeta | M\rangle$, but now lets the scatter in $\zeta$ correlate with the concentration. Following \cite{Baxter:2024:Concentration}, we use a value of $\beta = 0.1$ as measured in their simulated samples, but also mimic a much more extreme correlation, $\beta = 0.3$ for robustness. We then proceed with the same selection, $\xisz > 4$, as before.

The impact of assembly bias on recovered $\ln \Asz$ is shown in Figure \ref{fig:sim_validations}. Inducing a concentration-correlated scatter, and therefore a concentration-dependent selection, has no visible impact on the recovered $\ln \Asz$ value. This is consistent with the results of \citet{Baxter:2024:Concentration}, who found the predictions of cluster bias change by $\lesssim 1\%$ for a wide range of choices. Therefore, the SPT {tSZ-selected} clusters are robust to assembly bias effects.

\begin{figure}
    \centering
    \includegraphics[width=\linewidth]{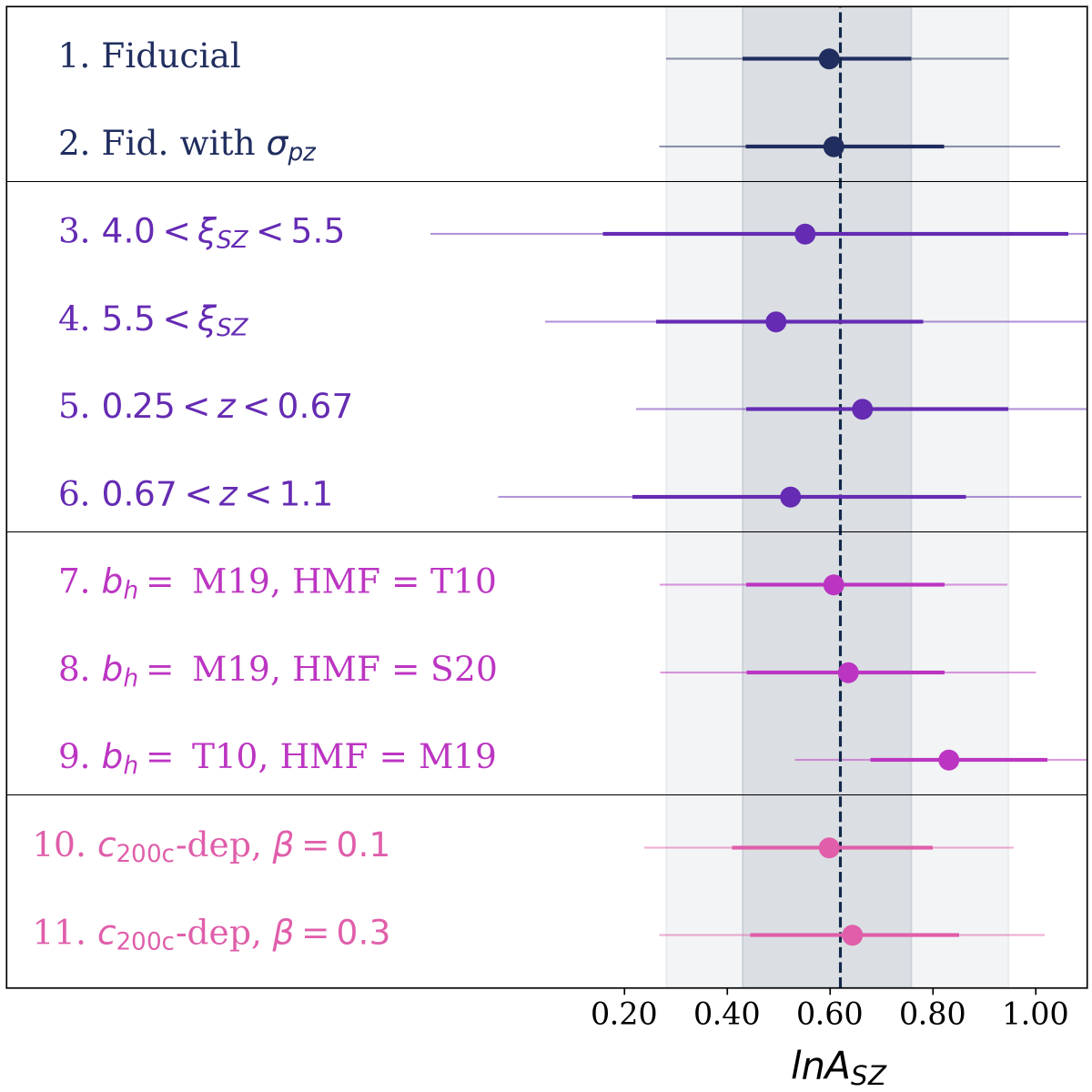}
    \caption{$\ln \Asz$ estimated from the average $\wrp$, estimated over the 18 independent mock catalogs, under various analysis setups. Shaded region shows the $1\sigma \textrm{ and } 2\sigma$ uncertainties of the fiducial mock measurement, while the thick {and} thin lines show the same for the other points. The dashed line indicates the input value of $\ln \Asz = 0.62$.}
    \label{fig:sim_validations}
\end{figure}

\section{SPT-3G Catalog Data Splits and Validations}\label{sec:data_validations}

Our checks thus far have validated the measurement and modeling pipelines of our work. However, when turning to the SPT-3G data, we also want to perform a set of null tests to ensure there are no hidden systematics in the data affecting our measurement and inference. We perform a variety of data splits in a blinded fashion and compare {two metrics: (a) the $p$-value of the residuals between data and the best-fit model being consistent with noise fluctuations, and (b)} the difference in $\ln \Asz$ constraints {between pairs of subsamples}. For the latter metric, {w}e can estimate the significance of parameter differences between two subsamples by normalizing the difference with the uncertainty,
\begin{align}
    \sigma_{\rm joint} & = \sqrt{\sigma_{A}^{2} + {\sigma_{B}^{2}}},  \nonumber\\
\end{align}
which is simply adding the uncertainties of the individual splits in quadrature. This is possible because our measurement is shot-noise dominated (see Appendix \ref{appx:CovMatrix}) and so two subsamples with no common clusters have uncorrelated measurements and therefore uncorrelated uncertainties. The upper and lower uncertainties of our constraints are generally symmetric, so we average them to obtain a single uncertainty per constraint and use it in the above expression.

In Figure {\ref{fig:validation_splits}} and Table {\ref{tab:data_validations}}, we show the constraints on $\ln\Asz$ as well as the
best-fit model $\chi^2$ and PTE for different splits and assumptions in our analysis of the SPT-3G
data set, which we will describe and discuss throughout this section. 

\subsection{{Measurement} Blinding}\label{sec:data_validations:blinding}

Throughout our data validations, we used a blinded version of our {measurement}. This was done using the methodology of \citet{Muir:2020:Blinding}. Briefly, we generate two theoretical measurements --- one at a fiducial value of the input parameters, and another at a value slightly offset from the fiducial. Then, the {measurement} can be blinded as,
\begin{equation}
    \wrp^{\rm data, blind} = \wrp^{\rm data} + (\wrp^{\rm th, offset} - \wrp^{\rm th, fid})
\end{equation}
Any inference performed on $\wrp^{\rm data, blind}$ will return parameter values that are offset by roughly the difference between the fiducial and offset parameter sets. If we randomly generate an offset parameter value, then we can blind the {measurement}.

For our work, we generate this offset using the $\ln \Asz$ parameter. We use a fiducial value of $\ln \Asz = \ln 2$ and randomly draw a different value to use in the blinding from a range $\ln \Asz = [0, \ln 3]$, which spans roughly the $2.5\sigma$ region of our expected constraints. We have extensively validated our blinding procedure on the mock catalogs of Section \ref{subsec:LJ_sims} and confirmed we can accurately recover the expected offset parameter value. 

For all measurements to follow, we performed the tests on the blinded {measurement}. However, for consistency, we present the inference made after unblinding. Our {criterion} for unblinding was $0.0015 < p < 0.9985$ consistency ($3\sigma$) between a given data split and its complement.

\subsection{$\xisz$ Splits}

{Our cluster catalog has high purity, and we expect any contamination to be localized to low-$\xisz$ objects.} In order to {check for effects from  contamination}, we perform a split in $\xisz$. Similar to the analogous simulation validation test, we select the split point to be $\xisz = 5.5$, {because above this significance the sample} is $\gg  99\%$ pure (\citetalias{Bleem:2026}). Thus, we choose $4 < \xisz < 5.5$ and $\xisz \geq 5.5$ as our two bins. There are $2799$ and $2673$ clusters in the low- and high-$\xisz$ samples, respectively. The model is well-fit with $p = 0.7{3}$ and $p= 0.{47}$ for the high- and low-$\xisz$ sample, respectively. The difference between the two best-fits is $\Delta_{\ln\Asz} = 0.{20}$ and is consistent at $0.{4}\sigma$. We find no indication that cluster purity is affecting our measurement.

\subsection{Redshift Splits}\label{subsec:data_z_val}

Similarly, the performance of the cluster catalog varies with redshift. The optical confirmation of galaxy clusters is less certain at higher redshifts, causing the catalog to become less pure and more incomplete. We once again follow the established choices made in our simulation-based validations, and split the sample into a low-$z$ ($0.25 \leq z \leq 0.67$) and a high-$z$ ($0.67 \leq z \leq 1.1$) subsample. 

The results of this split are shown as points 4 and 5 in Figure \ref{fig:validation_splits}. The model fits the low- and high-$z$ sample at $p = 0.005$ and $p = 0.94$. In both cases, the residuals are consistent with noise within $3\sigma$, and so we consider the model fit to be adequate. However, given the PTE values are close to our threshold for consistency, we performed an extensive set of checks. This includes confirming the performance of these redshift splits in simulations (Section \ref{sec:sims:splits}), the accuracy of our covariance model across a range of different redshift slices (Appendix \ref{appx:CovMatrix:Valid}), the lack of statistically significant contamination from stellar density or galactic extinction in the cluster clustering estimates (Appendix \ref{appx:syscorr}), etc. {As an additional, data-driven check of any mismatch in the covariance model's redshift dependence,} we subdivided the catalog into more redshift bins, and found the {PTE values} in the higher-$z$ bins did not asymptote to $p = 1$ and instead scattered around $p = 0.5$ as expected. {Thus, the $p$-values above are not pushed to the threshold due to some redshift-dependent bias in the covariance model.}

Thus, we consider the model to be adequately fitting these data. Focusing on the inferred values, we see the difference is $\Delta_{\ln\Asz} = {0.70}$ which corresponds to ${1.24}\sigma$. Once again, this is within our $3\sigma$ {criterion} and the test is considered to have passed.

\begin{figure}
    \centering
    \includegraphics[width=1\linewidth]{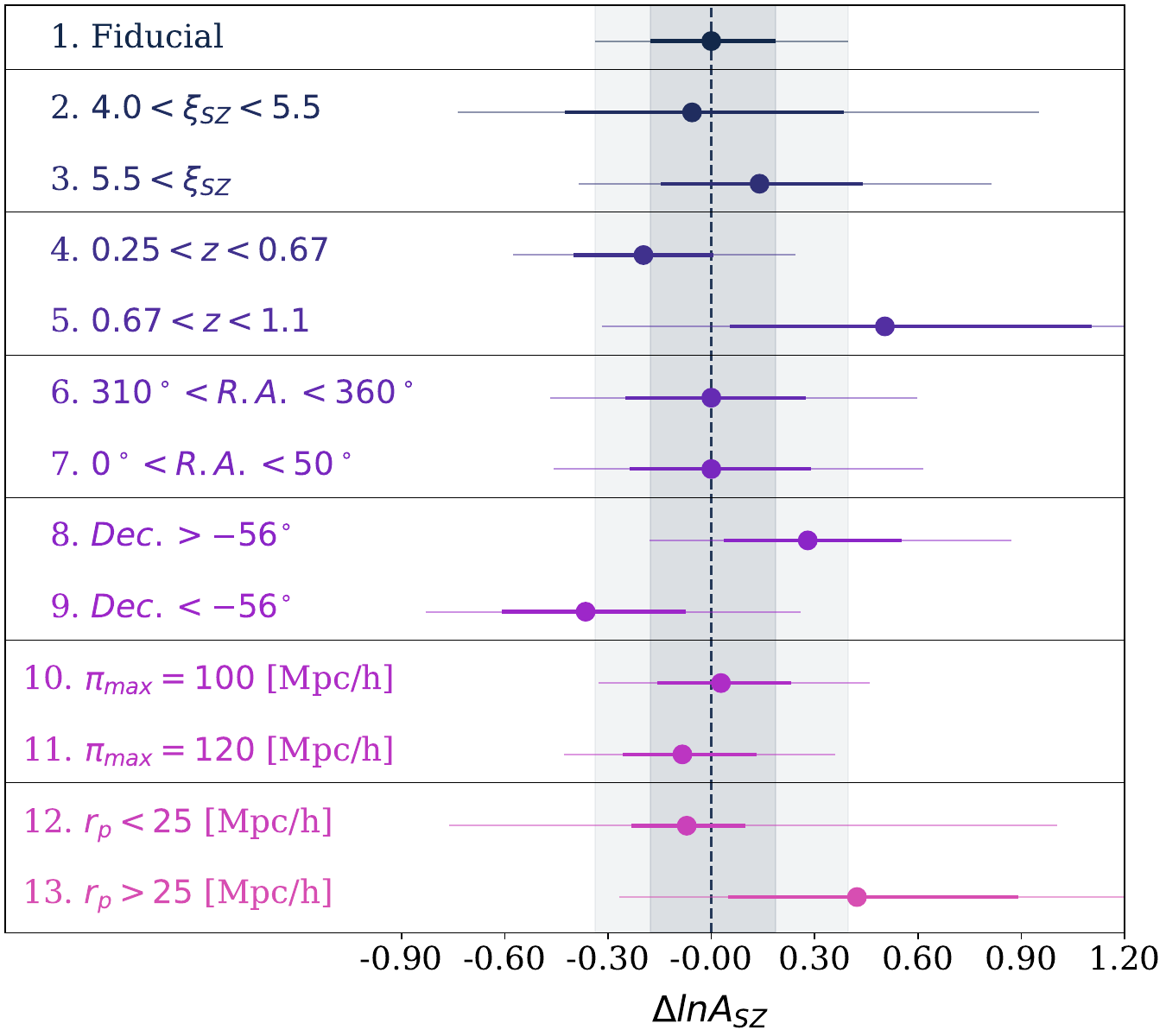}
    \caption{Difference in inferred $\ln{\Asz}$ from the fiducial analysis of the SPT-3G full-$z$ sample to those from the data variations used as part of our validation procedures; see Section \ref{sec:data_validations}. The constraints are accompanied by the $1\sigma$ and $2\sigma$ uncertainties.}
    \label{fig:validation_splits}
\end{figure}

\subsection{Sky Splits}
In clustering analyses, spatially localized effects can be particularly harmful by contaminating the measurement with non-cosmological correlations. To check this, we split the data into two halves based on \RA or \DEC 

SPT-3G observations are made by scanning the sky along lines of constant declination with fixed scanning speed in the \RA direction. As a result of the scanning pattern, the noise within the SPT-3G maps varies with \DEC and is fairly constant in \RA Due to this, a split in \RA is effectively a ``null test,'' two samples that should be statistically identical in the absence of extended structure or contaminants in one half of the map. To test this, we split the field at the center \RA  of the 3G-{M}ain footprint, which is at $\RA = 0^{\circ}$. The results of this split are shown as points 6 and 7 in Figure \ref{fig:validation_splits}. The models are good fits to the measurements, with $p > 0.05$, and the resulting constraints are consistent at better than $1\sigma$.

The other spatial split of relevance is in the \DEC direction due to the aforementioned \DEC dependence of SPT-3G's noise. We split the 3G-Main footprint at the central declination value of ${\rm \DEC} = -56^{\circ}$ and perform the comparison test. The results of this split are shown as points 8 and 9 in Figure \ref{fig:validation_splits}. {Once again, the measurements are sufficiently well-fit by the models, with $p = 0.008$ and $p = 0.73$ for the high- and low-declination patches, respectively. The difference in the resulting constraints has a significance of $1.7\sigma$. All probabilities are below our $3\sigma$ threshold, so the test is considered to have passed. We note that while it is below our threshold, this test does return the largest difference in constraints of our tests. However, as discussed in Appendix \ref{appx:syscorr}, we have found that spatial variations in the noise (where the variations are maximally $10\%$) do not induce a statistically significant spatial variation in the cluster sample. Taking this into account, we assume the declination dependence shown here to be a statistical fluctuation and do not consider it further in this work.}

\subsection{$\pimax$ Checks}

Our simulation checks found the optimal $\pimax$ value, given our specific survey configuration and redshift data quality, to be $80 \mpc/h$ as this maximizes our SNR. Our pipeline explicitly models the line-of-sight correlations including RSDs and photo-$z$ error contributions. In order to test our sensitivity to mismodeling of these terms, or to other observational systematics, we now perform parameter inference on measurements made using larger values of $\pimax$. In general, integrating over larger lines of sight suppresses any observational and modeling subtleties for the line-of-sight spatial correlations; in the limit of infinite projection, we obtain an angular correlation function which is essentially insensitive to such effects. Therefore, consistency in parameter inference across different $\pimax$ indicates robustness in our modeling of such line-of-sight effects.

We choose $\pimax = 100 \mpc/h$ and $\pimax = 120 \mpc/h$ as variations of our fiducial choice of $\pimax = 80 \mpc/h$. {In both cases, the data are adequately fit with $p = 0.18$ and $p = 0.35$ and} yield shifts of $\Delta_{\ln\Asz} = 0.027$ and $-0.084$, respectively, relative to the fiducial result. These correspond to significances of only $0.10\sigma$ and $0.31\sigma$, respectively. There is no clear evidence for effects missing from our modeling of the line-of-sight correlations.

\subsection{$\rp$ Scale Checks}

As described in Section \ref{sec:theory}, our analysis uses $\rp$ scales that can be reliably modeled without invoking non-linear bias and halo exclusion ingredients to the modeling. We check the impact of the smallest scales on our parameter inference by splitting our {measurement} into a small-scale ($\rp < 25 \mpc/h$) and large-scale ($\rp > 25 \mpc/h$) subset and comparing the subsequent parameter results. The results are shown as points 12 and 13 in Figure \ref{fig:validation_splits} and listed in Table \ref{tab:data_validations}. We find the $\Delta_{\ln\Asz} = {0.49}$ between the small and large scale measurements corresponds to a difference of ${1.09}\sigma$. Both measurements are well-fit with $p > 0.5$. {Therefore, non-linear effects do not appear to cause a statistically significant bias in our analysis}, and the modeling can be considered accurate across the full range of scales we consider here.

\begin{table}[]
\centering
\renewcommand{\arraystretch}{1.5}
\begin{tabular}{|c|c|c|c|c|}
\hline
Data Split & $\ln \Asz$ & $\Delta \ln \Asz / \sigma_{\rm joint}$ & $\chi^{2}$ & PTE \\[5pt]
\hline

$4 < \xisz < 5.5$ 
& $0.33^{+0.44}_{-0.37}$ 
& \multirow{2}{*}{$0.38\sigma$}
& 6.07/9 & 0.73 \\

$\xisz > 5.5$ 
& $0.52^{+0.30}_{-0.28}$ 
&
& 8.69/9 & 0.47 \\

\hline

$0.25 < z < 0.67 $ 
& $0.19^{+0.20}_{-0.20}$ 
& \multirow{2}{*}{$1.25\sigma$}
& 23.41/9 & 0.005 \\

$0.67 < z < 1.1 $ 
& $0.89^{+0.60}_{-0.45}$ 
&
& 3.44/9 & 0.94 \\

\hline

$310^\circ < \RA < 360^\circ $ 
& $0.38^{+0.28}_{-0.25}$ 
& \multirow{2}{*}{$0.002\sigma$}
& 16.08/9 & 0.07 \\

$0^\circ < \RA < 50^\circ $ 
& $0.38^{+0.29}_{-0.24}$ 
&
& 14.35/9 & 0.11 \\

\hline

$\DEC > -56^\circ $ 
& $0.66^{+0.27}_{-0.24}$ 
& \multirow{2}{*}{$1.74\sigma$}
& 22.40/9 & 0.008 \\

$\DEC < -56^\circ $ 
& $0.02^{+0.29}_{-0.24}$ 
&
& 6.08/9 & 0.73 \\

\hline

$\pimax = 100\mpc/h$ 
& $0.41^{+0.20}_{-0.18}$ 
& \multirow{2}{*}{---}
& 10.06/9 & 0.35 \\

$\pimax = 120\mpc/h$ 
& $0.30^{+0.22}_{-0.17}$ 
&
& 12.63/9 & 0.18 \\

\hline

$\rp < 25 \mpc/h$ 
& $0.31^{+0.17}_{-0.16}$ 
& \multirow{2}{*}{$1.11\sigma$}
& 4.30/9 & 0.89 \\

$\rp > 25 \mpc/h$ 
& $0.81^{+0.47}_{-0.37}$ 
&
& 11.43/9 & 0.73 \\

\hline
\end{tabular}

\caption{Inferred values of $\ln \Asz$ from the various data validation checks on the SPT-3G sample, the significance of the difference between the paired subsamples of a given category, and the associated goodness-of-fit statistics $\chi^{2}$ with the corresponding PTE evaluated for 9 degrees of freedom. We do not quote a difference significance for the $\pi_{\rm max}$ tests as the latter are not subsample/measurement splits in the other comparisons.}
\label{tab:data_validations}
\end{table}

\section{Clustering of tSZ Clusters}\label{sec:results}

After performing the data validation tests of Section \ref{sec:data_validations}, which all pass, we now present our unblinded results from the projected correlation of clusters in SPT-3G.

\subsection{Constraints on the Mass-Observable Relation}
\begin{figure*}
    \centering
    \includegraphics[width=1\linewidth]{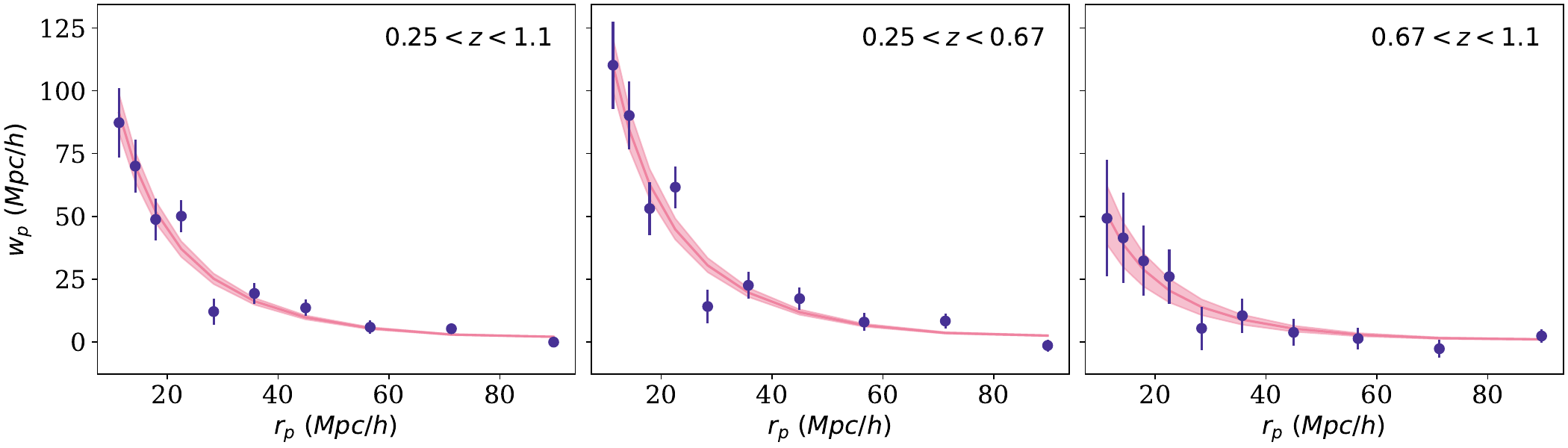}
    \caption{Projected correlation functions of the full-$z$ (left), low-$z$ (center), and high-$z$ (right) SPT-3G Main survey cluster samples. The solid line denotes the best-fit model with the $1\sigma$ uncertainty given by the shaded region.}
    \label{fig:wp_results}
\end{figure*}

We first show the unblinded $\wrp$ measurements in Figure \ref{fig:wp_results}. We detect clustering at high significance in the full-$z$ sample at ${\rm SNR} = 12.{8}${, which places this detection at $>2\rm\times$ the significance of the prior best tSZ-selected cluster clustering measurement given in \citet{Lesci:2023:PlanckClustering}}. Our best-fit model for this sample has an associated PTE of $p = 0.03{5}$, indicating that our data is {statistically consistent with} our model. From this best fit, we obtain a constraint on $\ln \Asz$ of $\ln \Asz = {0.43^{+0.17}_{-0.21}}$. 

We additionally detect clustering in the low- and high-$
z$ subsamples. The low-$z$ sample on its own gives ${\rm SNR} = 11.{9}$ while clustering in the high-$z$ sample is detected at lower significance (SNR $ = 4.{7}$) due to the significantly lower clustering amplitude, caused in part by the larger redshift errors. {While both subsamples have a similar number of clusters,} the uncertainties in the measurement from the  high-$z$ subsample are larger due to its larger comoving volume, resulting in a lower 3D number density and therefore, a larger measurement uncertainty. 

As discussed in Section \ref{sec:data_validations}, the best-fit model fits the low- and high-$z$ sample at $p = 0.005$ and $p = 0.94$ which are both consistent with our criteria of $3\sigma$ consistency. We also jointly analyze our high- and low-$z$ samples by concatenating the two $\wrp$ measurements and modeling a covariance matrix for this joint {measurement} {that includes any cross-covariance between the two}. {Note that our datavector contains both} two-point auto-correlations and ignores the one two-point cross-correlation {between the low- and high-$z$ redshift bin}; the latter is essentially zero as the redshift overlap between the two subsamples is negligible by sample definition. From the low- and high-$z$ joint analysis we obtain $\ln \Asz = {0.42^{+0.19}_{-0.21}}$. This result is consistent within $0.{04}\sigma$ to the constraint from the full-$z$ sample. 

Both of our inferred values for $\ln \Asz$ are also consistent with the abundance-estimated value from \citetalias{Bleem:2026}. While the latter has high precision ($\sigma_{\ln \Asz} = 0.02$) when fixing cosmology, we discuss in Section \ref{sec:results:discussion} the significant opportunities enabled by the addition of these cluster clustering measurements to existing tSZ cluster analyses.

We also measure the clustering for the ultra-high-$z$ subsample, although we do not fit it to a model due to the uncertainty of the $\sigma_z$ distribution at these redshifts. Additionally, due to the inclusion of $\sigma_z$ in the theoretical covariance model, we opt to use the jackknife-based covariance given by Equation~\ref{eq:cov} instead. While this method is still biased, it does not require any $\sigma_z$ modeling. Additionally, at the minimum redshift of this subsample of $z = 1.1$, the distance spanned by the angular area of the field is much larger, allowing for $\textrm{N}_{\textrm{patch}} = 100$. Given this covariance method, we measure clustering for this subsample at SNR $ = 3.7$. However, we stress that this is only a tentative SNR due to the biased nature of this covariance and as we have not validated the high-$z$ sample for clustering-related systematics.

\subsection{Halo Bias Constraints}

\begin{figure}
    \centering
    \includegraphics[width=1\linewidth]{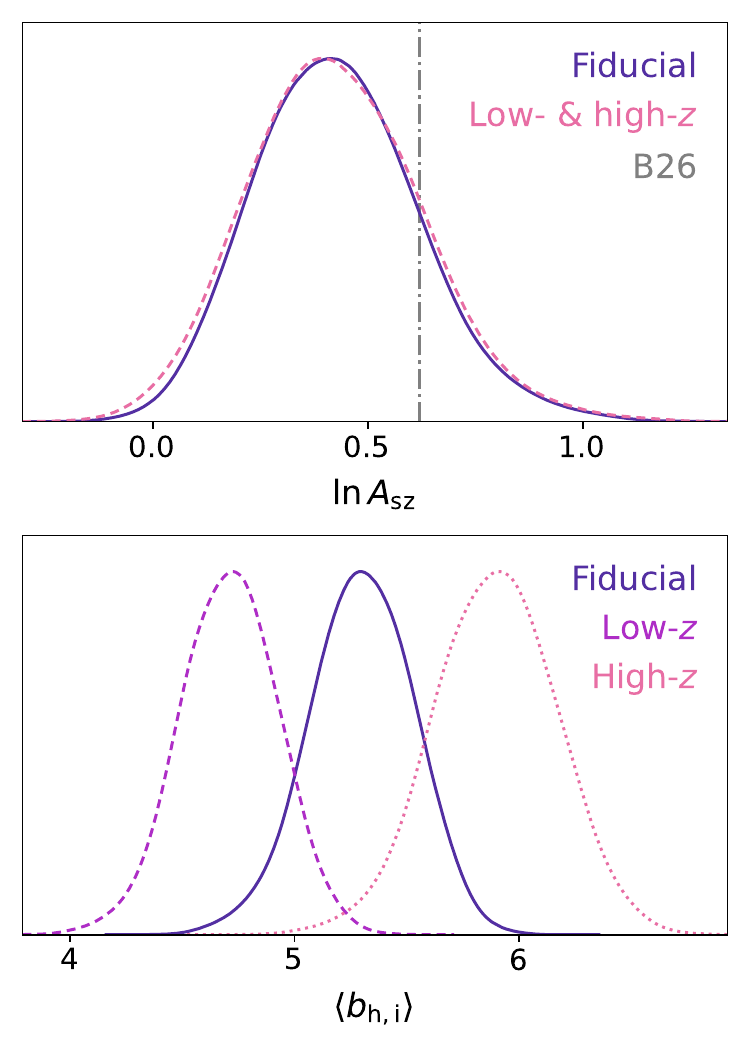}
    \caption{Top: the posteriors of $\ln \Asz$ from analyzing the full-$z$ sample (purple) and from jointly analyzing the low-$z$ and high-$z$ samples (pink). Bottom: corresponding posteriors of the average halo bias, $b_{\textrm{h}}$, for the full-$z$ sample (purple), and for the low-$z$ (dashed pink) and high-$z$ (dotted pink) bin in our two-bin joint analysis. {Note that it is expected for the halo bias posteriors to differ due to the different halo populations of each sample}.}
    \label{fig:ASZ_BIAS_results}
\end{figure}

\begin{figure*}
    \centering
    \includegraphics[width=0.9\linewidth]{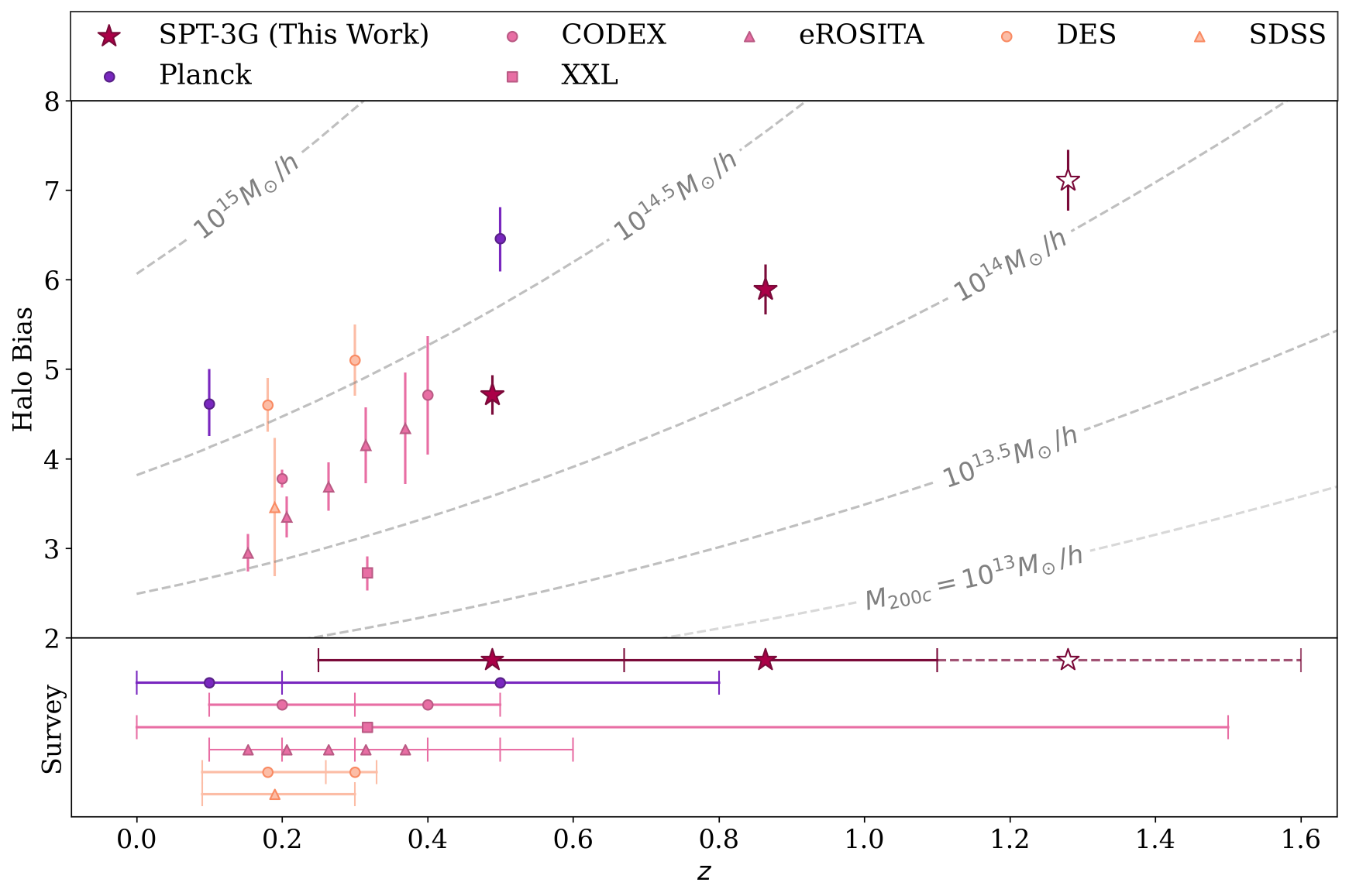}
    \caption{Comparison between the halo bias of the SPT-3G Main survey clusters (red stars) with those from other reported measurements of cluster samples. \footnote{{Some clustering measurements were not included as they did not report a halo bias constraint. In particular, \citet[][SDSS]{Fumagalli_2024} and \citet[][KiDs]{Romanello_2024} measured cluster clustering but did not report the bias measurement. \citet[][KiDs]{Ingoglia_2022} reported halo bias measurements from stacked weak lensing but were excluded as we limit our scope to only measurements from clustering.}} Other tSZ cluster samples are shown in purple, with \textit{Planck} denoted by circles \citep{Lesci:2023:PlanckClustering}. X-ray cluster samples are shown in pink, with CODEX \citep{lindholm_clustering_2021} marked by circles, XXL \citep{marulli_xxl_2018} marked with squares, and eROSITA \citep{seppi_srgerosita_2024} marked with triangles. Optical cluster samples are shown in peach, with DES \citep{baxter_constraining_2016} marked by circles and SDSS \citep{Estrada:2009:Clustering} marked with triangles. The redshift ranges of each sample are shown in the bottom panel, with caps indicating the start and end of each bin. Note that only eROSITA has overlapping redshift bins. We also predict the bias of SPT clusters with $1.1 < z < 1.5$ --- which is not used in our main analysis --- and show it as a hollow star and a dashed line. This prediction is done using the fiducial $\ln \Asz$ posterior in Figure \ref{fig:ASZ_BIAS_results}.}
    \label{fig:bias_comp}
\end{figure*}

Our constraints on $\ln\Asz$ can be translated into constraints on the halo bias at different redshifts. We present these in Figure \ref{fig:ASZ_BIAS_results} for the full-$z$ sample, as well as the low-$z$ and high-$z$ subsamples, previously presented in Figure \ref{fig:wp_results}. 

The full-$z$ sample gives an inferred halo bias of $b_{\textrm{h}} = {5.30\pm{0.24}}$, which corresponds to a $4.5\%$ accuracy in our bias measurement. In the $z$-subsamples, the inferred halo bias increases from $b_{\textrm{h}} = {4.71\pm 0.22}$ for the low-$z$ subsample to $b_{\textrm{h}} = {5.89\pm 0.28}$ for the high-$z$ subsample. In both subsamples, we infer the bias to $\sim 4.7\%$ accuracy.  

{In Figure \ref{fig:bias_comp}, we put these bias constraints into context by comparing them to other halo bias constraints for cluster samples measured through clustering. From this, two observations readily become clear:
\begin{itemize}
    \item This is the tightest measurement of halo bias from a tSZ-selected cluster sample.
    \item These are the best constraints of halo bias from a high-redshift ($\langle z\rangle > 0.6$) cluster (sub)sample selected from data of any wavelength.
\end{itemize}
}
{High-redshift measurements are particularly important for measuring the redshift evolution of $\Omega_{\rm m}$ and $\sigma_8$ but such measurements have typically been limited only to lower-bias objects such as galaxies and quasars \citep[e.g.,][]{Wang2018BOSSTomography,Ruggeri2019eBOSS, Tanidis2023BOSSHarmonic}.}

We also predict the halo bias for an ``ultra-high-$z$'' sample, which contains clusters with $1.1<z<1.5$ that are excluded from our main analysis due to unreliability in photometric redshift estimates. {We stress that we do not fit these measurements and instead are simply predicting the bias of this sample given its distribution of $\xisz$ and $z$ (the $\sigma_z$ is not required as we are predicting the bias and not $\wrp$)} This prediction is done by fixing $\ln \Asz$ to the joint low- and high-$z$ best-fit value of $\ln \Asz = {0.42^{+0.19}_{-0.21}}$. The resulting predicted halo bias for the ultra-high-$z$ subsample is found to be $b_{\textrm{h}} = {7.42\pm 0.35}$, making it a sample with one of the highest halo bias values to date. As previously mentioned, the highest-redshift tSZ clusters are the rarest collapsed structures in the Universe, and as a result have the highest clustering bias. With proper modeling enabled by more reliable redshifts, such as those made possible using the infrared photometric filters of the \textit{Euclid} space mission \citep{EUCLID_2025}, the inclusion of the high-bias and high-redshift regime in our inference will serve as a powerful lever arm for constraining redshift-dependent signals in cosmology and in the mass-observable scaling relations.

\subsection{Discussion}\label{sec:results:discussion}

\begin{figure*}
    \centering
    \includegraphics[width=2\columnwidth]{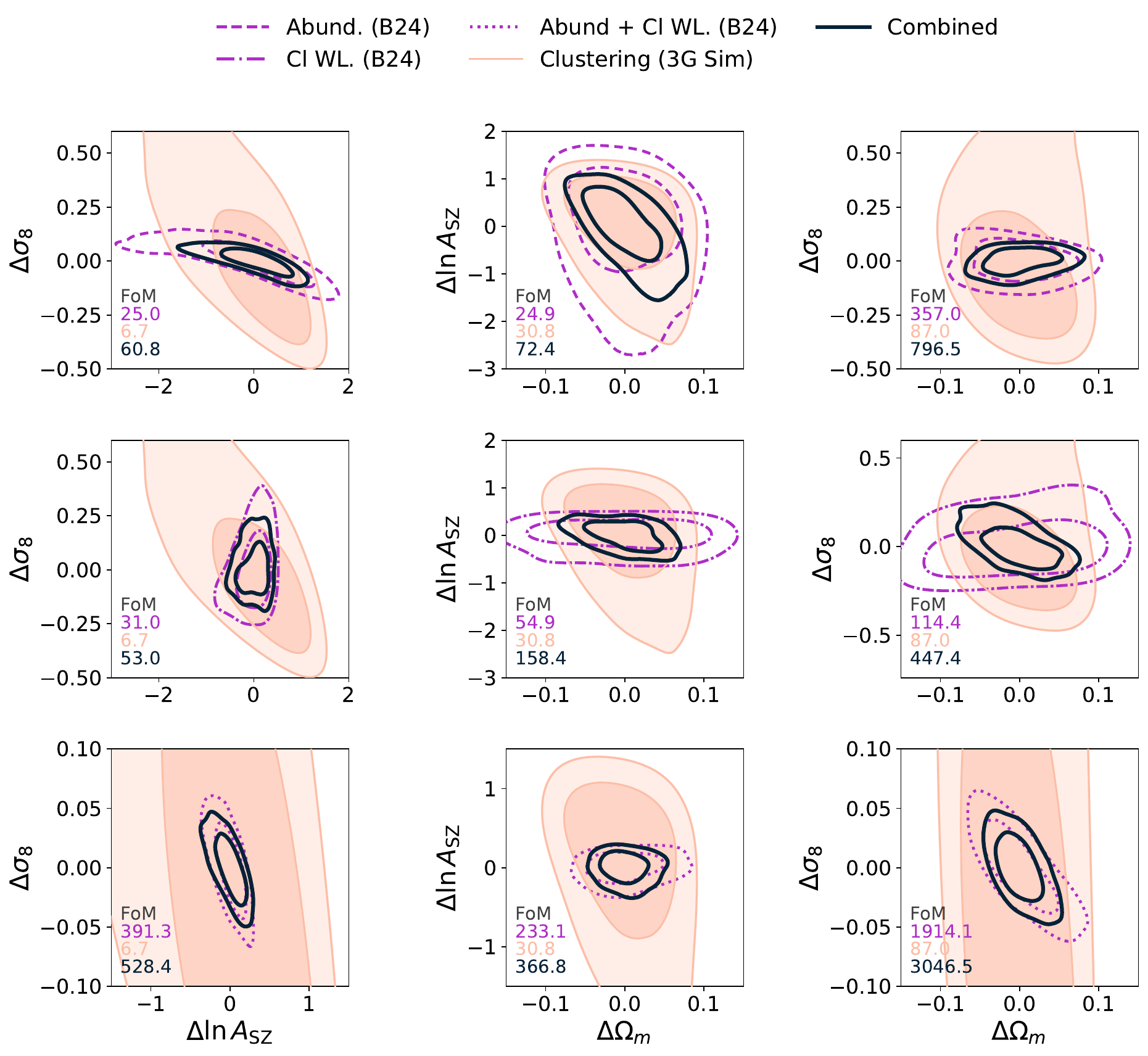}
    \caption{{Simulated constraints from cluster clustering (peach contours) compared to {constraints from cluster abundance (top) and cluster WL mass calibration (middle) from \citet{bocquet_cluster_2024b}}, as well as the combination of all three (bottom). {We use ``Cl WL.'' to abbreviate Cluster weak lensing.} The FoM is given by $1/\det[C]$ where $C$ is the parameter covariance of any two sets of parameters. The addition of cluster clustering provides a significant gain in {all three} cases. This comes about because cluster clustering provides competitive constraints on $\Omega_{\rm m}$ while also generally helping alleviate high-dimensional degeneracies.}}
    \label{fig:abund_vs_cluster}
\end{figure*}

We have thus far discussed the contribution of cluster clustering to a one-parameter analysis of $\ln \Asz$. However, this is a limited case that is useful as a demonstration of the measurement's utility, but is not representative of the different advantages provided by the clustering measurement. In general, we find three key advantages that cluster clustering provides to tSZ-based cluster analyses.

{To demonstrate these complementarities, we present in Figure \ref{fig:abund_vs_cluster} the latest available abundance constraints of \citet{bocquet_cluster_2024b} alongside a simulated analysis of 3G clustering using the Last Journey measurement.\footnote{{Constraints or forecasts from the 3G cluster abundance data (with WL mass calibration) were not available at the time of this work. The abundance-only constraints are prior-dominated and therefore identical across B24 and its 3G variant. The lensing-based constraints will improve notably when using the latter, though the precise improvement has not yet been quantified.}} We combine the chains by importance sampling at the likelihood level. This simpler combination --- compared to analyzing the data jointly --- is possible because the cross-covariance between cluster abundance and cluster clustering is negligible \citep{To:2021:6x2pt, Fumagalli_2026}. Thus the likelihoods can be treated as being independent. We importance sample the chains from \citet{bocquet_cluster_2024b} using the likelihood of our 3G cluster chain,
\begin{equation}
    P_{\rm comb}(\vec{\theta}) = P_{\rm B24}(\vec{\theta}) \times [P_{\rm clus,sim}(\vec{\theta}) / \pi(\vec{\theta})],
\end{equation}
where $\vec{\theta}$ are the parameters in question and $\pi(\vec{\theta})$ is the prior {used in the simulated clustering analysis. These are uniform priors with edges given by $\Asz \in [0.01, 6.0]$, $\Bsz \in [1, 2]$, $\Csz \in [0, 2]$, $\Omega_{\rm m} \in [0.1, 0.5]$, $\sigma_8 \in [0.6, 1.0]$, and $h \in [0.6, 0.8]$. We divide by this prior probability to convert $P_{\rm clus,sim}$ into the cluster clustering likelihood and ensure the priors are not double-counted in our estimate of $P_{\rm comb}(\vec{\theta})$ \citep{Taylor:2024:CombineHarvestor}.} We approximate the function $P_{\rm clus,sim}(\vec{\theta})$ by running a kernel-density estimate (KDE) on the $\approx 10^5$ samples generated during our MCMC sampling. We have resampled the KDE and checked that it accurately reproduces the contours of the original chain. The combined contours in Figure \ref{fig:abund_vs_cluster} are obtained through this technique. 
}

We now discuss the advantages of the clustering measurement. First, the parameter $\ln \Asz$ (as well as the other associated scaling-relation parameters) have direct degeneracies with cosmological parameters. This has been well-document{ed} in abundance-only {analyses} \citep{bocquet_cluster_2019, bocquet_cluster_2024b}. {Henceforth we will refer to the fiducial abundance-only analysis, which uses weak lensing (WL) measurements to calibrate the cluster masses, as ``Abund. {+ Cl} WL'' and the analysis without WL as {just} ``Abund.''}. The clustering-only result also has degeneracies, as $\ln \Asz$ controls the mean halo bias of our cluster sample, and constraints on the bias are directly degenerate with $\sigma_8$ as the amplitude of clustering scales as $(b_{\rm h} \sigma_8)^2$. While both abundance-only and clustering-only results have degeneracies with cosmology, the direction of these degeneracies is different across the two, and so their combination improves precision {\citep{Fumagalli_2024}}. {The top row of Figure \ref{fig:abund_vs_cluster} demonstrates this for the SPT analyses.} Therefore, while the abundance-only constraints deliver precise constraints on $\ln \Asz$ in a fixed-cosmology analysis, we require additional information to constrain both $\ln \Asz$ and cosmology. The clustering measurements provide one such synergistic addition.

Second, clustering analyses are more sensitive to $\Omega_{\rm m}$ (relative to abundance-only analyses) due to the dependence of our two-point correlation functions on the shape of the matter power spectrum, the redshift--distance relation, etc. Thus, cluster clustering can outperform abundance-only analyses when constraining $\Omega_{\rm m}$ {\citep{Fumagalli_2024, Zhan_2026}}. The addition of clustering{ to the ``Abund + Cl WL'' analysis} enables better constraints on all parameters as a result. {The bottom row of Figure \ref{fig:abund_vs_cluster} shows that the {figure of merit} (FoM) increases by 25\% to 30\% in various 2D parameter projections.} A similar synergy has been taken advantage of in galaxy surveys using the combination of weak lensing and galaxy clustering; the latter plays the same role as cluster clustering here, by constraining $\Omega_{\rm m}$ more precisely than the other probes in the analysis \citep{DESY3, More:2026:HSCY3_3x2pt, DESY6}.

Finally, and closely related to the above two points, the clustering measurement is vital in pursuing a millimeter-wave cluster analysis that is completely independent of galaxy weak lensing. {Establishing such a lensing-independent analysis is important for two reasons. First, it would allow tSZ cluster measurements to provide a completely independent cross-check of cosmological constraints from the primary CMB and galaxy weak-lensing surveys, including of any residual differences among these probes.} Second, cluster analyses of all wavelengths --- from X-ray \citep{Grandis:2024:eROSITAMass,Kleinebreil:2025:eROSITAMass}, to optical \citep{McClintock:2019:DESMass}, to tSZ \citep{bocquet_cluster_2019,bocquet_cluster_2024b} --- {currently utilize weak lensing to calibrate the masses of their cluster samples. Future analyses will rely on the same Rubin and \textit{Euclid} \citep{EUCLID_2025} lensing samples for weak-lensing mass calibration as well, introducing shared data and systematic uncertainties across otherwise distinct cluster samples. An alternative calibration route is therefore useful to preserve the independence and redundancy of these probes.} By leveraging cluster clustering, together with other {mass calibration} techniques such as CMB lensing \citep{Baxter:2018:DESY1CMBlensing,Raghunathan:2019:SPTpolMassCalibration,Ansarinejad:2024:SPT3GMassCalibration}, we can instead perform a tSZ cluster analysis that is independent of galaxy lensing. Clustering provides a particularly apt replacement for weak lensing, as it not only helps break degeneracies between cosmology and the mass--observable relation, but also provides additional constraining power on $\Omega_{\rm m}$.

While we focus our discussion above on constraining $\Lambda$CDM parameters, the clustering measurements can provide unique opportunities for beyond-$\Lambda$CDM science as well. In particular, it is a unique measurement of the spatial clustering of biased tracers, both because the sample has one of the highest average halo biases (Figure \ref{fig:bias_comp}), and because the data is insensitive to spatial systematics (Figure \ref{fig:num_dense}) that otherwise significantly impact traditional galaxy datasets \citep{rodriguez-monroy_dark_2022}. The former is possible given the high number density of high-$z$ clusters in our sample, while the latter is enabled by the homogeneous noise properties of observations made from the South Pole and the simplicity of tSZ cluster detection. As an example, such properties make clustering measurements particularly interesting for measuring signatures of primordial non-Gaussianity (PNG) from the early Universe \citep[e.g.,][]{Dalal:2008:PNGs, Anbajagane:2026:Colliders, Anbajagane:2026:ColliderBias} and may even be used to constrain axion dark matter models \citep{London_2026}.

In Section \ref{sec:conclusions:future} below, we detail prospects for improving our measurements and SNR using existing and upcoming datasets.

\section{Conclusions}\label{sec:conclusions}

\subsection{Summary}\label{sec:conclusions:summary}

Galaxy clusters are uniquely sensitive to the origin and evolution of structure in our Universe. While their abundances have been beneficial in aiding our understanding of the Universe, their spatial clustering is also highly informative. This has been demonstrated in many analyses focusing on the astrophysical nature of clusters and some on their cosmological nature as well \citep{To:2021:6x2pt, DES:2025:6x2pt_Y6}. In this work, we measure cluster clustering using the tSZ-selected cluster catalog from SPT-3G, which has the highest number-density to date of any tSZ-selected cluster catalog. We build the pipelines both for optimally measuring this clustering from data and for analyzing the measurement to constrain cosmology and the mass-observable relation parameters. We provide a demonstrative analysis focusing on the latter, but comment on the former, particularly in its synergy with tSZ cluster abundance analyses. Our model incorporates a wide variety of effects, including redshift uncertainties, RSDs, mass-observable scalings, etc., and we demonstrate it can precisely recover input parameters from simulated datasets and {measurements}.

In addition to validating our pipelines on simulations, we perform a set of validations on blinded measurements (Section \ref{sec:data_validations}) and upon passing all tests, we unblind the constraints from the SPT-3G cluster clustering measurements. The main results of our work are:
\begin{itemize}
    \item We detect cluster clustering at 12.8$\sigma$ across $0.25 < z < 1.1$. This is  to date the highest-significance detection of spatial clustering in a {tSZ-selected} sample.
    
    \item For cluster redshifts with precision of $\sigma_z / (1 + z) \lesssim 0.015$, it is advantageous --- from a signal-to-noise perspective --- to utilize the projected correlation function over the angular one, as the former discards pairs that are far apart (i.e., uncorrelated) in their line-of-sight distance but are close in transverse distance. We find the former estimate improves our signal-to-noise by a factor of two over the latter.

    \item {With an inferred halo bias of $b_{\textrm{h}} = {5.30\pm{0.24}}$ for the full-$z$ sample,} we measure the tightest constraints of halo bias to date of any cluster sample with $\langle z\rangle > 0.5$, and also the tightest constraints on halo bias of any tSZ-selected sample. 

    \item {We place a constraint on the amplitude of the mass-observable scaling relation in a fixed cosmology of $\ln \Asz = {0.43^{+0.17}_{-0.21}}$.} 

    \item Using simulation-derived {measurements}, we show the synergy between cluster clustering and cluster abundances. When considering the FoM in different projected posteriors, we find including clustering improves on the latest abundance-based constraints from SPT by 30\%. When dropping WL mass calibration from the latter, the improvement increases to a factor of 2 to 3.
\end{itemize}

\subsection{Future Outlook}\label{sec:conclusions:future}

This work provides a pathfinder analysis for cluster clustering with tSZ-selected clusters. The current results are limited in three areas, each with their own prospects for overcoming them. We now detail these, with an eye towards what can be achieved in the near future.

The first and biggest limitation is the number density of the cluster sample. While the high bias of clusters, and the precision of cluster photometric redshifts (compared to galaxy redshifts), helps make up for the lower number density, the measurement is still limited by shot noise and will benefit from having more objects. While future catalogs from SPT's next-generation camera, SPT-3G+, will eventually provide such an improvement in the tSZ cluster sample \citep{Natoli_2026}, we can easily remedy this in the near term by cross-correlating the cluster data with a different tracer that is less {shot noise} limited, like an optical galaxy sample. Such a cross-correlation will significantly reduce the effective shot noise while still retaining the benefits of the high bias and the clean selection of the SPT-3G SZ clusters. 

Furthermore, the tSZ cluster sample could also be cross-correlated with weak gravitational lensing, which will further help increase the SNR of our cluster-based correlations and break parameter degeneracies. Such a combination of clustering measurements has already been successfully demonstrated in \citet{To:2021:6x2pt} for optically selected clusters and is now a standard practice in optical surveys \citep{DES:2025:6x2pt_Y6}. We can pursue the same approach but using a tSZ-selected cluster sample rather than an optically selected one, and while using a projected correlation function for the cluster clustering to boost our SNR.

Repeating such a measurement with a tSZ-selected cluster sample is promising for several reasons. One being that the cluster clustering measurement presented in this work is a remarkably clean estimate of spatial correlations. The SPT-3G main survey is very homogeneous, with noise levels consistent to $10\%$ across the survey area and a footprint that largely avoids observing any extended, contaminating structure from the Milky Way. Additionally, the SPT-3G cluster detection --- being determined by the SZ effect --- is significantly less impacted by systematics like stellar density and dust extinction (Appendix \ref{appx:syscorr}). These systematics commonly impact optical detection of extragalactic objects and complicate the interpretation of their spatial correlations on large angular scales \citep{rodriguez-monroy_dark_2022}. As a result, tSZ-selected clusters enable a measurement of clustering that is highly insensitive to the dominant systematics in galaxy clustering analyses.

The second limitation in our analysis comes from the small area spanned by the sample, which increases our measurement uncertainties. These uncertainties scale as the {inverse square root} sky-fraction of the survey, and are also particularly relevant for precisely measuring correlations on the largest $\rp$ scales, as such correlations are more significantly affected by cosmic variance rather than {shot noise} (Appendix \ref{appx:CovMatrix}). While the latter point is less relevant for SZ {mass calibration}\footnote{The precision of our $\ln \Asz$ constraints are informed predominantly by our small-scale measurements, as they have far higher signal-to-noise.} these scales are particularly useful for cosmological analyses focusing on modified gravity \citep{Ferreira:2019:MG}, primordial non-Gaussianities \citep{Dalal:2008:PNGs, Anbajagane:2026:ColliderBias}, etc. For this reason, the upcoming cluster catalog from the SPT-3G Extended-10k, a combination of the Main, Summer, and Wide surveys, will be particularly fruitful for this type of analysis. With $10,000\textrm{ deg}^{2}$, the extended survey has $>6\times$ more area than that spanned by the catalog used in this work. The combined sample will add thousands of additional clusters to the analysis and improve precision significantly. Additionally, the future cluster sample from the Simon's Observatory is projected to contain $\sim 33,000$ clusters over $60\%$ of the sky \citep{Abitbol_2025} and will be a good candidate for clustering measurements.   

Finally, as discussed in Section \ref{sec:results}, we are limited by the availability of accurate and precision photometric redshifts on the high-$z$ end; in particular, above $z \gtrsim 1.1$ where common spectral features used for redshift estimation no longer lie in wavelength bands covered by DES and rely largely on redshifts from the \textit{Spitzer}/IRAC \citep{Fazio_2004_IRAC} and WISE \citep{Wright_2010_WISE, Meisner_2022_WISE} surveys. Improved redshifts from near-infrared filters in \textit{Euclid} \citep{EUCLID_2025} and LSST \citep{LSST_2018} will enable precision analyses with our highest redshift and highest bias cluster sample and give us a powerful lever arm for the high-redshift {mass calibration}s and for cosmology.

The prospects for cluster clustering with {tSZ-selected} clusters will only continue to grow from here, with the wide-field datasets from SPT and with the availability of next-generation optical datasets for calibration. This will unlock new science opportunities, using the statistics of the rarest halos that can form, and provide a unique glimpse at the nature of structure in our Universe.

\section*{Acknowledgments}

We thank Gillian Beltz-Mohrmann, Chun-Hao To, and the late Eric Baxter for their consultation and support during this project. We thank Katrin Heitmann and Steve Rangel for their work on the development of the Last Journey simulation which we made extensive use of in this work. DA acknowledges that support for this work was provided by NASA through the NASA Hubble Fellowship grant \#HST-HF2-51596.001 awarded by the Space Telescope Science Institute, which is operated by the
Association of Universities for Research in Astronomy, Inc., for NASA, under contract
NAS5-26555.

%SPT
The South Pole Telescope program is supported by the National Science Foundation (NSF) through awards OPP-1852617 and OPP-2332483. Partial support is also provided by the Kavli Institute of Cosmological Physics at the University of Chicago.  
%Argonne
Work at Argonne National Laboratory, NERSC, and
ALCF was supported by the U.S. Department of Energy, Office of High Energy Physics. Argonne, a U.S. Department of Energy Office of Science Laboratory, is operated by UChicago Argonne, LLC under Contract No. DE-AC02-06CH11357. 
% Davis
The UC Davis group acknowledges support from Michael and Ester Vaida. 
% Fermilab
This document was prepared by the SPT-3G collaboration using the resources of the Fermi National Accelerator Laboratory (Fermilab), a U.S. Department of Energy, Office of Science, Office of High Energy Physics HEP User Facility. Fermilab is managed by Fermi Forward Discovery Group, LLC, acting under Contract No. 89243024CSC000002.
% Melbourne
The Melbourne authors acknowledge support from the Australian Research Council’s Discovery Project scheme (No. DP260100705). 
% Paris
The Paris group has received funding from the European Research Council (ERC) under the European Union’s Horizon 2020 research and innovation program (grant agreement No 101001897), and funding from the Centre National d’Etudes Spatiales. 
% SLAC
The SLAC group is supported in part by the Department of Energy at SLAC National Accelerator Laboratory, under contract DE-AC02-76SF00515.

This work made use of the Core Cosmology Library software. This software can be found at \href{https://github.com/LSSTDESC/CCL}{https://github.com/LSSTDESC/CCL}.

This work additionally made use of the CosmoSIS framework which can be found at \href{https://cosmosis.readthedocs.io/}{https://cosmosis.readthedocs.io/}.

%\nocite{*}

\newpage

\bibliographystyle{aasjournal}
\bibliography{clustering.bib}% Produces the bibliography via BibTeX.

@ARTICLE{Baxter:2018:DESY1CMBlensing,
       author = {{Baxter}, E.~J. and {Raghunathan}, S. and {Crawford}, T.~M. and
                 {Fosalba}, P. and {Hou}, Z. and {Holder}, G.~P. and
                 {Omori}, Y. and {Patil}, S. and {Rozo}, E. and
                 {Abbott}, T.~M.~C. and others},
        title = "{A measurement of CMB cluster lensing with SPT and DES year 1 data}",
      journal = {\mnras},
         year = 2018,
        month = may,
       volume = {476},
       number = {2},
        pages = {2674-2688},
          doi = {10.1093/mnras/sty305},
archivePrefix = {arXiv},
       eprint = {1708.01360},
 primaryClass = {astro-ph.CO},
       adsurl = {https://ui.adsabs.harvard.edu/abs/2018MNRAS.476.2674B}
}

@ARTICLE{Raghunathan:2019:SPTpolMassCalibration,
       author = {{Raghunathan}, S. and {Patil}, S. and {Baxter}, E. and
                 {Benson}, B.~A. and {Bleem}, L.~E. and {Chou}, T.~L. and
                 {Crawford}, T.~M. and {Holder}, G.~P. and {McClintock}, T. and
                 {Reichardt}, C.~L. and {Rozo}, E. and {Varga}, T.~N. and others},
        title = "{Mass Calibration of Optically Selected DES Clusters Using a Measurement of CMB-Cluster Lensing with SPTpol Data}",
      journal = {\apj},
         year = 2019,
        month = feb,
       volume = {872},
       number = {2},
          eid = {170},
        pages = {170},
          doi = {10.3847/1538-4357/ab01ca},
archivePrefix = {arXiv},
       eprint = {1810.10998},
 primaryClass = {astro-ph.CO},
       adsurl = {https://ui.adsabs.harvard.edu/abs/2019ApJ...872..170R}
}

@ARTICLE{Ansarinejad:2024:SPT3GMassCalibration,
       author = {{Ansarinejad}, B. and {Raghunathan}, S. and {Abbott}, T.~M.~C. and
                 {Ade}, P.~A.~R. and {Aguena}, M. and {Alves}, O. and
                 {Anderson}, A.~J. and {Andrade-Oliveira}, F. and
                 {Archipley}, M. and {Balkenhol}, L. and {Benabed}, K. and
                 {Bender}, A.~N. and {Benson}, B.~A. and {Bertin}, E. and
                 {Bianchini}, F. and {Bleem}, L.~E. and others},
        title = "{Mass calibration of DES Year-3 clusters via SPT-3G CMB cluster lensing}",
      journal = {\jcap},
         year = 2024,
        month = jul,
       volume = {2024},
       number = {07},
          eid = {024},
        pages = {024},
          doi = {10.1088/1475-7516/2024/07/024},
archivePrefix = {arXiv},
       eprint = {2404.02153},
 primaryClass = {astro-ph.CO},
       adsurl = {https://ui.adsabs.harvard.edu/abs/2024JCAP...07..024A}
}

@ARTICLE{McClintock:2019:DESMass,
       author = {{McClintock}, T. and {Varga}, T.~N. and {Gruen}, D. and
                 {Rozo}, E. and {Rykoff}, E.~S. and {Shin}, T. and
                 {Melchior}, P. and {DeRose}, J. and {Seitz}, S. and
                 {Dietrich}, J.~P. and {Sheldon}, E. and {Zhang}, Y. and
                 {von der Linden}, A. and {Jeltema}, T. and {Mantz}, A.~B. and
                 {Romer}, A.~K. and others},
        title = "{Dark Energy Survey Year 1 results: weak lensing mass calibration of redMaPPer galaxy clusters}",
      journal = {\mnras},
         year = 2019,
        month = jan,
       volume = {482},
       number = {1},
        pages = {1352-1378},
          doi = {10.1093/mnras/sty2711},
archivePrefix = {arXiv},
       eprint = {1805.00039},
 primaryClass = {astro-ph.CO},
       adsurl = {https://ui.adsabs.harvard.edu/abs/2019MNRAS.482.1352M}
}

@ARTICLE{Grandis:2024:eROSITAMass,
       author = {{Grandis}, S. and {Ghirardini}, V. and {Bocquet}, S. and
                 {Garrel}, C. and {Mohr}, J.~J. and {Liu}, A. and
                 {Kluge}, M. and {Kimmig}, L. and {Reiprich}, T.~H. and others},
        title = "{The SRG/eROSITA All-Sky Survey. Dark Energy Survey year 3 weak gravitational lensing by eRASS1 selected galaxy clusters}",
      journal = {\aap},
         year = 2024,
        month = jul,
       volume = {687},
          eid = {A178},
        pages = {A178},
          doi = {10.1051/0004-6361/202348615},
archivePrefix = {arXiv},
       eprint = {2402.08455},
 primaryClass = {astro-ph.CO},
       adsurl = {https://ui.adsabs.harvard.edu/abs/2024A&A...687A.178G}
}

@ARTICLE{Kleinebreil:2025:eROSITAMass,
       author = {{Kleinebreil}, F. and {Grandis}, S. and {Schrabback}, T. and
                 {Ghirardini}, V. and {Chiu}, I. and {Liu}, A. and
                 {Kluge}, M. and {Reiprich}, T.~H. and {Artis}, E. and
                 {Bahar}, Y.~E. and {Balzer}, F. and {Bulbul}, E. and
                 {Clerc}, N. and {Comparat}, J. and {Garrel}, C. and
                 {Gruen}, D. and {Li}, X. and {Miyatake}, H. and
                 {Miyazaki}, S. and {Ramos-Ceja}, M.~E. and {Sanders}, J. and
                 {Seppi}, R. and {Okabe}, N. and {Zhang}, X.},
        title = "{The SRG/eROSITA All-Sky Survey: Weak lensing of eRASS1 galaxy clusters in KiDS-1000 and consistency checks with DES Y3 and HSC-Y3}",
      journal = {\aap},
         year = 2025,
        month = mar,
       volume = {695},
          eid = {A216},
        pages = {A216},
          doi = {10.1051/0004-6361/202449599},
archivePrefix = {arXiv},
       eprint = {2402.08456},
 primaryClass = {astro-ph.CO},
       adsurl = {https://ui.adsabs.harvard.edu/abs/2025A&A...695A.216K}
}

@ARTICLE{Dalal:2008:PNGs,
       author = {{Dalal}, Neal and {Dor{\'e}}, Olivier and {Huterer}, Dragan and {Shirokov}, Alexander},
        title = "{Imprints of primordial non-Gaussianities on large-scale structure: Scale-dependent bias and abundance of virialized objects}",
      journal = {\prd},
         year = 2008,
        month = jun,
       volume = {77},
       number = {12},
          eid = {123514},
        pages = {123514},
          doi = {10.1103/PhysRevD.77.123514},
archivePrefix = {arXiv},
       eprint = {0710.4560},
 primaryClass = {astro-ph},
       adsurl = {https://ui.adsabs.harvard.edu/abs/2008PhRvD..77l3514D}
}

@ARTICLE{More:2026:HSCY3_3x2pt,
       author = {{More}, Surhud and {Sugiyama}, Sunao and {Miyatake}, Hironao and {Rau}, Markus Michael and {Shirasaki}, Masato and {Li}, Xiangchong and {Nishizawa}, Atsushi J. and {Osato}, Ken and {Zhang}, Tianqing and {Takada}, Masahiro and {Hamana}, Takashi and {Takahashi}, Ryuichi and {Dalal}, Roohi and {Mandelbaum}, Rachel and {Strauss}, Michael A. and {Kobayashi}, Yosuke and {Nishimichi}, Takahiro and {Oguri}, Masamune and {Luo}, Wentao and {Kannawadi}, Arun and {Hsieh}, Bau-Ching and {Armstrong}, Robert and {Bosch}, James and {Komiyama}, Yutaka and {Lupton}, Robert H. and {Lust}, Nate B. and {MacArthur}, Lauren A. and {Miyazaki}, Satoshi and {Murayama}, Hitoshi and {Okura}, Yuki and {Price}, Paul A. and {Tait}, Philip J. and {Tanaka}, Masayuki and {Wang}, Shiang-Yu},
        title = "{Hyper Suprime-Cam Year 3 results: Measurements of clustering of SDSS-BOSS galaxies, galaxy-galaxy lensing, and cosmic shear}",
      journal = {\prd},
         year = 2023,
        month = dec,
       volume = {108},
       number = {12},
          eid = {123520},
        pages = {123520},
          doi = {10.1103/PhysRevD.108.123520},
archivePrefix = {arXiv},
       eprint = {2304.00703},
 primaryClass = {astro-ph.CO},
       adsurl = {https://ui.adsabs.harvard.edu/abs/2023PhRvD.108l3520M}
}

@ARTICLE{DESY6,
       author = {{DES Collaboration} and {Abbott}, T.~M.~C. and {Adamow}, M. and {Aguena}, M. and {Alarcon}, A. and {Allam}, S.~S. and {Alves}, O. and {Amon}, A. and {Anbajagane}, D. and {Andrade-Oliveira}, F. and {Avila}, S. and {Bacon}, D. and {Baxter}, E.~J. and {Beas-Gonzalez}, J. and {Bechtol}, K. and {Becker}, M.~R. and {Bernstein}, G.~M. and {Bertin}, E. and {Blazek}, J. and {Bocquet}, S. and {Brooks}, D. and {Brout}, D. and {Camacho}, H. and {Camacho-Ciurana}, G. and {Camilleri}, R. and {Campailla}, G. and {Campos}, A. and {Carnero Rosell}, A. and {Carrasco Kind}, M. and {Carretero}, J. and {Carrilho}, P. and {Castander}, F.~J. and {Cawthon}, R. and {Chang}, C. and {Choi}, A. and {Coloma-Nadal}, J.~M. and {Costanzi}, M. and {Crocce}, M. and {d'Assignies}, W. and {da Costa}, L.~N. and {da Silva Pereira}, M.~E. and {Davis}, T.~M. and {De Vicente}, J. and {DeRose}, J. and {Diehl}, H.~T. and {Dodelson}, S. and {Doel}, P. and {Doux}, C. and {Drlica-Wagner}, A. and {Eifler}, T.~F. and {Elvin-Poole}, J. and {Estrada}, J. and {Everett}, S. and {Evrard}, A.~E. and {Fang}, J. and {Farahi}, A. and {Fert{\'e}}, A. and {Flaugher}, B. and {Fosalba}, P. and {Frieman}, J. and {Garc{\'\i}a-Bellido}, J. and {Gatti}, M. and {Gaztanaga}, E. and {Giannini}, G. and {Giles}, P. and {Glazebrook}, K. and {Gorsuch}, M. and {Gruen}, D. and {Gruendl}, R.~A. and {Gschwend}, J. and {Gutierrez}, G. and {Harrison}, I. and {Hartley}, W.~G. and {Henning}, E. and {Herner}, K. and {Hinton}, S.~R. and {Hollowood}, D.~L. and {Honscheid}, K. and {Huff}, E.~M. and {Huterer}, D. and {Jain}, B. and {James}, D.~J. and {Jarvis}, M. and {Jeffrey}, N. and {Jeltema}, T. and {Kacprzak}, T. and {Kent}, S. and {Kovacs}, A. and {Krause}, E. and {Kron}, R. and {Kuehn}, K. and {Lahav}, O. and {Lee}, S. and {Legnani}, E. and {Lidman}, C. and {Lin}, H. and {MacCrann}, N. and {Manera}, M. and {Manning}, T. and {Marshall}, J.~L. and {Mau}, S. and {McCullough}, J. and {Mena-Fern{\'a}ndez}, J. and {Menanteau}, F. and {Miquel}, R. and {Mohr}, J.~J. and {Muir}, J. and {Myles}, J. and {Nichol}, R.~C. and {Nord}, B. and {O'Donnell}, J.~H. and {Ogando}, R.~L.~C. and {Palmese}, A. and {Paterno}, M. and {Peoples}, J. and {Percival}, W.~J. and {Petravick}, D. and {Pieres}, A. and {Plazas Malag{\'o}n}, A.~A. and {Porredon}, A. and {Pourtsidou}, A. and {Prat}, J. and {Preston}, C. and {Raveri}, M. and {Riquelme}, W. and {Rodriguez-Monroy}, M. and {Rogozenski}, P. and {Romer}, A.~K. and {Roodman}, A. and {Rosenfeld}, R. and {Ross}, A.~J. and {Rozo}, E. and {Rykoff}, E.~S. and {Samuroff}, S. and {S{\'a}nchez}, C. and {Sanchez}, E. and {Sanchez Cid}, D. and {Schutt}, T. and {Sevilla-Noarbe}, I. and {Sheldon}, E. and {Sherman}, N. and {Shin}, T. and {Smith}, M. and {Soares-Santos}, M. and {Suchyta}, E. and {Swanson}, M.~E.~C. and {Tabbutt}, M. and {Tarle}, G. and {Thomas}, D. and {To}, C. and {Tong}, A. and {Toribio San Cipriano}, L. and {Troxel}, M.~A. and {Tsedrik}, M. and {Tucker}, D.~L. and {Vikram}, V. and {Walker}, A.~R. and {Weaverdyck}, N. and {Wechsler}, R.~H. and {Weinberg}, D.~H. and {Weller}, J. and {Wetzell}, V. and {Whyley}, A. and {Wilkinson}, R.~D. and {Wiseman}, P. and {Wu}, H.-Y. and {Yamamoto}, M. and {Yanny}, B. and {Yin}, B. and {Zacharegkas}, G. and {Zhang}, Y. and {Zuntz}, J.},
        title = "{Dark Energy Survey Year 6 Results: Cosmological Constraints from Galaxy Clustering and Weak Lensing}",
      journal = {arXiv e-prints},
         year = 2026,
        month = jan,
          eid = {arXiv:2601.14559},
        pages = {arXiv:2601.14559},
          doi = {10.48550/arXiv.2601.14559},
archivePrefix = {arXiv},
       eprint = {2601.14559},
 primaryClass = {astro-ph.CO},
       adsurl = {https://ui.adsabs.harvard.edu/abs/2026arXiv260114559D}
}

@ARTICLE{DESY3,
       author = {{Abbott}, T.~M.~C. and {Aguena}, M. and {Alarcon}, A. and {Allam}, S. and {Alves}, O. and {Amon}, A. and {Andrade-Oliveira}, F. and {Annis}, J. and {Avila}, S. and {Bacon}, D. and {Baxter}, E. and {Bechtol}, K. and {Becker}, M.~R. and {Bernstein}, G.~M. and {Bhargava}, S. and {Birrer}, S. and {Blazek}, J. and {Brandao-Souza}, A. and {Bridle}, S.~L. and {Brooks}, D. and {Buckley-Geer}, E. and {Burke}, D.~L. and {Camacho}, H. and {Campos}, A. and {Carnero Rosell}, A. and {Carrasco Kind}, M. and {Carretero}, J. and {Castander}, F.~J. and {Cawthon}, R. and {Chang}, C. and {Chen}, A. and {Chen}, R. and {Choi}, A. and {Conselice}, C. and {Cordero}, J. and {Costanzi}, M. and {Crocce}, M. and {da Costa}, L.~N. and {da Silva Pereira}, M.~E. and {Davis}, C. and {Davis}, T.~M. and {De Vicente}, J. and {DeRose}, J. and {Desai}, S. and {Di Valentino}, E. and {Diehl}, H.~T. and {Dietrich}, J.~P. and {Dodelson}, S. and {Doel}, P. and {Doux}, C. and {Drlica-Wagner}, A. and {Eckert}, K. and {Eifler}, T.~F. and {Elsner}, F. and {Elvin-Poole}, J. and {Everett}, S. and {Evrard}, A.~E. and {Fang}, X. and {Farahi}, A. and {Fernandez}, E. and {Ferrero}, I. and {Fert{\'e}}, A. and {Fosalba}, P. and {Friedrich}, O. and {Frieman}, J. and {Garc{\'\i}a-Bellido}, J. and {Gatti}, M. and {Gaztanaga}, E. and {Gerdes}, D.~W. and {Giannantonio}, T. and {Giannini}, G. and {Gruen}, D. and {Gruendl}, R.~A. and {Gschwend}, J. and {Gutierrez}, G. and {Harrison}, I. and {Hartley}, W.~G. and {Herner}, K. and {Hinton}, S.~R. and {Hollowood}, D.~L. and {Honscheid}, K. and {Hoyle}, B. and {Huff}, E.~M. and {Huterer}, D. and {Jain}, B. and {James}, D.~J. and {Jarvis}, M. and {Jeffrey}, N. and {Jeltema}, T. and {Kovacs}, A. and {Krause}, E. and {Kron}, R. and {Kuehn}, K. and {Kuropatkin}, N. and {Lahav}, O. and {Leget}, P.-F. and {Lemos}, P. and {Liddle}, A.~R. and {Lidman}, C. and {Lima}, M. and {Lin}, H. and {MacCrann}, N. and {Maia}, M.~A.~G. and {Marshall}, J.~L. and {Martini}, P. and {McCullough}, J. and {Melchior}, P. and {Mena-Fern{\'a}ndez}, J. and {Menanteau}, F. and {Miquel}, R. and {Mohr}, J.~J. and {Morgan}, R. and {Muir}, J. and {Myles}, J. and {Nadathur}, S. and {Navarro-Alsina}, A. and {Nichol}, R.~C. and {Ogando}, R.~L.~C. and {Omori}, Y. and {Palmese}, A. and {Pandey}, S. and {Park}, Y. and {Paz-Chinch{\'o}n}, F. and {Petravick}, D. and {Pieres}, A. and {Plazas Malag{\'o}n}, A.~A. and {Porredon}, A. and {Prat}, J. and {Raveri}, M. and {Rodriguez-Monroy}, M. and {Rollins}, R.~P. and {Romer}, A.~K. and {Roodman}, A. and {Rosenfeld}, R. and {Ross}, A.~J. and {Rykoff}, E.~S. and {Samuroff}, S. and {S{\'a}nchez}, C. and {Sanchez}, E. and {Sanchez}, J. and {Sanchez Cid}, D. and {Scarpine}, V. and {Schubnell}, M. and {Scolnic}, D. and {Secco}, L.~F. and {Serrano}, S. and {Sevilla-Noarbe}, I. and {Sheldon}, E. and {Shin}, T. and {Smith}, M. and {Soares-Santos}, M. and {Suchyta}, E. and {Swanson}, M.~E.~C. and {Tabbutt}, M. and {Tarle}, G. and {Thomas}, D. and {To}, C. and {Troja}, A. and {Troxel}, M.~A. and {Tucker}, D.~L. and {Tutusaus}, I. and {Varga}, T.~N. and {Walker}, A.~R. and {Weaverdyck}, N. and {Wechsler}, R. and {Weller}, J. and {Yanny}, B. and {Yin}, B. and {Zhang}, Y. and {Zuntz}, J. and {DES Collaboration}},
        title = "{Dark Energy Survey Year 3 results: Cosmological constraints from galaxy clustering and weak lensing}",
      journal = {\prd},
         year = 2022,
        month = jan,
       volume = {105},
       number = {2},
          eid = {023520},
        pages = {023520},
          doi = {10.1103/PhysRevD.105.023520},
archivePrefix = {arXiv},
       eprint = {2105.13549},
 primaryClass = {astro-ph.CO},
       adsurl = {https://ui.adsabs.harvard.edu/abs/2022PhRvD.105b3520A}
}

@ARTICLE{Bleem:2026,
       author = {{Bleem}, L.~E. and {Klein}, M. and {Kornoelje}, K. and {Bocquet}, S. and {Sobrin}, J.~A. and {Aguena}, M. and {Anderes}, E. and {Anderson}, A.~J. and {Andrade-Oliveira}, F. and {Ansarinejad}, B. and {Archipley}, M. and {Balkenhol}, L. and {Barron}, D.~R. and {Barry}, P.~S. and {Bayliss}, M. and {Benabed}, K. and {Bender}, A.~N. and {Benson}, B.~A. and {Bianchini}, F. and {Bouchet}, F.~R. and {Brooks}, D. and {Burke}, D.~L. and {Calzadilla}, M. and {Camilleri}, R. and {Camphuis}, E. and {Campitiello}, M.~G. and {Carlstrom}, J.~E. and {Carnero Rosell}, A. and {Carretero}, J. and {Carron}, J. and {Chang}, C.~L. and {Chichura}, P.~M. and {Chokshi}, A. and {Chou}, T.-L. and {Coerver}, A. and {Costanzi}, M. and {Crawford}, T.~M. and {da Costa}, L.~N. and {Daley}, C. and {Davis}, T.~M. and {de Haan}, T. and {De Vicente}, J. and {Desai}, S. and {Dibert}, K.~R. and {Diehl}, H.~T. and {Dobbs}, M.~A. and {Doohan}, M. and {Dutcher}, D. and {Everett}, S. and {Evrard}, G. and {Feng}, C. and {Ferguson}, K.~R. and {Ferree}, N.~C. and {Fichman}, K. and {Flaugher}, B. and {Floyd}, B. and {Fosdick}, K. and {Foster}, A. and {Galli}, S. and {Gambrel}, A.~E. and {Gao}, A.~K. and {Garc{\'\i}a-Bellido}, J. and {Ge}, F. and {Gladders}, M.~D. and {Grandis}, S. and {Guidi}, F. and {Guns}, S. and {Gutierrez}, G. and {Halverson}, N.~W. and {Hinton}, S.~R. and {Hivon}, E. and {Holder}, G.~P. and {Hollowood}, D.~L. and {Holzapfel}, W.~L. and {Hood}, J.~C. and {Hryciuk}, A. and {Huang}, N. and {Jhaveri}, T. and {Kent}, S. and {K{\'e}ruzor{\'e}}, F. and {Khalife}, A.~R. and {Khullar}, G. and {Knox}, L. and {Kuehn}, K. and {Kuo}, C.-L. and {Lahav}, O. and {Levy}, K. and {Li}, Y. and {Lowitz}, A.~E. and {Lu}, C. and {Lynch}, G.~P. and {Maccarone}, T.~J. and {Mahler}, G. and {Maniyar}, A.~S. and {Marshall}, J.~L. and {Martsen}, E.~S. and {McDonald}, M. and {Mena-Fern{\'a}ndez}, J. and {Menanteau}, F. and {Millea}, M. and {Miquel}, R. and {Mohr}, J.~J. and {Montgomery}, J. and {Myles}, J. and {Nakato}, Y. and {Natoli}, T. and {Nichol}, R.~C. and {Novosad}, V. and {Ogando}, R.~L.~C. and {Omori}, Y. and {Ouellette}, A. and {Pan}, Z. and {Phadke}, K.~A. and {Plazas Malag{\'o}n}, A.~A. and {Pollak}, A.~W. and {Porredon}, A. and {Prabhu}, K. and {Prat}, J. and {Quan}, W. and {Raghunathan}, S. and {Rahimi}, M. and {Rahlin}, A. and {Reichardt}, C.~L. and {Romer}, A.~K. and {Rouble}, M. and {Ruhl}, J.~E. and {Sanchez}, E. and {Sanchez Cid}, D. and {Schrabback}, T. and {Sevilla-Noarbe}, I. and {Silva Oliveira}, A.~C. and {Simpson}, A. and {Somboonpanyakul}, T. and {Stark}, A.~A. and {Suchyta}, E. and {Swanson}, M.~E.~C. and {Tandoi}, C. and {To}, C. and {Trendafilova}, C. and {Vieira}, J.~D. and {Vieregg}, A.~G. and {Vikram}, V. and {Vitrier}, A. and {Wan}, Y. and {Weaverdyck}, N. and {Weller}, J. and {Whitehorn}, N. and {Wu}, W.~L.~K. and {Young}, M.~R. and {Zebrowski}, J.~A. and {Zhan}, J.},
        title = "{Galaxy Clusters Selected via the Sunyaev-Zel'dovich Effect in 5 year data from the SPT-3G Main Survey}",
      journal = {arXiv e-prints},
         year = 2026,
        month = jul,
          eid = {arXiv:2607.01175},
        pages = {arXiv:2607.01175},
          doi = {10.48550/arXiv.2607.01175},
archivePrefix = {arXiv},
       eprint = {2607.01175},
 primaryClass = {astro-ph.CO},
       adsurl = {https://ui.adsabs.harvard.edu/abs/2026arXiv260701175B}
}

@ARTICLE{Lesci:2023:PlanckClustering,
       author = {{Lesci}, G.~F. and {Veropalumbo}, A. and {Sereno}, M. and {Marulli}, F. and {Moscardini}, L. and {Giocoli}, C.},
        title = "{Mass bias and cosmological constraints from Planck cluster clustering}",
      journal = {\aap},
         year = 2023,
        month = jun,
       volume = {674},
          eid = {A80},
        pages = {A80},
          doi = {10.1051/0004-6361/202346261},
archivePrefix = {arXiv},
       eprint = {2302.14074},
 primaryClass = {astro-ph.CO},
       adsurl = {https://ui.adsabs.harvard.edu/abs/2023A&A...674A..80L}
}

@ARTICLE{Cooray:2002:HaloModel,
       author = {{Cooray}, Asantha and {Sheth}, Ravi},
        title = "{Halo models of large scale structure}",
      journal = {\physrep},
         year = 2002,
        month = dec,
       volume = {372},
       number = {1},
        pages = {1-129},
          doi = {10.1016/S0370-1573(02)00276-4},
archivePrefix = {arXiv},
       eprint = {astro-ph/0206508},
 primaryClass = {astro-ph},
       adsurl = {https://ui.adsabs.harvard.edu/abs/2002PhR...372....1C}
}

@ARTICLE{Schlegel:1998:DustMaps,
       author = {{Schlegel}, David J. and {Finkbeiner}, Douglas P. and {Davis}, Marc},
        title = "{Maps of Dust Infrared Emission for Use in Estimation of Reddening and Cosmic Microwave Background Radiation Foregrounds}",
      journal = {\apj},
         year = 1998,
        month = jun,
       volume = {500},
       number = {2},
        pages = {525-553},
          doi = {10.1086/305772},
archivePrefix = {arXiv},
       eprint = {astro-ph/9710327},
 primaryClass = {astro-ph},
       adsurl = {https://ui.adsabs.harvard.edu/abs/1998ApJ...500..525S}
}

@ARTICLE{Baxter:2024:Concentration,
       author = {{Baxter}, Eric J. and {Pandey}, Shivam and {Adhikari}, Susmita and {Cui}, Weiguang and {Shin}, Tae-hyeon and {Li}, Qingyang and {Rasia}, Elena},
        title = "{The impact of halo concentration on the Sunyaev Zel'dovich effect signal from massive galaxy clusters}",
      journal = {\mnras},
         year = 2024,
        month = jan,
       volume = {527},
       number = {3},
        pages = {7847-7860},
          doi = {10.1093/mnras/stad3704},
archivePrefix = {arXiv},
       eprint = {2304.08731},
 primaryClass = {astro-ph.CO},
       adsurl = {https://ui.adsabs.harvard.edu/abs/2024MNRAS.527.7847B}
}

@ARTICLE{Estrada:2009:Clustering,
       author = {{Estrada}, Juan and {Sefusatti}, Emiliano and {Frieman}, Joshua A.},
        title = "{The Correlation Function of Optically Selected Galaxy Clusters in the Sloan Digital Sky Survey}",
      journal = {\apj},
         year = 2009,
        month = feb,
       volume = {692},
       number = {1},
        pages = {265-282},
          doi = {10.1088/0004-637X/692/1/265},
archivePrefix = {arXiv},
       eprint = {0801.3485},
 primaryClass = {astro-ph},
       adsurl = {https://ui.adsabs.harvard.edu/abs/2009ApJ...692..265E}
}

@ARTICLE{To:2021:6x2pt,
       author = {{To}, C. and {Krause}, E. and {Rozo}, E. and {Wu}, H. and {Gruen}, D. and {Wechsler}, R.~H. and {Eifler}, T.~F. and {Rykoff}, E.~S. and {Costanzi}, M. and {Becker}, M.~R. and {Bernstein}, G.~M. and {Blazek}, J. and {Bocquet}, S. and {Bridle}, S.~L. and {Cawthon}, R. and {Choi}, A. and {Crocce}, M. and {Davis}, C. and {DeRose}, J. and {Drlica-Wagner}, A. and {Elvin-Poole}, J. and {Fang}, X. and {Farahi}, A. and {Friedrich}, O. and {Gatti}, M. and {Gaztanaga}, E. and {Giannantonio}, T. and {Hartley}, W.~G. and {Hoyle}, B. and {Jarvis}, M. and {MacCrann}, N. and {McClintock}, T. and {Miranda}, V. and {Pereira}, M.~E.~S. and {Park}, Y. and {Porredon}, A. and {Prat}, J. and {Rau}, M.~M. and {Ross}, A.~J. and {Samuroff}, S. and {S{\'a}nchez}, C. and {Sevilla-Noarbe}, I. and {Sheldon}, E. and {Troxel}, M.~A. and {Varga}, T.~N. and {Vielzeuf}, P. and {Zhang}, Y. and {Zuntz}, J. and {Abbott}, T.~M.~C. and {Aguena}, M. and {Amon}, A. and {Annis}, J. and {Avila}, S. and {Bertin}, E. and {Bhargava}, S. and {Brooks}, D. and {Burke}, D.~L. and {Carnero Rosell}, A. and {Carrasco Kind}, M. and {Carretero}, J. and {Chang}, C. and {Conselice}, C. and {da Costa}, L.~N. and {Davis}, T.~M. and {Desai}, S. and {Diehl}, H.~T. and {Dietrich}, J.~P. and {Everett}, S. and {Evrard}, A.~E. and {Ferrero}, I. and {Flaugher}, B. and {Fosalba}, P. and {Frieman}, J. and {Garc{\'\i}a-Bellido}, J. and {Gruendl}, R.~A. and {Gutierrez}, G. and {Hinton}, S.~R. and {Hollowood}, D.~L. and {Honscheid}, K. and {Huterer}, D. and {James}, D.~J. and {Jeltema}, T. and {Kron}, R. and {Kuehn}, K. and {Kuropatkin}, N. and {Lima}, M. and {Maia}, M.~A.~G. and {Marshall}, J.~L. and {Menanteau}, F. and {Miquel}, R. and {Morgan}, R. and {Muir}, J. and {Myles}, J. and {Palmese}, A. and {Paz-Chinch{\'o}n}, F. and {Plazas}, A.~A. and {Romer}, A.~K. and {Roodman}, A. and {Sanchez}, E. and {Santiago}, B. and {Scarpine}, V. and {Serrano}, S. and {Smith}, M. and {Suchyta}, E. and {Swanson}, M.~E.~C. and {Tarle}, G. and {Thomas}, D. and {Tucker}, D.~L. and {Weller}, J. and {Wester}, W. and {Wilkinson}, R.~D. and {DES Collaboration}},
        title = "{Dark Energy Survey Year 1 Results: Cosmological Constraints from Cluster Abundances, Weak Lensing, and Galaxy Correlations}",
      journal = {\prl},
         year = 2021,
        month = apr,
       volume = {126},
       number = {14},
          eid = {141301},
        pages = {141301},
          doi = {10.1103/PhysRevLett.126.141301},
archivePrefix = {arXiv},
       eprint = {2010.01138},
 primaryClass = {astro-ph.CO},
       adsurl = {https://ui.adsabs.harvard.edu/abs/2021PhRvL.126n1301T}
}

@ARTICLE{DES:2025:6x2pt_Y6,
       author = {{Abbott}, T.~M.~C. and {Aguena}, M. and {Alarcon}, A. and {Amon}, A. and {Anbajagane}, D. and {Andrade-Oliveira}, F. and {Avila}, S. and {Allam}, S.~S. and {Bacon}, D. and {Becker}, M.~R. and {Bhargava}, S. and {Blazek}, J. and {Bocquet}, S. and {Brooks}, D. and {Rosell}, A. Carnero and {Carretero}, J. and {Castander}, F.~J. and {Chang}, C. and {Choi}, A. and {Conselice}, C. and {Costanzi}, M. and {Crocce}, M. and {da Costa}, L.~N. and {Pereira}, M.~E.~S. and {Doux}, C. and {Davis}, T.~M. and {Desai}, S. and {Diehl}, H.~T. and {Dodelson}, S. and {Doel}, P. and {Elvin-Poole}, J. and {Esteves}, J. and {Everett}, S. and {Farahi}, A. and {Fert{\'e}}, A. and {Flaugher}, B. and {Garc{\'\i}a-Bellido}, J. and {Gatti}, M. and {Giannini}, G. and {Giles}, P. and {Grandis}, S. and {Gruen}, D. and {Gruendl}, R.~A. and {Gutierrez}, G. and {Harrison}, I. and {Hinton}, S.~R. and {Hollowood}, D.~L. and {Honscheid}, K. and {Jeffrey}, N. and {Jeltema}, T. and {Krause}, E. and {Lahav}, O. and {Lee}, S. and {Lidman}, C. and {Lima}, M. and {Lin}, H. and {Mohr}, J.~J. and {Marshall}, J.~L. and {McCullough}, J. and {Mena-Fern'andez}, J. and {Miquel}, R. and {Muir}, J. and {Myles}, J. and {Ogando}, R.~L.~C. and {Palmese}, A. and {Paterno}, M. and {Malag{\'o}n}, A.~A. Plazas and {Porredon}, A. and {Prat}, J. and {Romer}, A.~K. and {Roodman}, A. and {Rozo}, E. and {Rykoff}, E.~S. and {Rosenfeld}, R. and {Sanchez}, E. and {Cid}, D. Sanchez and {Sevilla-Noarbe}, I. and {Smith}, M. and {Suchyta}, E. and {Shin}, Tae-hyeon and {Tarle}, G. and {Thomas}, D. and {To}, Chun-Hao and {Troxel}, M.~A. and {Vikram}, V. and {Walker}, A.~R. and {Weinberg}, David H. and {Weaverdyck}, N. and {Wechsler}, R.~H. and {Weller}, J. and {Wu}, H.-Y. and {Yamamoto}, M. and {Yanny}, B. and {Zhang}, Y. and {Zhou}, C. and {DES Collaboration}},
        title = "{Dark energy survey year 3 results: Cosmological constraints from cluster abundances, weak lensing, and galaxy clustering}",
      journal = {\prd},
         year = 2025,
        month = oct,
       volume = {112},
       number = {8},
          eid = {083535},
        pages = {083535},
          doi = {10.1103/3dzh-d8f5},
archivePrefix = {arXiv},
       eprint = {2503.13632},
 primaryClass = {astro-ph.CO},
       adsurl = {https://ui.adsabs.harvard.edu/abs/2025PhRvD.112h3535A}
}

@ARTICLE{Jarvis:2004:Treecorr,
       author = {{Jarvis}, M. and {Bernstein}, G. and {Jain}, B.},
        title = "{The skewness of the aperture mass statistic}",
      journal = {\mnras},
         year = 2004,
        month = jul,
       volume = {352},
       number = {1},
        pages = {338-352},
          doi = {10.1111/j.1365-2966.2004.07926.x},
archivePrefix = {arXiv},
       eprint = {astro-ph/0307393},
 primaryClass = {astro-ph},
       adsurl = {https://ui.adsabs.harvard.edu/abs/2004MNRAS.352..338J}
}

@ARTICLE{abbott21,
       author = {{Abbott}, T.~M.~C. and {Adam{\'o}w}, M. and {Aguena}, M. and {Allam}, S. and {Amon}, A. and {Annis}, J. and {Avila}, S. and {Bacon}, D. and {Banerji}, M. and {Bechtol}, K. and {Becker}, M.~R. and {Bernstein}, G.~M. and {Bertin}, E. and {Bhargava}, S. and {Bridle}, S.~L. and {Brooks}, D. and {Burke}, D.~L. and {Carnero Rosell}, A. and {Carrasco Kind}, M. and {Carretero}, J. and {Castander}, F.~J. and {Cawthon}, R. and {Chang}, C. and {Choi}, A. and {Conselice}, C. and {Costanzi}, M. and {Crocce}, M. and {da Costa}, L.~N. and {Davis}, T.~M. and {De Vicente}, J. and {DeRose}, J. and {Desai}, S. and {Diehl}, H.~T. and {Dietrich}, J.~P. and {Drlica-Wagner}, A. and {Eckert}, K. and {Elvin-Poole}, J. and {Everett}, S. and {Evrard}, A.~E. and {Ferrero}, I. and {Fert{\'e}}, A. and {Flaugher}, B. and {Fosalba}, P. and {Friedel}, D. and {Frieman}, J. and {Garc{\'\i}a-Bellido}, J. and {Gaztanaga}, E. and {Gelman}, L. and {Gerdes}, D.~W. and {Giannantonio}, T. and {Gill}, M.~S.~S. and {Gruen}, D. and {Gruendl}, R.~A. and {Gschwend}, J. and {Gutierrez}, G. and {Hartley}, W.~G. and {Hinton}, S.~R. and {Hollowood}, D.~L. and {Honscheid}, K. and {Huterer}, D. and {James}, D.~J. and {Jeltema}, T. and {Johnson}, M.~D. and {Kent}, S. and {Kron}, R. and {Kuehn}, K. and {Kuropatkin}, N. and {Lahav}, O. and {Li}, T.~S. and {Lidman}, C. and {Lin}, H. and {MacCrann}, N. and {Maia}, M.~A.~G. and {Manning}, T.~A. and {Maloney}, J.~D. and {March}, M. and {Marshall}, J.~L. and {Martini}, P. and {Melchior}, P. and {Menanteau}, F. and {Miquel}, R. and {Morgan}, R. and {Myles}, J. and {Neilsen}, E. and {Ogando}, R.~L.~C. and {Palmese}, A. and {Paz-Chinch{\'o}n}, F. and {Petravick}, D. and {Pieres}, A. and {Plazas}, A.~A. and {Pond}, C. and {Rodriguez-Monroy}, M. and {Romer}, A.~K. and {Roodman}, A. and {Rykoff}, E.~S. and {Sako}, M. and {Sanchez}, E. and {Santiago}, B. and {Scarpine}, V. and {Serrano}, S. and {Sevilla-Noarbe}, I. and {Smith}, J. Allyn and {Smith}, M. and {Soares-Santos}, M. and {Suchyta}, E. and {Swanson}, M.~E.~C. and {Tarle}, G. and {Thomas}, D. and {To}, C. and {Tremblay}, P.~E. and {Troxel}, M.~A. and {Tucker}, D.~L. and {Turner}, D.~J. and {Varga}, T.~N. and {Walker}, A.~R. and {Wechsler}, R.~H. and {Weller}, J. and {Wester}, W. and {Wilkinson}, R.~D. and {Yanny}, B. and {Zhang}, Y. and {Nikutta}, R. and {Fitzpatrick}, M. and {Jacques}, A. and {Scott}, A. and {Olsen}, K. and {Huang}, L. and {Herrera}, D. and {Juneau}, S. and {Nidever}, D. and {Weaver}, B.~A. and {Adean}, C. and {Correia}, V. and {de Freitas}, M. and {Freitas}, F.~N. and {Singulani}, C. and {Vila-Verde}, G. and {Linea Science Server}},
        title = "{The Dark Energy Survey Data Release 2}",
      journal = {\apjs},
         year = 2021,
        month = aug,
       volume = {255},
       number = {2},
          eid = {20},
        pages = {20},
          doi = {10.3847/1538-4365/ac00b3},
archivePrefix = {arXiv},
       eprint = {2101.05765},
 primaryClass = {astro-ph.IM},
       adsurl = {https://ui.adsabs.harvard.edu/abs/2021ApJS..255...20A}
}

@ARTICLE{Muir:2020:Blinding,
       author = {{Muir}, J. and {Bernstein}, G.~M. and {Huterer}, D. and {Elsner}, F. and {Krause}, E. and {Roodman}, A. and {Allam}, S. and {Annis}, J. and {Avila}, S. and {Bechtol}, K. and {Bertin}, E. and {Brooks}, D. and {Buckley-Geer}, E. and {Burke}, D.~L. and {Carnero Rosell}, A. and {Carrasco Kind}, M. and {Carretero}, J. and {Cawthon}, R. and {Costanzi}, M. and {da Costa}, L.~N. and {De Vicente}, J. and {Desai}, S. and {Dietrich}, J.~P. and {Doel}, P. and {Eifler}, T.~F. and {Everett}, S. and {Fosalba}, P. and {Frieman}, J. and {Garc{\'\i}a-Bellido}, J. and {Gerdes}, D.~W. and {Gruen}, D. and {Gruendl}, R.~A. and {Gschwend}, J. and {Hartley}, W.~G. and {Hollowood}, D.~L. and {James}, D.~J. and {Jarvis}, M. and {Kuehn}, K. and {Kuropatkin}, N. and {Lahav}, O. and {March}, M. and {Marshall}, J.~L. and {Melchior}, P. and {Menanteau}, F. and {Miquel}, R. and {Ogando}, R.~L.~C. and {Palmese}, A. and {Paz-Chinch{\'o}n}, F. and {Plazas}, A.~A. and {Romer}, A.~K. and {Sanchez}, E. and {Scarpine}, V. and {Schubnell}, M. and {Serrano}, S. and {Sevilla-Noarbe}, I. and {Smith}, M. and {Suchyta}, E. and {Tarle}, G. and {Thomas}, D. and {Troxel}, M.~A. and {Walker}, A.~R. and {Weller}, J. and {Wester}, W. and {Zuntz}, J. and {DES Collaboration}},
        title = "{Blinding multiprobe cosmological experiments}",
      journal = {\mnras},
         year = 2020,
        month = may,
       volume = {494},
       number = {3},
        pages = {4454-4470},
          doi = {10.1093/mnras/staa965},
archivePrefix = {arXiv},
       eprint = {1911.05929},
 primaryClass = {astro-ph.CO},
       adsurl = {https://ui.adsabs.harvard.edu/abs/2020MNRAS.494.4454M}
}

@ARTICLE{McClintock:2019:Aemulusbias,
       author = {{McClintock}, Thomas and {Rozo}, Eduardo and {Banerjee}, Arka and {Becker}, Matthew R. and {DeRose}, Joseph and {McLaughlin}, Sean and {Tinker}, Jeremy L. and {Wechsler}, Risa H. and {Zhai}, Zhongxu},
        title = "{The Aemulus Project IV: Emulating Halo Bias}",
      journal = {arXiv e-prints},
         year = 2019,
        month = jul,
          eid = {arXiv:1907.13167},
        pages = {arXiv:1907.13167},
          doi = {10.48550/arXiv.1907.13167},
archivePrefix = {arXiv},
       eprint = {1907.13167},
 primaryClass = {astro-ph.CO},
       adsurl = {https://ui.adsabs.harvard.edu/abs/2019arXiv190713167M}
}

@ARTICLE{Fumagalli:2024:CovEuclid,
       author = {{Euclid Collaboration} and {Fumagalli}, A. and {Saro}, A. and {Borgani}, S. and {Castro}, T. and {Costanzi}, M. and {Monaco}, P. and {Munari}, E. and {Sefusatti}, E. and {Le Brun}, A.~M.~C. and {Aghanim}, N. and {Auricchio}, N. and {Baldi}, M. and {Bodendorf}, C. and {Bonino}, D. and {Branchini}, E. and {Brescia}, M. and {Brinchmann}, J. and {Camera}, S. and {Capobianco}, V. and {Carbone}, C. and {Carretero}, J. and {Castander}, F.~J. and {Castellano}, M. and {Cavuoti}, S. and {Cledassou}, R. and {Congedo}, G. and {Conselice}, C.~J. and {Conversi}, L. and {Copin}, Y. and {Corcione}, L. and {Courbin}, F. and {Cropper}, M. and {Da Silva}, A. and {Degaudenzi}, H. and {Dubath}, F. and {Dupac}, X. and {Dusini}, S. and {Farrens}, S. and {Ferriol}, S. and {Frailis}, M. and {Franceschi}, E. and {Franzetti}, P. and {Galeotta}, S. and {Garilli}, B. and {Gillard}, W. and {Gillis}, B. and {Giocoli}, C. and {Grazian}, A. and {Grupp}, F. and {Haugan}, S.~V.~H. and {Holmes}, W. and {Hornstrup}, A. and {Hudelot}, P. and {Jahnke}, K. and {K{\"u}mmel}, M. and {Kermiche}, S. and {Kiessling}, A. and {Kilbinger}, M. and {Kitching}, T. and {Kunz}, M. and {Kurki-Suonio}, H. and {Ligori}, S. and {Lilje}, P.~B. and {Lloro}, I. and {Mansutti}, O. and {Marggraf}, O. and {Markovic}, K. and {Marulli}, F. and {Massey}, R. and {Maurogordato}, S. and {Medinaceli}, E. and {Mei}, S. and {Meneghetti}, M. and {Meylan}, G. and {Moresco}, M. and {Moscardini}, L. and {Niemi}, S.-M. and {Padilla}, C. and {Paltani}, S. and {Pasian}, F. and {Pedersen}, K. and {Percival}, W.~J. and {Pettorino}, V. and {Pires}, S. and {Polenta}, G. and {Poncet}, M. and {Raison}, F. and {Rebolo-Lopez}, R. and {Renzi}, A. and {Rhodes}, J. and {Riccio}, G. and {Romelli}, E. and {Roncarelli}, M. and {Saglia}, R. and {Sapone}, D. and {Sartoris}, B. and {Schneider}, P. and {Secroun}, A. and {Seidel}, G. and {Sirignano}, C. and {Sirri}, G. and {Stanco}, L. and {Tallada-Cresp{\'\i}}, P. and {Taylor}, A.~N. and {Tereno}, I. and {Toledo-Moreo}, R. and {Torradeflot}, F. and {Tutusaus}, I. and {Valenziano}, L. and {Vassallo}, T. and {Wang}, Y. and {Weller}, J. and {Zacchei}, A. and {Zamorani}, G. and {Zoubian}, J. and {Andreon}, S. and {Bardelli}, S. and {Boucaud}, A. and {Bozzo}, E. and {Colodro-Conde}, C. and {Di Ferdinando}, D. and {Fabbian}, G. and {Farina}, M. and {Lindholm}, V. and {Maino}, D. and {Mauri}, N. and {Neissner}, C. and {Scottez}, V. and {Zucca}, E. and {Baccigalupi}, C. and {Balaguera-Antol{\'\i}nez}, A. and {Ballardini}, M. and {Bernardeau}, F. and {Biviano}, A. and {Blanchard}, A. and {Borlaff}, A.~S. and {Burigana}, C. and {Cabanac}, R. and {Carvalho}, C.~S. and {Casas}, S. and {Castignani}, G. and {Chambers}, K. and {Cooray}, A.~R. and {Coupon}, J. and {Courtois}, H.~M. and {Davini}, S. and {de la Torre}, S. and {Desprez}, G. and {Dole}, H. and {Escartin}, J.~A. and {Escoffier}, S. and {Ferreira}, P.~G. and {Finelli}, F. and {Garcia-Bellido}, J. and {George}, K. and {Gozaliasl}, G. and {Hildebrandt}, H. and {Hook}, I. and {Jimenez Mu{\~n}oz}, A. and {Joachimi}, B. and {Kansal}, V. and {Keih{\"a}nen}, E. and {Kirkpatrick}, C.~C. and {Loureiro}, A. and {Magliocchetti}, M. and {Maoli}, R. and {Marcin}, S. and {Martinelli}, M. and {Martinet}, N. and {Matthew}, S. and {Maturi}, M. and {Maurin}, L. and {Metcalf}, R.~B. and {Morgante}, G. and {Nadathur}, S. and {Nucita}, A.~A. and {Patrizii}, L. and {Pollack}, J.~E. and {Popa}, V. and {Porciani}, C. and {Potter}, D. and {Pourtsidou}, A. and {P{\"o}ntinen}, M. and {S{\'a}nchez}, A.~G. and {Sakr}, Z. and {Schirmer}, M. and {Sereno}, M. and {Spurio Mancini}, A. and {Stadel}, J. and {Steinwagner}, J. and {Valieri}, C. and {Valiviita}, J. and {Veropalumbo}, A. and {Viel}, M.},
        title = "{Euclid preparation. XXXV. Covariance model validation for the two-point correlation function of galaxy clusters}",
      journal = {\aap},
         year = 2024,
        month = mar,
       volume = {683},
          eid = {A253},
        pages = {A253},
          doi = {10.1051/0004-6361/202245540},
archivePrefix = {arXiv},
       eprint = {2211.12965},
 primaryClass = {astro-ph.CO},
       adsurl = {https://ui.adsabs.harvard.edu/abs/2024A&A...683A.253E}
}

@article{baxter_constraining_2016,
	title = {Constraining the mass-richness relationship of {redMaPPer} clusters with angular clustering},
	volume = {463},
	issn = {0035-8711},
	url = {https://ui.adsabs.harvard.edu/abs/2016MNRAS.463..205B},
	doi = {10.1093/mnras/stw1939},
	urldate = {2023-06-13},
	journal = {Monthly Notices of the Royal Astronomical Society},
	author = {Baxter, Eric J. and Rozo, Eduardo and Jain, Bhuvnesh and Rykoff, Eli and Wechsler, Risa H.},
	month = nov,
	year = {2016},
	note = {ADS Bibcode: 2016MNRAS.463..205B},
	pages = {205--221},
}

@article{majumdar_self-calibration_2004,
	title = {Self-{Calibration} in {Cluster} {Studies} of {Dark} {Energy}: {Combining} the {Cluster} {Redshift} {Distribution}, the {Power} {Spectrum}, and {Mass} {Measurements}},
	volume = {613},
	issn = {0004-637X},
	shorttitle = {Self-{Calibration} in {Cluster} {Studies} of {Dark} {Energy}},
	url = {https://ui.adsabs.harvard.edu/abs/2004ApJ...613...41M},
	doi = {10.1086/422829},
	urldate = {2023-06-19},
	journal = {The Astrophysical Journal},
	author = {Majumdar, Subhabrata and Mohr, Joseph J.},
	month = sep,
	year = {2004},
	note = {ADS Bibcode: 2004ApJ...613...41M},
	pages = {41--50},
}

@article{bocquet_cluster_2019,
	title = {Cluster {Cosmology} {Constraints} from the 2500 deg\${\textasciicircum}2\$ {SPT}-{SZ} {Survey}: {Inclusion} of {Weak} {Gravitational} {Lensing} {Data} from {Magellan} and the {Hubble} {Space} {Telescope}},
	volume = {878},
	issn = {1538-4357},
	shorttitle = {Cluster {Cosmology} {Constraints} from the 2500 deg\${\textasciicircum}2\$ {SPT}-{SZ} {Survey}},
	url = {http://arxiv.org/abs/1812.01679},
	doi = {10.3847/1538-4357/ab1f10},
	number = {1},
	urldate = {2023-06-21},
	journal = {The Astrophysical Journal},
	author = {Bocquet, S. and Dietrich, J. P. and Schrabback, T. and Bleem, L. E. and Klein, M. and Allen, S. W. and Applegate, D. E. and Ashby, M. L. N. and Bautz, M. and Bayliss, M. and Benson, B. A. and Brodwin, M. and Bulbul, E. and Canning, R. E. A. and Capasso, R. and Carlstrom, J. E. and Chang, C. L. and Chiu, I. and Cho, H.-M. and Clocchiatti, A. and Crawford, T. M. and Crites, A. T. and de Haan, T. and Desai, S. and Dobbs, M. A. and Foley, R. J. and Forman, W. R. and Garmire, G. P. and George, E. M. and Gladders, M. D. and Gonzalez, A. H. and Grandis, S. and Gupta, N. and Halverson, N. W. and Hlavacek-Larrondo, J. and Hoekstra, H. and Holder, G. P. and Holzapfel, W. L. and Hou, Z. and Hrubes, J. D. and Huang, N. and Jones, C. and Khullar, G. and Knox, L. and Kraft, R. and Lee, A. T. and von der Linden, A. and Luong-Van, D. and Mantz, A. and Marrone, D. P. and McDonald, M. and McMahon, J. J. and Meyer, S. S. and Mocanu, L. M. and Mohr, J. J. and Morris, R. G. and Padin, S. and Patil, S. and Pryke, C. and Rapetti, D. and Reichardt, C. L. and Rest, A. and Ruhl, J. E. and Saliwanchik, B. R. and Saro, A. and Sayre, J. T. and Schaffer, K. K. and Shirokoff, E. and Stalder, B. and Stanford, S. A. and Staniszewski, Z. and Stark, A. A. and Story, K. T. and Strazzullo, V. and Stubbs, C. W. and Vanderlinde, K. and Vieira, J. D. and Vikhlinin, A. and Williamson, R. and Zenteno, A.},
	month = jun,
	year = {2019},
	note = {arXiv:1812.01679 [astro-ph]},
	pages = {55},
}

@article{sdss_collaboration_angular_2002,
	title = {The {Angular} {Correlation} {Function} of {Galaxies} from {Early} {SDSS} {Data}},
	volume = {579},
	issn = {0004-637X, 1538-4357},
	url = {http://arxiv.org/abs/astro-ph/0107417},
	doi = {10.1086/342787},
	number = {1},
	urldate = {2023-06-21},
	journal = {The Astrophysical Journal},
	author = {{SDSS~Collaboration} and Connolly, Andrew and Scranton, Ryan and Johnston, David and Dodelson, Scott and Eisenstein, Daniel and Frieman, Joshua A. and Gunn, James E. and Hui, Lam and Jain, Bhuvnesh and Kent, Stephen and Loveday, Jon and Narayanan, Vijay and Nichol, Robert C. and O'Connell, Liam and Postman, Marc and Scoccimarro, Roman and Sheth, Ravi K. and Stebbins, Albert and Strauss, Michael A. and Szalay, Alexander S. and Szapudi, Istvan and Tegmark, Max and Vogeley, Michael and Zehavi, Idit},
	month = nov,
	year = {2002},
	note = {arXiv:astro-ph/0107417},
	pages = {42--47},
}

@article{lima_self-calibration_2005,
	title = {Self-calibration of cluster dark energy studies: {Observable}-mass distribution},
	volume = {72},
	issn = {1550-7998, 1550-2368},
	shorttitle = {Self-calibration of cluster dark energy studies},
	url = {https://link.aps.org/doi/10.1103/PhysRevD.72.043006},
	doi = {10.1103/PhysRevD.72.043006},
	language = {en},
	number = {4},
	urldate = {2023-08-25},
	journal = {Physical Review D},
	author = {Lima, Marcos and Hu, Wayne},
	month = aug,
	year = {2005},
	pages = {043006},
}

@article{lindholm_clustering_2021,
	title = {Clustering of {CODEX} clusters},
	volume = {646},
	issn = {0004-6361, 1432-0746},
	url = {https://www.aanda.org/10.1051/0004-6361/202038807},
	doi = {10.1051/0004-6361/202038807},
	language = {en},
	urldate = {2023-11-30},
	journal = {Astronomy \& Astrophysics},
	author = {Lindholm, V. and Finoguenov, A. and Comparat, J. and Kirkpatrick, C. C. and Rykoff, E. and Clerc, N. and Collins, C. and Damsted, S. and Ider Chitham, J. and Padilla, N.},
	month = feb,
	year = {2021},
	pages = {A8},
}

@ARTICLE{seppi_srgerosita_2024,
       author = {{Seppi}, R. and {Comparat}, J. and {Ghirardini}, V. and {Garrel}, C. and {Artis}, E. and {S{\'a}nchez}, A.~G. and {Liu}, A. and {Clerc}, N. and {Bulbul}, E. and {Grandis}, S. and {Kluge}, M. and {Reiprich}, T.~H. and {Merloni}, A. and {Zhang}, X. and {Bahar}, Y.~E. and {Shreeram}, S. and {Sanders}, J. and {Ramos-Ceja}, M. and {Krumpe}, M.},
        title = "{The SRG/eROSITA All-Sky Survey. Tracing the large-scale structure with a clustering study of galaxy clusters}",
      journal = {\aap},
         year = 2024,
        month = jun,
       volume = {686},
          eid = {A196},
        pages = {A196},
          doi = {10.1051/0004-6361/202348843},
archivePrefix = {arXiv},
       eprint = {2402.08460},
 primaryClass = {astro-ph.CO},
       adsurl = {https://ui.adsabs.harvard.edu/abs/2024A&A...686A.196S}
}

@article{weinberg_observational_2013,
	title = {Observational probes of cosmic acceleration},
	volume = {530},
	issn = {0370-1573},
	url = {https://ui.adsabs.harvard.edu/abs/2013PhR...530...87W},
	doi = {10.1016/j.physrep.2013.05.001},
	urldate = {2024-04-11},
	journal = {Physics Reports},
	author = {Weinberg, David H. and Mortonson, Michael J. and Eisenstein, Daniel J. and Hirata, Christopher and Riess, Adam G. and Rozo, Eduardo},
	month = sep,
	year = {2013},
	note = {ADS Bibcode: 2013PhR...530...87W},
	pages = {87--255},
}

@article{allen_cosmological_2011,
	title = {Cosmological {Parameters} from {Observations} of {Galaxy} {Clusters}},
	volume = {49},
	issn = {0066-4146},
	url = {https://ui.adsabs.harvard.edu/abs/2011ARA&A..49..409A},
	doi = {10.1146/annurev-astro-081710-102514},
	urldate = {2024-04-11},
	journal = {Annual Review of Astronomy and Astrophysics},
	author = {Allen, Steven W. and Evrard, August E. and Mantz, Adam B.},
	month = sep,
	year = {2011},
	note = {ADS Bibcode: 2011ARA\&A..49..409A},
	pages = {409--470},
}

@article{lacey_merger_1993,
	title = {Merger rates in hierarchical models of galaxy formation},
	volume = {262},
	issn = {0035-8711},
	url = {https://ui.adsabs.harvard.edu/abs/1993MNRAS.262..627L},
	doi = {10.1093/mnras/262.3.627},
	urldate = {2024-04-12},
	journal = {Monthly Notices of the Royal Astronomical Society},
	author = {Lacey, Cedric and Cole, Shaun},
	month = jun,
	year = {1993},
	note = {ADS Bibcode: 1993MNRAS.262..627L},
	pages = {627--649},
}

@article{kravtsov_formation_2012,
	title = {Formation of {Galaxy} {Clusters}},
	volume = {50},
	issn = {0066-4146},
	url = {https://ui.adsabs.harvard.edu/abs/2012ARA&A..50..353K},
	doi = {10.1146/annurev-astro-081811-125502},
	urldate = {2024-04-12},
	journal = {Annual Review of Astronomy and Astrophysics},
	author = {Kravtsov, Andrey V. and Borgani, Stefano},
	month = sep,
	year = {2012},
	note = {ADS Bibcode: 2012ARA\&A..50..353K},
	pages = {353--409},
}

@article{mantz_weighing_2015,
	title = {Weighing the giants – {IV}. {Cosmology} and neutrino mass},
	volume = {446},
	issn = {0035-8711},
	url = {https://doi.org/10.1093/mnras/stu2096},
	doi = {10.1093/mnras/stu2096},
	number = {3},
	urldate = {2024-04-12},
	journal = {Monthly Notices of the Royal Astronomical Society},
	author = {Mantz, Adam B. and von der Linden, Anja and Allen, Steven W. and Applegate, Douglas E. and Kelly, Patrick L. and Morris, R. Glenn and Rapetti, David A. and Schmidt, Robert W. and Adhikari, Saroj and Allen, Mark T. and Burchat, Patricia R. and Burke, David L. and Cataneo, Matteo and Donovan, David and Ebeling, Harald and Shandera, Sarah and Wright, Adam},
	month = jan,
	year = {2015},
	pages = {2205--2225},
}

@article{benson_cosmological_2013,
	title = {Cosmological {Constraints} from {Sunyaev}-{Zel}'dovich-selected {Clusters} with {X}-{Ray} {Observations} in the {First} 178 deg2 of the {South} {Pole} {Telescope} {Survey}},
	volume = {763},
	issn = {0004-637X},
	url = {https://ui.adsabs.harvard.edu/abs/2013ApJ...763..147B},
	doi = {10.1088/0004-637X/763/2/147},
	urldate = {2024-04-12},
	journal = {The Astrophysical Journal},
	author = {Benson, B. A. and de Haan, T. and Dudley, J. P. and Reichardt, C. L. and Aird, K. A. and Andersson, K. and Armstrong, R. and Ashby, M. L. N. and Bautz, M. and Bayliss, M. and Bazin, G. and Bleem, L. E. and Brodwin, M. and Carlstrom, J. E. and Chang, C. L. and Cho, H. M. and Clocchiatti, A. and Crawford, T. M. and Crites, A. T. and Desai, S. and Dobbs, M. A. and Foley, R. J. and Forman, W. R. and George, E. M. and Gladders, M. D. and Gonzalez, A. H. and Halverson, N. W. and Harrington, N. and High, F. W. and Holder, G. P. and Holzapfel, W. L. and Hoover, S. and Hrubes, J. D. and Jones, C. and Joy, M. and Keisler, R. and Knox, L. and Lee, A. T. and Leitch, E. M. and Liu, J. and Lueker, M. and Luong-Van, D. and Mantz, A. and Marrone, D. P. and McDonald, M. and McMahon, J. J. and Mehl, J. and Meyer, S. S. and Mocanu, L. and Mohr, J. J. and Montroy, T. E. and Murray, S. S. and Natoli, T. and Padin, S. and Plagge, T. and Pryke, C. and Rest, A. and Ruel, J. and Ruhl, J. E. and Saliwanchik, B. R. and Saro, A. and Sayre, J. T. and Schaffer, K. K. and Shaw, L. and Shirokoff, E. and Song, J. and Spieler, H. G. and Stalder, B. and Staniszewski, Z. and Stark, A. A. and Story, K. and Stubbs, C. W. and Suhada, R. and van Engelen, A. and Vanderlinde, K. and Vieira, J. D. and Vikhlinin, A. and Williamson, R. and Zahn, O. and Zenteno, A.},
	month = feb,
	year = {2013},
	note = {ADS Bibcode: 2013ApJ...763..147B},
	pages = {147},
}

@article{marulli_xxl_2018,
	title = {The {XXL} {Survey}. {XVI}. {The} clustering of {X}-ray selected galaxy clusters at z   0.3},
	volume = {620},
	issn = {0004-6361},
	url = {https://ui.adsabs.harvard.edu/abs/2018A&A...620A...1M},
	doi = {10.1051/0004-6361/201833238},
	urldate = {2024-04-12},
	journal = {Astronomy and Astrophysics},
	author = {Marulli, F. and Veropalumbo, A. and Sereno, M. and Moscardini, L. and Pacaud, F. and Pierre, M. and Plionis, M. and Cappi, A. and Adami, C. and Alis, S. and Altieri, B. and Birkinshaw, M. and Ettori, S. and Faccioli, L. and Gastaldello, F. and Koulouridis, E. and Lidman, C. and Le Fèvre, J. -P. and Maurogordato, S. and Poggianti, B. and Pompei, E. and Sadibekova, T. and Valtchanov, I.},
	month = nov,
	year = {2018},
	note = {ADS Bibcode: 2018A\&A...620A...1M},
	pages = {A1},
}

@article{vikhlinin_chandra_2009,
	title = {Chandra {Cluster} {Cosmology} {Project} {III}: {Cosmological} {Parameter} {Constraints}},
	volume = {692},
	issn = {0004-637X},
	shorttitle = {Chandra {Cluster} {Cosmology} {Project} {III}},
	url = {https://ui.adsabs.harvard.edu/abs/2009ApJ...692.1060V},
	doi = {10.1088/0004-637X/692/2/1060},
	urldate = {2024-04-12},
	journal = {The Astrophysical Journal},
	author = {Vikhlinin, A. and Kravtsov, A. V. and Burenin, R. A. and Ebeling, H. and Forman, W. R. and Hornstrup, A. and Jones, C. and Murray, S. S. and Nagai, D. and Quintana, H. and Voevodkin, A.},
	month = feb,
	year = {2009},
	note = {ADS Bibcode: 2009ApJ...692.1060V},
	pages = {1060--1074},
}

@article{sunyaev_observations_1972,
	title = {The {Observations} of {Relic} {Radiation} as a {Test} of the {Nature} of {X}-{Ray} {Radiation} from the {Clusters} of {Galaxies}},
	volume = {4},
	issn = {0010-26790146-2970},
	url = {https://ui.adsabs.harvard.edu/abs/1972CoASP...4..173S},
	urldate = {2024-04-12},
	journal = {Comments on Astrophysics and Space Physics},
	author = {Sunyaev, R. A. and Zeldovich, Ya. B.},
	month = nov,
	year = {1972},
	note = {ADS Bibcode: 1972CoASP...4..173S},
	pages = {173},
}

@article{bleem_galaxy_2015,
	title = {Galaxy {Clusters} {Discovered} via the {Sunyaev}-{Zel}'dovich {Effect} in the 2500-{Square}-{Degree} {SPT}-{SZ} {Survey}},
	volume = {216},
	issn = {0067-0049},
	url = {https://ui.adsabs.harvard.edu/abs/2015ApJS..216...27B},
	doi = {10.1088/0067-0049/216/2/27},
	urldate = {2024-04-12},
	journal = {The Astrophysical Journal Supplement Series},
	author = {Bleem, L. E. and Stalder, B. and de Haan, T. and Aird, K. A. and Allen, S. W. and Applegate, D. E. and Ashby, M. L. N. and Bautz, M. and Bayliss, M. and Benson, B. A. and Bocquet, S. and Brodwin, M. and Carlstrom, J. E. and Chang, C. L. and Chiu, I. and Cho, H. M. and Clocchiatti, A. and Crawford, T. M. and Crites, A. T. and Desai, S. and Dietrich, J. P. and Dobbs, M. A. and Foley, R. J. and Forman, W. R. and George, E. M. and Gladders, M. D. and Gonzalez, A. H. and Halverson, N. W. and Hennig, C. and Hoekstra, H. and Holder, G. P. and Holzapfel, W. L. and Hrubes, J. D. and Jones, C. and Keisler, R. and Knox, L. and Lee, A. T. and Leitch, E. M. and Liu, J. and Lueker, M. and Luong-Van, D. and Mantz, A. and Marrone, D. P. and McDonald, M. and McMahon, J. J. and Meyer, S. S. and Mocanu, L. and Mohr, J. J. and Murray, S. S. and Padin, S. and Pryke, C. and Reichardt, C. L. and Rest, A. and Ruel, J. and Ruhl, J. E. and Saliwanchik, B. R. and Saro, A. and Sayre, J. T. and Schaffer, K. K. and Schrabback, T. and Shirokoff, E. and Song, J. and Spieler, H. G. and Stanford, S. A. and Staniszewski, Z. and Stark, A. A. and Story, K. T. and Stubbs, C. W. and Vanderlinde, K. and Vieira, J. D. and Vikhlinin, A. and Williamson, R. and Zahn, O. and Zenteno, A.},
	month = feb,
	year = {2015},
	note = {ADS Bibcode: 2015ApJS..216...27B},
	pages = {27},
}

@article{benson_spt-3g_2014,
	title = {{SPT}-{3G}: a next-generation cosmic microwave background polarization experiment on the {South} {Pole} telescope},
	volume = {9153},
	shorttitle = {{SPT}-{3G}},
	url = {https://ui.adsabs.harvard.edu/abs/2014SPIE.9153E..1PB},
	doi = {10.1117/12.2057305},
	urldate = {2024-04-12},
	author = {Benson, B. A. and Ade, P. A. R. and Ahmed, Z. and Allen, S. W. and Arnold, K. and Austermann, J. E. and Bender, A. N. and Bleem, L. E. and Carlstrom, J. E. and Chang, C. L. and Cho, H. M. and Cliche, J. F. and Crawford, T. M. and Cukierman, A. and de Haan, T. and Dobbs, M. A. and Dutcher, D. and Everett, W. and Gilbert, A. and Halverson, N. W. and Hanson, D. and Harrington, N. L. and Hattori, K. and Henning, J. W. and Hilton, G. C. and Holder, G. P. and Holzapfel, W. L. and Irwin, K. D. and Keisler, R. and Knox, L. and Kubik, D. and Kuo, C. L. and Lee, A. T. and Leitch, E. M. and Li, D. and McDonald, M. and Meyer, S. S. and Montgomery, J. and Myers, M. and Natoli, T. and Nguyen, H. and Novosad, V. and Padin, S. and Pan, Z. and Pearson, J. and Reichardt, C. and Ruhl, J. E. and Saliwanchik, B. R. and Simard, G. and Smecher, G. and Sayre, J. T. and Shirokoff, E. and Stark, A. A. and Story, K. and Suzuki, A. and Thompson, K. L. and Tucker, C. and Vanderlinde, K. and Vieira, J. D. and Vikhlinin, A. and Wang, G. and Yefremenko, V. and Yoon, K. W.},
	month = jul,
	year = {2014},
	note = {Conference Name: Millimeter, Submillimeter, and Far-Infrared Detectors and Instrumentation for Astronomy VII
Place: eprint: arXiv:1407.2973
ADS Bibcode: 2014SPIE.9153E..1PB},
	pages = {91531P},
}

@article{rodriguez-monroy_dark_2022,
	title = {Dark {Energy} {Survey} {Year} 3 results: galaxy clustering and systematics treatment for lens galaxy samples},
	volume = {511},
	issn = {0035-8711},
	shorttitle = {Dark {Energy} {Survey} {Year} 3 results},
	url = {https://ui.adsabs.harvard.edu/abs/2022MNRAS.511.2665R},
	doi = {10.1093/mnras/stac104},
	urldate = {2024-04-12},
	journal = {Monthly Notices of the Royal Astronomical Society},
	author = {Rodríguez-Monroy, M. and Weaverdyck, N. and Elvin-Poole, J. and Crocce, M. and Carnero Rosell, A. and Andrade-Oliveira, F. and Avila, S. and Bechtol, K. and Bernstein, G. M. and Blazek, J. and Camacho, H. and Cawthon, R. and De Vicente, J. and DeRose, J. and Dodelson, S. and Everett, S. and Fang, X. and Ferrero, I. and Ferté, A. and Friedrich, O. and Gaztanaga, E. and Giannini, G. and Gruendl, R. A. and Hartley, W. G. and Herner, K. and Huff, E. M. and Jarvis, M. and Krause, E. and MacCrann, N. and Mena-Fernández, J. and Muir, J. and Pandey, S. and Park, Y. and Porredon, A. and Prat, J. and Rosenfeld, R. and Ross, A. J. and Rozo, E. and Rykoff, E. S. and Sanchez, E. and Sanchez Cid, D. and Sevilla-Noarbe, I. and Tabbutt, M. and To, C. and Wagoner, E. L. and Wechsler, R. H. and Aguena, M. and Allam, S. and Amon, A. and Annis, J. and Bacon, D. and Baxter, E. and Bertin, E. and Bhargava, S. and Brooks, D. and Burke, D. L. and Carrasco Kind, M. and Carretero, J. and Castander, F. J. and Choi, A. and Conselice, C. and Costanzi, M. and da Costa, L. N. and Pereira, M. E. S. and Desai, S. and Diehl, H. T. and Flaugher, B. and Fosalba, P. and Frieman, J. and García-Bellido, J. and Giannantonio, T. and Gruen, D. and Gschwend, J. and Gutierrez, G. and Hinton, S. R. and Hollowood, D. L. and Honscheid, K. and Huterer, D. and Jain, B. and James, D. J. and Kuehn, K. and Kuropatkin, N. and Lima, M. and Maia, M. A. G. and March, M. and Marshall, J. L. and Melchior, P. and Menanteau, F. and Miller, C. J. and Miquel, R. and Mohr, J. J. and Morgan, R. and Palmese, A. and Paz-Chinchón, F. and Pieres, A. and Plazas Malagón, A. A. and Roodman, A. and Scarpine, V. and Serrano, S. and Smith, M. and Soares-Santos, M. and Suchyta, E. and Tarle, G. and Thomas, D. and Varga, T. N. and {DES Collaboration}},
	month = apr,
	year = {2022},
	note = {ADS Bibcode: 2022MNRAS.511.2665R},
	pages = {2665--2687},
}

@article{dvornik_kids-1000_2023,
	title = {{KiDS}-1000: {Combined} halo-model cosmology constraints from galaxy abundance, galaxy clustering and galaxy-galaxy lensing},
	volume = {675},
	issn = {0004-6361, 1432-0746},
	shorttitle = {{KiDS}-1000},
	url = {http://arxiv.org/abs/2210.03110},
	doi = {10.1051/0004-6361/202245158},
	urldate = {2024-04-12},
	journal = {Astronomy \& Astrophysics},
	author = {Dvornik, Andrej and Heymans, Catherine and Asgari, Marika and Mahony, Constance and Joachimi, Benjamin and Bilicki, Maciej and Chisari, Elisa and Hildebrandt, Hendrik and Hoekstra, Henk and Johnston, Harry and Kuijken, Konrad and Mead, Alexander and Miyatake, Hironao and Nishimichi, Takahiro and Reischke, Robert and Unruh, Sandra and Wright, Angus H.},
	month = jul,
	year = {2023},
	note = {arXiv:2210.03110 [astro-ph]},
	pages = {A189},
}

@ARTICLE{prabhu_2024,
       author = {{Prabhu}, K. and {Raghunathan}, S. and {Millea}, M. and {Lynch}, G.~P. and {Ade}, P.~A.~R. and {Anderes}, E. and {Anderson}, A.~J. and {Ansarinejad}, B. and {Archipley}, M. and {Balkenhol}, L. and {Benabed}, K. and {Bender}, A.~N. and {Benson}, B.~A. and {Bianchini}, F. and {Bleem}, L.~E. and {Bouchet}, F.~R. and {Bryant}, L. and {Camphuis}, E. and {Carlstrom}, J.~E. and {Cecil}, T.~W. and {Chang}, C.~L. and {Chaubal}, P. and {Chichura}, P.~M. and {Chokshi}, A. and {Chou}, T. -L. and {Coerver}, A. and {Crawford}, T.~M. and {Cukierman}, A. and {Daley}, C. and {de Haan}, T. and {Dibert}, K.~R. and {Dobbs}, M.~A. and {Doussot}, A. and {Dutcher}, D. and {Everett}, W. and {Feng}, C. and {Ferguson}, K.~R. and {Fichman}, K. and {Foster}, A. and {Galli}, S. and {Gambrel}, A.~E. and {Gardner}, R.~W. and {Ge}, F. and {Goeckner-Wald}, N. and {Gualtieri}, R. and {Guidi}, F. and {Guns}, S. and {Halverson}, N.~W. and {Hivon}, E. and {Holder}, G.~P. and {Holzapfel}, W.~L. and {Hood}, J.~C. and {Hryciuk}, A. and {Huang}, N. and {K{\'e}ruzor{\'e}}, F. and {Knox}, L. and {Korman}, M. and {Kornoelje}, K. and {Kuo}, C. -L. and {Lee}, A.~T. and {Levy}, K. and {Lowitz}, A.~E. and {Lu}, C. and {Maniyar}, A. and {Menanteau}, F. and {Montgomery}, J. and {Nakato}, Y. and {Natoli}, T. and {Noble}, G.~I. and {Novosad}, V. and {Omori}, Y. and {Padin}, S. and {Pan}, Z. and {Paschos}, P. and {Phadke}, K.~A. and {Pollak}, A.~W. and {Quan}, W. and {Rahimi}, M. and {Rahlin}, A. and {Reichardt}, C.~L. and {Rouble}, M. and {Ruhl}, J.~E. and {Schiappucci}, E. and {Smecher}, G. and {Sobrin}, J.~A. and {Stark}, A.~A. and {Stephen}, J. and {Suzuki}, A. and {Tandoi}, C. and {Thompson}, K.~L. and {Thorne}, B. and {Trendafilova}, C. and {Tucker}, C. and {Umilta}, C. and {Vitrier}, A. and {Vieira}, J.~D. and {Wan}, Y. and {Wang}, G. and {Whitehorn}, N. and {Wu}, W.~L.~K. and {Yefremenko}, V. and {Young}, M.~R. and {Zebrowski}, J.~A.},
        title = "{Testing the {\ensuremath{\Lambda}}CDM Cosmological Model with Forthcoming Measurements of the Cosmic Microwave Background with SPT-3G}",
      journal = {\apj},
         year = 2024,
        month = sep,
       volume = {973},
       number = {1},
          eid = {4},
        pages = {4},
          doi = {10.3847/1538-4357/ad5ff1},
archivePrefix = {arXiv},
       eprint = {2403.17925},
 primaryClass = {astro-ph.CO},
       adsurl = {https://ui.adsabs.harvard.edu/abs/2024ApJ...973....4P}
}

@ARTICLE{dey19,
       author = {{Dey}, Arjun and {Schlegel}, David J. and {Lang}, Dustin and {Blum}, Robert and {Burleigh}, Kaylan and {Fan}, Xiaohui and {Findlay}, Joseph R. and {Finkbeiner}, Doug and {Herrera}, David and {Juneau}, St{\'e}phanie and {Landriau}, Martin and {Levi}, Michael and {McGreer}, Ian and {Meisner}, Aaron and {Myers}, Adam D. and {Moustakas}, John and {Nugent}, Peter and {Patej}, Anna and {Schlafly}, Edward F. and {Walker}, Alistair R. and {Valdes}, Francisco and {Weaver}, Benjamin A. and {Y{\`e}che}, Christophe and {Zou}, Hu and {Zhou}, Xu and {Abareshi}, Behzad and {Abbott}, T.~M.~C. and {Abolfathi}, Bela and {Aguilera}, C. and {Alam}, Shadab and {Allen}, Lori and {Alvarez}, A. and {Annis}, James and {Ansarinejad}, Behzad and {Aubert}, Marie and {Beechert}, Jacqueline and {Bell}, Eric F. and {BenZvi}, Segev Y. and {Beutler}, Florian and {Bielby}, Richard M. and {Bolton}, Adam S. and {Brice{\~n}o}, C{\'e}sar and {Buckley-Geer}, Elizabeth J. and {Butler}, Karen and {Calamida}, Annalisa and {Carlberg}, Raymond G. and {Carter}, Paul and {Casas}, Ricard and {Castander}, Francisco J. and {Choi}, Yumi and {Comparat}, Johan and {Cukanovaite}, Elena and {Delubac}, Timoth{\'e}e and {DeVries}, Kaitlin and {Dey}, Sharmila and {Dhungana}, Govinda and {Dickinson}, Mark and {Ding}, Zhejie and {Donaldson}, John B. and {Duan}, Yutong and {Duckworth}, Christopher J. and {Eftekharzadeh}, Sarah and {Eisenstein}, Daniel J. and {Etourneau}, Thomas and {Fagrelius}, Parker A. and {Farihi}, Jay and {Fitzpatrick}, Mike and {Font-Ribera}, Andreu and {Fulmer}, Leah and {G{\"a}nsicke}, Boris T. and {Gaztanaga}, Enrique and {George}, Koshy and {Gerdes}, David W. and {Gontcho}, Satya Gontcho A. and {Gorgoni}, Claudio and {Green}, Gregory and {Guy}, Julien and {Harmer}, Diane and {Hernandez}, M. and {Honscheid}, Klaus and {Huang}, Lijuan Wendy and {James}, David J. and {Jannuzi}, Buell T. and {Jiang}, Linhua and {Joyce}, Richard and {Karcher}, Armin and {Karkar}, Sonia and {Kehoe}, Robert and {Kneib}, Jean-Paul and {Kueter-Young}, Andrea and {Lan}, Ting-Wen and {Lauer}, Tod R. and {Le Guillou}, Laurent and {Le Van Suu}, Auguste and {Lee}, Jae Hyeon and {Lesser}, Michael and {Perreault Levasseur}, Laurence and {Li}, Ting S. and {Mann}, Justin L. and {Marshall}, Robert and {Mart{\'\i}nez-V{\'a}zquez}, C.~E. and {Martini}, Paul and {du Mas des Bourboux}, H{\'e}lion and {McManus}, Sean and {Meier}, Tobias Gabriel and {M{\'e}nard}, Brice and {Metcalfe}, Nigel and {Mu{\~n}oz-Guti{\'e}rrez}, Andrea and {Najita}, Joan and {Napier}, Kevin and {Narayan}, Gautham and {Newman}, Jeffrey A. and {Nie}, Jundan and {Nord}, Brian and {Norman}, Dara J. and {Olsen}, Knut A.~G. and {Paat}, Anthony and {Palanque-Delabrouille}, Nathalie and {Peng}, Xiyan and {Poppett}, Claire L. and {Poremba}, Megan R. and {Prakash}, Abhishek and {Rabinowitz}, David and {Raichoor}, Anand and {Rezaie}, Mehdi and {Robertson}, A.~N. and {Roe}, Natalie A. and {Ross}, Ashley J. and {Ross}, Nicholas P. and {Rudnick}, Gregory and {Safonova}, Sasha and {Saha}, Abhijit and {S{\'a}nchez}, F. Javier and {Savary}, Elodie and {Schweiker}, Heidi and {Scott}, Adam and {Seo}, Hee-Jong and {Shan}, Huanyuan and {Silva}, David R. and {Slepian}, Zachary and {Soto}, Christian and {Sprayberry}, David and {Staten}, Ryan and {Stillman}, Coley M. and {Stupak}, Robert J. and {Summers}, David L. and {Sien Tie}, Suk and {Tirado}, H. and {Vargas-Maga{\~n}a}, Mariana and {Vivas}, A. Katherina and {Wechsler}, Risa H. and {Williams}, Doug and {Yang}, Jinyi and {Yang}, Qian and {Yapici}, Tolga and {Zaritsky}, Dennis and {Zenteno}, A. and {Zhang}, Kai and {Zhang}, Tianmeng and {Zhou}, Rongpu and {Zhou}, Zhimin},
        title = "{Overview of the DESI Legacy Imaging Surveys}",
      journal = {\aj},
         year = 2019,
        month = may,
       volume = {157},
       number = {5},
          eid = {168},
        pages = {168},
          doi = {10.3847/1538-3881/ab089d},
archivePrefix = {arXiv},
       eprint = {1804.08657},
 primaryClass = {astro-ph.IM},
       adsurl = {https://ui.adsabs.harvard.edu/abs/2019AJ....157..168D}
}

@ARTICLE{Omori_2024_Agora,
       author = {{Omori}, Yuuki},
        title = "{AGORA: Multicomponent simulation for cross-survey science}",
      journal = {\mnras},
         year = 2024,
        month = jun,
       volume = {530},
       number = {4},
        pages = {5030-5068},
          doi = {10.1093/mnras/stae1031},
archivePrefix = {arXiv},
       eprint = {2212.07420},
 primaryClass = {astro-ph.CO},
       adsurl = {https://ui.adsabs.harvard.edu/abs/2024MNRAS.530.5030O}
}

@ARTICLE{bocquet_cluster_2024a,
       author = {{Bocquet}, S. and {Grandis}, S. and {Bleem}, L.~E. and {Klein}, M. and {Mohr}, J.~J. and {Aguena}, M. and {Alarcon}, A. and {Allam}, S. and {Allen}, S.~W. and {Alves}, O. and {Amon}, A. and {Ansarinejad}, B. and {Bacon}, D. and {Bayliss}, M. and {Bechtol}, K. and {Becker}, M.~R. and {Benson}, B.~A. and {Bernstein}, G.~M. and {Brodwin}, M. and {Brooks}, D. and {Campos}, A. and {Canning}, R.~E.~A. and {Carlstrom}, J.~E. and {Carnero Rosell}, A. and {Carrasco Kind}, M. and {Carretero}, J. and {Cawthon}, R. and {Chang}, C. and {Chen}, R. and {Choi}, A. and {Cordero}, J. and {Costanzi}, M. and {da Costa}, L.~N. and {Pereira}, M.~E.~S. and {Davis}, C. and {DeRose}, J. and {Desai}, S. and {de Haan}, T. and {De Vicente}, J. and {Diehl}, H.~T. and {Dodelson}, S. and {Doel}, P. and {Doux}, C. and {Drlica-Wagner}, A. and {Eckert}, K. and {Elvin-Poole}, J. and {Everett}, S. and {Ferrero}, I. and {Fert{\'e}}, A. and {Flores}, A.~M. and {Frieman}, J. and {Garc{\'\i}a-Bellido}, J. and {Gatti}, M. and {Giannini}, G. and {Gladders}, M.~D. and {Gruen}, D. and {Gruendl}, R.~A. and {Harrison}, I. and {Hartley}, W.~G. and {Herner}, K. and {Hinton}, S.~R. and {Hollowood}, D.~L. and {Holzapfel}, W.~L. and {Honscheid}, K. and {Huang}, N. and {Huff}, E.~M. and {James}, D.~J. and {Jarvis}, M. and {Khullar}, G. and {Kim}, K. and {Kraft}, R. and {Kuehn}, K. and {Kuropatkin}, N. and {K{\'e}ruzor{\'e}}, F. and {Lee}, S. and {Leget}, P.-F. and {MacCrann}, N. and {Mahler}, G. and {Mantz}, A. and {Marshall}, J.~L. and {McCullough}, J. and {McDonald}, M. and {Mena-Fern{\'a}ndez}, J. and {Miquel}, R. and {Myles}, J. and {Navarro-Alsina}, A. and {Ogando}, R.~L.~C. and {Palmese}, A. and {Pandey}, S. and {Pieres}, A. and {Plazas Malag{\'o}n}, A.~A. and {Prat}, J. and {Raveri}, M. and {Reichardt}, C.~L. and {Roberson}, J. and {Rollins}, R.~P. and {Romer}, A.~K. and {Romero}, C. and {Roodman}, A. and {Ross}, A.~J. and {Rykoff}, E.~S. and {Salvati}, L. and {S{\'a}nchez}, C. and {Sanchez}, E. and {Sanchez Cid}, D. and {Saro}, A. and {Schrabback}, T. and {Schubnell}, M. and {Secco}, L.~F. and {Sevilla-Noarbe}, I. and {Sharon}, K. and {Sheldon}, E. and {Shin}, T. and {Smith}, M. and {Somboonpanyakul}, T. and {Stalder}, B. and {Stark}, A.~A. and {Strazzullo}, V. and {Suchyta}, E. and {Swanson}, M.~E.~C. and {Tarle}, G. and {To}, C. and {Troxel}, M.~A. and {Tutusaus}, I. and {Varga}, T.~N. and {von der Linden}, A. and {Weaverdyck}, N. and {Weller}, J. and {Wiseman}, P. and {Yanny}, B. and {Yin}, B. and {Young}, M. and {Zhang}, Y. and {Zuntz}, J. and {(The DES} and {SPT Collaborations)}},
        title = "{SPT clusters with DES and HST weak lensing. I. Cluster lensing and Bayesian population modeling of multiwavelength cluster datasets}",
      journal = {\prd},
         year = 2024,
        month = oct,
       volume = {110},
       number = {8},
          eid = {083509},
        pages = {083509},
          doi = {10.1103/PhysRevD.110.083509},
archivePrefix = {arXiv},
       eprint = {2310.12213},
 primaryClass = {astro-ph.CO},
       adsurl = {https://ui.adsabs.harvard.edu/abs/2024PhRvD.110h3509B}
}

@ARTICLE{bocquet_cluster_2024b,
       author = {{Bocquet}, S. and {Grandis}, S. and {Bleem}, L.~E. and {Klein}, M. and {Mohr}, J.~J. and {Schrabback}, T. and {Abbott}, T.~M.~C. and {Ade}, P.~A.~R. and {Aguena}, M. and {Alarcon}, A. and {Allam}, S. and {Allen}, S.~W. and {Alves}, O. and {Amon}, A. and {Anderson}, A.~J. and {Annis}, J. and {Ansarinejad}, B. and {Austermann}, J.~E. and {Avila}, S. and {Bacon}, D. and {Bayliss}, M. and {Beall}, J.~A. and {Bechtol}, K. and {Becker}, M.~R. and {Bender}, A.~N. and {Benson}, B.~A. and {Bernstein}, G.~M. and {Bhargava}, S. and {Bianchini}, F. and {Brodwin}, M. and {Brooks}, D. and {Bryant}, L. and {Campos}, A. and {Canning}, R.~E.~A. and {Carlstrom}, J.~E. and {Carnero Rosell}, A. and {Carrasco Kind}, M. and {Carretero}, J. and {Castander}, F.~J. and {Cawthon}, R. and {Chang}, C.~L. and {Chang}, C. and {Chaubal}, P. and {Chen}, R. and {Chiang}, H.~C. and {Choi}, A. and {Chou}, T.-L. and {Citron}, R. and {Corbett Moran}, C. and {Cordero}, J. and {Costanzi}, M. and {Crawford}, T.~M. and {Crites}, A.~T. and {da Costa}, L.~N. and {Pereira}, M.~E.~S. and {Davis}, C. and {Davis}, T.~M. and {DeRose}, J. and {Desai}, S. and {de Haan}, T. and {Diehl}, H.~T. and {Dobbs}, M.~A. and {Dodelson}, S. and {Doux}, C. and {Drlica-Wagner}, A. and {Eckert}, K. and {Elvin-Poole}, J. and {Everett}, S. and {Everett}, W. and {Ferrero}, I. and {Fert{\'e}}, A. and {Flores}, A.~M. and {Frieman}, J. and {Gallicchio}, J. and {Garc{\'\i}a-Bellido}, J. and {Gatti}, M. and {George}, E.~M. and {Giannini}, G. and {Gladders}, M.~D. and {Gruen}, D. and {Gruendl}, R.~A. and {Gupta}, N. and {Gutierrez}, G. and {Halverson}, N.~W. and {Harrison}, I. and {Hartley}, W.~G. and {Herner}, K. and {Hinton}, S.~R. and {Holder}, G.~P. and {Hollowood}, D.~L. and {Holzapfel}, W.~L. and {Honscheid}, K. and {Hrubes}, J.~D. and {Huang}, N. and {Hubmayr}, J. and {Huff}, E.~M. and {Huterer}, D. and {Irwin}, K.~D. and {James}, D.~J. and {Jarvis}, M. and {Khullar}, G. and {Kim}, K. and {Knox}, L. and {Kraft}, R. and {Krause}, E. and {Kuehn}, K. and {Kuropatkin}, N. and {K{\'e}ruzor{\'e}}, F. and {Lahav}, O. and {Lee}, A.~T. and {Leget}, P.-F. and {Li}, D. and {Lin}, H. and {Lowitz}, A. and {MacCrann}, N. and {Mahler}, G. and {Mantz}, A. and {Marshall}, J.~L. and {McCullough}, J. and {McDonald}, M. and {McMahon}, J.~J. and {Mena-Fern{\'a}ndez}, J. and {Menanteau}, F. and {Meyer}, S.~S. and {Miquel}, R. and {Montgomery}, J. and {Myles}, J. and {Natoli}, T. and {Navarro-Alsina}, A. and {Nibarger}, J.~P. and {Noble}, G.~I. and {Novosad}, V. and {Ogando}, R.~L.~C. and {Omori}, Y. and {Padin}, S. and {Pandey}, S. and {Paschos}, P. and {Patil}, S. and {Pieres}, A. and {Plazas Malag{\'o}n}, A.~A. and {Porredon}, A. and {Prat}, J. and {Pryke}, C. and {Raveri}, M. and {Reichardt}, C.~L. and {Roberson}, J. and {Rollins}, R.~P. and {Romero}, C. and {Roodman}, A. and {Ruhl}, J.~E. and {Rykoff}, E.~S. and {Saliwanchik}, B.~R. and {Salvati}, L. and {S{\'a}nchez}, C. and {Sanchez}, E. and {Sanchez Cid}, D. and {Saro}, A. and {Schaffer}, K.~K. and {Secco}, L.~F. and {Sevilla-Noarbe}, I. and {Sharon}, K. and {Sheldon}, E. and {Shin}, T. and {Sievers}, C. and {Smecher}, G. and {Smith}, M. and {Somboonpanyakul}, T. and {Sommer}, M. and {Stalder}, B. and {Stark}, A.~A. and {Stephen}, J. and {Strazzullo}, V. and {Suchyta}, E. and {Tarle}, G. and {To}, C. and {Troxel}, M.~A. and {Tucker}, C. and {Tutusaus}, I. and {Varga}, T.~N. and {Veach}, T. and {Vieira}, J.~D. and {Vikhlinin}, A. and {von der Linden}, A. and {Wang}, G. and {Weaverdyck}, N. and {Weller}, J. and {Whitehorn}, N. and {Wu}, W.~L.~K. and {Yanny}, B. and {Yefremenko}, V. and {Yin}, B. and {Young}, M. and {Zebrowski}, J.~A. and {Zhang}, Y. and {Zohren}, H. and {Zuntz}, J. and {(SPT} and {DES Collaborations)}},
        title = "{SPT clusters with DES and HST weak lensing. II. Cosmological constraints from the abundance of massive halos}",
      journal = {\prd},
         year = 2024,
        month = oct,
       volume = {110},
       number = {8},
          eid = {083510},
        pages = {083510},
          doi = {10.1103/PhysRevD.110.083510},
archivePrefix = {arXiv},
       eprint = {2401.02075},
 primaryClass = {astro-ph.CO},
       adsurl = {https://ui.adsabs.harvard.edu/abs/2024PhRvD.110h3510B}
}

@ARTICLE{landy_szalay,
       author = {{Landy}, Stephen D. and {Szalay}, Alexander S.},
        title = "{Bias and Variance of Angular Correlation Functions}",
      journal = {\apj},
         year = 1993,
        month = jul,
       volume = {412},
        pages = {64},
          doi = {10.1086/172900},
       adsurl = {https://ui.adsabs.harvard.edu/abs/1993ApJ...412...64L}
}

@ARTICLE{sinha_2020_corrfunc,
    author = {{Sinha}, Manodeep and {Garrison}, Lehman H.},
    title = "{CORRFUNC - a suite of blazing fast correlation functions on
    the CPU}",
    journal = {\mnras},
    year = "2020",
    month = "Jan",
    volume = {491},
    number = {2},
    pages = {3022-3041},
    doi = {10.1093/mnras/stz3157},
    adsurl =
    {https://ui.adsabs.harvard.edu/abs/2020MNRAS.491.3022S}
}

@ARTICLE{hamilton_1992,
       author = {{Hamilton}, A.~J.~S.},
        title = "{Measuring Omega and the Real Correlation Function from the Redshift Correlation Function}",
      journal = {\apjl},
         year = 1992,
        month = jan,
       volume = {385},
        pages = {L5},
          doi = {10.1086/186264},
       adsurl = {https://ui.adsabs.harvard.edu/abs/1992ApJ...385L...5H}
}

@ARTICLE{chisari_2019_ccl,
       author = {{Chisari}, Nora Elisa and {Alonso}, David and {Krause}, Elisabeth and {Leonard}, C. Danielle and {Bull}, Philip and {Neveu}, J{\'e}r{\'e}my and {Villarreal}, Antonia Sierra and {Singh}, Sukhdeep and {McClintock}, Thomas and {Ellison}, John and {Du}, Zilong and {Zuntz}, Joe and {Mead}, Alexander and {Joudaki}, Shahab and {Lorenz}, Christiane S. and {Tr{\"o}ster}, Tilman and {Sanchez}, Javier and {Lanusse}, Francois and {Ishak}, Mustapha and {Hlozek}, Ren{\'e}e and {Blazek}, Jonathan and {Campagne}, Jean-Eric and {Almoubayyed}, Husni and {Eifler}, Tim and {Kirby}, Matthew and {Kirkby}, David and {Plaszczynski}, St{\'e}phane and {Slosar}, An{\v{z}}e and {Vrastil}, Michal and {Wagoner}, Erika L. and {LSST Dark Energy Science Collaboration}},
        title = "{Core Cosmology Library: Precision Cosmological Predictions for LSST}",
      journal = {\apjs},
         year = 2019,
        month = may,
       volume = {242},
       number = {1},
          eid = {2},
        pages = {2},
          doi = {10.3847/1538-4365/ab1658},
archivePrefix = {arXiv},
       eprint = {1812.05995},
 primaryClass = {astro-ph.CO},
       adsurl = {https://ui.adsabs.harvard.edu/abs/2019ApJS..242....2C}
}

@ARTICLE{tinker_2010,
       author = {{Tinker}, Jeremy L. and {Robertson}, Brant E. and {Kravtsov}, Andrey V. and {Klypin}, Anatoly and {Warren}, Michael S. and {Yepes}, Gustavo and {Gottl{\"o}ber}, Stefan},
        title = "{The Large-scale Bias of Dark Matter Halos: Numerical Calibration and Model Tests}",
      journal = {\apj},
         year = 2010,
        month = dec,
       volume = {724},
       number = {2},
        pages = {878-886},
          doi = {10.1088/0004-637X/724/2/878},
archivePrefix = {arXiv},
       eprint = {1001.3162},
 primaryClass = {astro-ph.CO},
       adsurl = {https://ui.adsabs.harvard.edu/abs/2010ApJ...724..878T}
}

@ARTICLE{cromer_2019,
       author = {{Cromer}, Dylan and {Battaglia}, Nicholas and {Madhavacheril}, Mathew S.},
        title = "{Improving constraints on fundamental physics parameters with the clustering of Sunyaev-Zeldovich selected galaxy clusters}",
      journal = {\prd},
         year = 2019,
        month = sep,
       volume = {100},
       number = {6},
          eid = {063529},
        pages = {063529},
          doi = {10.1103/PhysRevD.100.063529},
archivePrefix = {arXiv},
       eprint = {1903.00976},
 primaryClass = {astro-ph.CO},
       adsurl = {https://ui.adsabs.harvard.edu/abs/2019PhRvD.100f3529C}
}

@ARTICLE{ghirardini_2024_erosita,
       author = {{Ghirardini}, V. and {Bulbul}, E. and {Artis}, E. and {Clerc}, N. and {Garrel}, C. and {Grandis}, S. and {Kluge}, M. and {Liu}, A. and {Bahar}, Y.~E. and {Balzer}, F. and {Chiu}, I. and {Comparat}, J. and {Gruen}, D. and {Kleinebreil}, F. and {Krippendorf}, S. and {Merloni}, A. and {Nandra}, K. and {Okabe}, N. and {Pacaud}, F. and {Predehl}, P. and {Ramos-Ceja}, M.~E. and {Reiprich}, T.~H. and {Sanders}, J.~S. and {Schrabback}, T. and {Seppi}, R. and {Zelmer}, S. and {Zhang}, X. and {Bornemann}, W. and {Brunner}, H. and {Burwitz}, V. and {Coutinho}, D. and {Dennerl}, K. and {Freyberg}, M. and {Friedrich}, S. and {Gaida}, R. and {Gueguen}, A. and {Haberl}, F. and {Kink}, W. and {Lamer}, G. and {Li}, X. and {Liu}, T. and {Maitra}, C. and {Meidinger}, N. and {Mueller}, S. and {Miyatake}, H. and {Miyazaki}, S. and {Robrade}, J. and {Schwope}, A. and {Stewart}, I.},
        title = "{The SRG/eROSITA all-sky survey: Cosmology constraints from cluster abundances in the western Galactic hemisphere}",
      journal = {\aap},
         year = 2024,
        month = sep,
       volume = {689},
          eid = {A298},
        pages = {A298},
          doi = {10.1051/0004-6361/202348852},
archivePrefix = {arXiv},
       eprint = {2402.08458},
 primaryClass = {astro-ph.CO},
       adsurl = {https://ui.adsabs.harvard.edu/abs/2024A&A...689A.298G}
}

@ARTICLE{Sobrin_2022,
       author = {{Sobrin}, J.~A. and {Anderson}, A.~J. and {Bender}, A.~N. and {Benson}, B.~A. and {Dutcher}, D. and {Foster}, A. and {Goeckner-Wald}, N. and {Montgomery}, J. and {Nadolski}, A. and {Rahlin}, A. and {Ade}, P.~A.~R. and {Ahmed}, Z. and {Anderes}, E. and {Archipley}, M. and {Austermann}, J.~E. and {Avva}, J.~S. and {Aylor}, K. and {Balkenhol}, L. and {Barry}, P.~S. and {Thakur}, R. Basu and {Benabed}, K. and {Bianchini}, F. and {Bleem}, L.~E. and {Bouchet}, F.~R. and {Bryant}, L. and {Byrum}, K. and {Carlstrom}, J.~E. and {Carter}, F.~W. and {Cecil}, T.~W. and {Chang}, C.~L. and {Chaubal}, P. and {Chen}, G. and {Cho}, H. -M. and {Chou}, T. -L. and {Cliche}, J. -F. and {Crawford}, T.~M. and {Cukierman}, A. and {Daley}, C. and {de Haan}, T. and {Denison}, E.~V. and {Dibert}, K. and {Ding}, J. and {Dobbs}, M.~A. and {Everett}, W. and {Feng}, C. and {Ferguson}, K.~R. and {Fu}, J. and {Galli}, S. and {Gambrel}, A.~E. and {Gardner}, R.~W. and {Gualtieri}, R. and {Guns}, S. and {Gupta}, N. and {Guyser}, R. and {Halverson}, N.~W. and {Harke-Hosemann}, A.~H. and {Harrington}, N.~L. and {Henning}, J.~W. and {Hilton}, G.~C. and {Hivon}, E. and {Holder}, G.~P. and {Holzapfel}, W.~L. and {Hood}, J.~C. and {Howe}, D. and {Huang}, N. and {Irwin}, K.~D. and {Jeong}, O.~B. and {Jonas}, M. and {Jones}, A. and {Khaire}, T.~S. and {Knox}, L. and {Kofman}, A.~M. and {Korman}, M. and {Kubik}, D.~L. and {Kuhlmann}, S. and {Kuo}, C. -L. and {Lee}, A.~T. and {Leitch}, E.~M. and {Lowitz}, A.~E. and {Lu}, C. and {Meyer}, S.~S. and {Michalik}, D. and {Millea}, M. and {Natoli}, T. and {Nguyen}, H. and {Noble}, G.~I. and {Novosad}, V. and {Omori}, Y. and {Padin}, S. and {Pan}, Z. and {Paschos}, P. and {Pearson}, J. and {Posada}, C.~M. and {Prabhu}, K. and {Quan}, W. and {Reichardt}, C.~L. and {Riebel}, D. and {Riedel}, B. and {Rouble}, M. and {Ruhl}, J.~E. and {Saliwanchik}, B. and {Sayre}, J.~T. and {Schiappucci}, E. and {Shirokoff}, E. and {Smecher}, G. and {Stark}, A.~A. and {Stephen}, J. and {Story}, K.~T. and {Suzuki}, A. and {Tandoi}, C. and {Thompson}, K.~L. and {Thorne}, B. and {Tucker}, C. and {Umilta}, C. and {Vale}, L.~R. and {Vanderlinde}, K. and {Vieira}, J.~D. and {Wang}, G. and {Whitehorn}, N. and {Wu}, W.~L.~K. and {Yefremenko}, V. and {Yoon}, K.~W. and {Young}, M.~R.},
        title = "{The Design and Integrated Performance of SPT-3G}",
      journal = {\apjs},
         year = 2022,
        month = feb,
       volume = {258},
       number = {2},
          eid = {42},
        pages = {42},
          doi = {10.3847/1538-4365/ac374f},
archivePrefix = {arXiv},
       eprint = {2106.11202},
 primaryClass = {astro-ph.IM},
       adsurl = {https://ui.adsabs.harvard.edu/abs/2022ApJS..258...42S}
}

@ARTICLE{Heitmann2021,
       author = {{Heitmann}, Katrin and {Frontiere}, Nicholas and {Rangel}, Esteban and {Larsen}, Patricia and {Pope}, Adrian and {Sultan}, Imran and {Uram}, Thomas and {Habib}, Salman and {Finkel}, Hal and {Korytov}, Danila and {Kovacs}, Eve and {Rizzi}, Silvio and {Insley}, Joe and {Knowles}, Janet Y.~K.},
        title = "{The Last Journey. I. An Extreme-scale Simulation on the Mira Supercomputer}",
      journal = {\apjs},
         year = 2021,
        month = feb,
       volume = {252},
       number = {2},
          eid = {19},
        pages = {19},
          doi = {10.3847/1538-4365/abcc67},
archivePrefix = {arXiv},
       eprint = {2006.01697},
 primaryClass = {astro-ph.CO},
       adsurl = {https://ui.adsabs.harvard.edu/abs/2021ApJS..252...19H}
}

@ARTICLE{habib2016,
       author = {{Habib}, Salman and {Pope}, Adrian and {Finkel}, Hal and {Frontiere}, Nicholas and {Heitmann}, Katrin and {Daniel}, David and {Fasel}, Patricia and {Morozov}, Vitali and {Zagaris}, George and {Peterka}, Tom and {Vishwanath}, Venkatram and {Luki{\'c}}, Zarija and {Sehrish}, Saba and {Liao}, Wei-keng},
        title = "{HACC: Simulating sky surveys on state-of-the-art supercomputing architectures}",
      journal = {\na},
         year = 2016,
        month = jan,
       volume = {42},
        pages = {49-65},
          doi = {10.1016/j.newast.2015.06.003},
archivePrefix = {arXiv},
       eprint = {1410.2805},
 primaryClass = {astro-ph.IM},
       adsurl = {https://ui.adsabs.harvard.edu/abs/2016NewA...42...49H}
}

@ARTICLE{planck2020,
       author = {{Planck Collaboration} and {Aghanim}, N. and {Akrami}, Y. and {Ashdown}, M. and {Aumont}, J. and {Baccigalupi}, C. and {Ballardini}, M. and {Banday}, A.~J. and {Barreiro}, R.~B. and {Bartolo}, N. and {Basak}, S. and {Battye}, R. and {Benabed}, K. and {Bernard}, J.-P. and {Bersanelli}, M. and {Bielewicz}, P. and {Bock}, J.~J. and {Bond}, J.~R. and {Borrill}, J. and {Bouchet}, F.~R. and {Boulanger}, F. and {Bucher}, M. and {Burigana}, C. and {Butler}, R.~C. and {Calabrese}, E. and {Cardoso}, J.-F. and {Carron}, J. and {Challinor}, A. and {Chiang}, H.~C. and {Chluba}, J. and {Colombo}, L.~P.~L. and {Combet}, C. and {Contreras}, D. and {Crill}, B.~P. and {Cuttaia}, F. and {de Bernardis}, P. and {de Zotti}, G. and {Delabrouille}, J. and {Delouis}, J.-M. and {Di Valentino}, E. and {Diego}, J.~M. and {Dor{\'e}}, O. and {Douspis}, M. and {Ducout}, A. and {Dupac}, X. and {Dusini}, S. and {Efstathiou}, G. and {Elsner}, F. and {En{\ss}lin}, T.~A. and {Eriksen}, H.~K. and {Fantaye}, Y. and {Farhang}, M. and {Fergusson}, J. and {Fernandez-Cobos}, R. and {Finelli}, F. and {Forastieri}, F. and {Frailis}, M. and {Fraisse}, A.~A. and {Franceschi}, E. and {Frolov}, A. and {Galeotta}, S. and {Galli}, S. and {Ganga}, K. and {G{\'e}nova-Santos}, R.~T. and {Gerbino}, M. and {Ghosh}, T. and {Gonz{\'a}lez-Nuevo}, J. and {G{\'o}rski}, K.~M. and {Gratton}, S. and {Gruppuso}, A. and {Gudmundsson}, J.~E. and {Hamann}, J. and {Handley}, W. and {Hansen}, F.~K. and {Herranz}, D. and {Hildebrandt}, S.~R. and {Hivon}, E. and {Huang}, Z. and {Jaffe}, A.~H. and {Jones}, W.~C. and {Karakci}, A. and {Keih{\"a}nen}, E. and {Keskitalo}, R. and {Kiiveri}, K. and {Kim}, J. and {Kisner}, T.~S. and {Knox}, L. and {Krachmalnicoff}, N. and {Kunz}, M. and {Kurki-Suonio}, H. and {Lagache}, G. and {Lamarre}, J.-M. and {Lasenby}, A. and {Lattanzi}, M. and {Lawrence}, C.~R. and {Le Jeune}, M. and {Lemos}, P. and {Lesgourgues}, J. and {Levrier}, F. and {Lewis}, A. and {Liguori}, M. and {Lilje}, P.~B. and {Lilley}, M. and {Lindholm}, V. and {L{\'o}pez-Caniego}, M. and {Lubin}, P.~M. and {Ma}, Y.-Z. and {Mac{\'\i}as-P{\'e}rez}, J.~F. and {Maggio}, G. and {Maino}, D. and {Mandolesi}, N. and {Mangilli}, A. and {Marcos-Caballero}, A. and {Maris}, M. and {Martin}, P.~G. and {Martinelli}, M. and {Mart{\'\i}nez-Gonz{\'a}lez}, E. and {Matarrese}, S. and {Mauri}, N. and {McEwen}, J.~D. and {Meinhold}, P.~R. and {Melchiorri}, A. and {Mennella}, A. and {Migliaccio}, M. and {Millea}, M. and {Mitra}, S. and {Miville-Desch{\^e}nes}, M.-A. and {Molinari}, D. and {Montier}, L. and {Morgante}, G. and {Moss}, A. and {Natoli}, P. and {N{\o}rgaard-Nielsen}, H.~U. and {Pagano}, L. and {Paoletti}, D. and {Partridge}, B. and {Patanchon}, G. and {Peiris}, H.~V. and {Perrotta}, F. and {Pettorino}, V. and {Piacentini}, F. and {Polastri}, L. and {Polenta}, G. and {Puget}, J.-L. and {Rachen}, J.~P. and {Reinecke}, M. and {Remazeilles}, M. and {Renzi}, A. and {Rocha}, G. and {Rosset}, C. and {Roudier}, G. and {Rubi{\~n}o-Mart{\'\i}n}, J.~A. and {Ruiz-Granados}, B. and {Salvati}, L. and {Sandri}, M. and {Savelainen}, M. and {Scott}, D. and {Shellard}, E.~P.~S. and {Sirignano}, C. and {Sirri}, G. and {Spencer}, L.~D. and {Sunyaev}, R. and {Suur-Uski}, A.-S. and {Tauber}, J.~A. and {Tavagnacco}, D. and {Tenti}, M. and {Toffolatti}, L. and {Tomasi}, M. and {Trombetti}, T. and {Valenziano}, L. and {Valiviita}, J. and {Van Tent}, B. and {Vibert}, L. and {Vielva}, P. and {Villa}, F. and {Vittorio}, N. and {Wandelt}, B.~D. and {Wehus}, I.~K. and {White}, M. and {White}, S.~D.~M. and {Zacchei}, A. and {Zonca}, A.},
        title = "{Planck 2018 results. VI. Cosmological parameters}",
      journal = {\aap},
         year = 2020,
        month = sep,
       volume = {641},
          eid = {A6},
        pages = {A6},
          doi = {10.1051/0004-6361/201833910},
archivePrefix = {arXiv},
       eprint = {1807.06209},
 primaryClass = {astro-ph.CO},
       adsurl = {https://ui.adsabs.harvard.edu/abs/2020A&A...641A...6P}
}

@ARTICLE{Child_2018,
       author = {{Child}, Hillary L. and {Habib}, Salman and {Heitmann}, Katrin and {Frontiere}, Nicholas and {Finkel}, Hal and {Pope}, Adrian and {Morozov}, Vitali},
        title = "{Halo Profiles and the Concentration-Mass Relation for a {\ensuremath{\Lambda}}CDM Universe}",
      journal = {\apj},
         year = 2018,
        month = may,
       volume = {859},
       number = {1},
          eid = {55},
        pages = {55},
          doi = {10.3847/1538-4357/aabf95},
archivePrefix = {arXiv},
       eprint = {1804.10199},
 primaryClass = {astro-ph.CO},
       adsurl = {https://ui.adsabs.harvard.edu/abs/2018ApJ...859...55C}
}

@ARTICLE{NFW_1996,
       author = {{Navarro}, Julio F. and {Frenk}, Carlos S. and {White}, Simon D.~M.},
        title = "{The Structure of Cold Dark Matter Halos}",
      journal = {\apj},
         year = 1996,
        month = may,
       volume = {462},
        pages = {563},
          doi = {10.1086/177173},
archivePrefix = {arXiv},
       eprint = {astro-ph/9508025},
 primaryClass = {astro-ph},
       adsurl = {https://ui.adsabs.harvard.edu/abs/1996ApJ...462..563N}
}

@ARTICLE{NFW_1997,
       author = {{Navarro}, Julio F. and {Frenk}, Carlos S. and {White}, Simon D.~M.},
        title = "{A Universal Density Profile from Hierarchical Clustering}",
      journal = {\apj},
         year = 1997,
        month = dec,
       volume = {490},
       number = {2},
        pages = {493-508},
          doi = {10.1086/304888},
archivePrefix = {arXiv},
       eprint = {astro-ph/9611107},
 primaryClass = {astro-ph},
       adsurl = {https://ui.adsabs.harvard.edu/abs/1997ApJ...490..493N}
}

@ARTICLE{wright10,
   author = {{Wright}, E.~L. and {Eisenhardt}, P.~R.~M. and {Mainzer}, A.~K. and 
	{Ressler}, M.~E. and {Cutri}, R.~M. and {Jarrett}, T. and {Kirkpatrick}, J.~D. and 
	{Padgett}, D. and {McMillan}, R.~S. and {Skrutskie}, M. and 
	{Stanford}, S.~A. and {Cohen}, M. and {Walker}, R.~G. and {Mather}, J.~C. and 
	{Leisawitz}, D. and {Gautier}, III, T.~N. and {McLean}, I. and 
	{Benford}, D. and {Lonsdale}, C.~J. and {Blain}, A. and {Mendez}, B. and 
	{Irace}, W.~R. and {Duval}, V. and {Liu}, F. and {Royer}, D. and 
	{Heinrichsen}, I. and {Howard}, J. and {Shannon}, M. and {Kendall}, M. and 
	{Walsh}, A.~L. and {Larsen}, M. and {Cardon}, J.~G. and {Schick}, S. and 
	{Schwalm}, M. and {Abid}, M. and {Fabinsky}, B. and {Naes}, L. and 
	{Tsai}, C.-W.},
    title = "{The Wide-field Infrared Survey Explorer (WISE): Mission Description and Initial On-orbit Performance}",
  journal = {\aj},
archivePrefix = "arXiv",
   eprint = {1008.0031},
 primaryClass = "astro-ph.IM",
     year = 2010,
    month = dec,
   volume = 140,
      eid = {1868},
    pages = {1868-1881},
      doi = {10.1088/0004-6256/140/6/1868},
   adsurl = {http://adsabs.harvard.edu/abs/2010AJ....140.1868W}
}

@ARTICLE{Ingoglia_2022,
       author = {{Ingoglia}, Lorenzo and {Covone}, Giovanni and {Sereno}, Mauro and {Giocoli}, Carlo and {Bardelli}, Sandro and {Bellagamba}, Fabio and {Castignani}, Gianluca and {Farrens}, Samuel and {Hildebrandt}, Hendrik and {Joudaki}, Shahab and {Jullo}, Eric and {Lanzieri}, Denise and {Lesci}, Giorgio F. and {Marulli}, Federico and {Maturi}, Matteo and {Moscardini}, Lauro and {Nanni}, Lorenza and {Puddu}, Emanuela and {Radovich}, Mario and {Roncarelli}, Mauro and {Sapio}, Feliciana and {Schimd}, Carlo},
        title = "{AMICO galaxy clusters in KiDS-DR3: measurement of the halo bias and power spectrum normalization from a stacked weak lensing analysis}",
      journal = {\mnras},
         year = 2022,
        month = mar,
       volume = {511},
       number = {1},
        pages = {1484-1501},
          doi = {10.1093/mnras/stac046},
archivePrefix = {arXiv},
       eprint = {2201.01545},
 primaryClass = {astro-ph.CO},
       adsurl = {https://ui.adsabs.harvard.edu/abs/2022MNRAS.511.1484I}
}

@ARTICLE{Kornoelje_2025,
       author = {{Kornoelje}, K. and {Bleem}, L.~E. and {Rykoff}, E.~S. and {Abbott}, T.~M.~C. and {Ade}, P.~A.~R. and {Aguena}, M. and {Alves}, O. and {Anderson}, A.~J. and {Andrade-Oliveira}, F. and {Ansarinejad}, B. and {Archipley}, M. and {Ashby}, M.~L.~N. and {Austermann}, J.~E. and {Bacon}, D. and {Balkenhol}, L. and {Bayliss}, Matthew B. and {Beall}, J.~A. and {Benabed}, K. and {Bender}, A.~N. and {Benson}, B.~A. and {Bianchini}, F. and {Bocquet}, S. and {Bouchet}, F.~R. and {Brooks}, D. and {Burke}, D.~L. and {Calzadilla}, M. and {Campitiello}, M.~G. and {Camphuis}, E. and {Carlstrom}, J.~E. and {Carnero Rosell}, A. and {Carretero}, J. and {Chang}, C.~L. and {Chaubal}, P. and {Chiang}, H.~C. and {Chichura}, P.~M. and {Chokshi}, A. and {Chou}, T.-L. and {Citron}, R. and {Coerver}, A. and {Moran}, C. Corbett and {Costanzi}, M. and {Crawford}, T.~M. and {Crites}, A.~T. and {da Costa}, L.~N. and {Daley}, C. and {de Haan}, T. and {De Vicente}, J. and {Desai}, S. and {Dibert}, K.~R. and {Dobbs}, M.~A. and {Doel}, P. and {Doohan}, M. and {Doussot}, A. and {Dutcher}, D. and {Everett}, W. and {Everett}, S. and {Feng}, C. and {Ferguson}, K.~R. and {Mena-Fern{\'a}ndez}, J. and {Ferrero}, I. and {Fichman}, K. and {Flaugher}, B. and {Floyd}, B. and {Foster}, A. and {Friedel}, D. and {Frieman}, J. and {Galli}, S. and {Gallicchio}, J. and {Gambrel}, A.~E. and {Garc{\'\i}a-Bellido}, J. and {Gardner}, R.~W. and {Gassis}, R. and {Gatti}, M. and {Ge}, F. and {George}, E.~M. and {Giannini}, G. and {Goeckner-Wald}, N. and {Grandis}, S. and {Gruen}, D. and {Gruendl}, R.~A. and {Gualtieri}, R. and {Guidi}, F. and {Guns}, S. and {Gupta}, N. and {Gutierrez}, G. and {Halverson}, N.~W. and {Hinton}, S.~R. and {Hivon}, E. and {Holder}, G.~P. and {Hollowood}, D.~L. and {Holzapfel}, W.~L. and {Honscheid}, K. and {Hood}, J.~C. and {Hrubes}, J.~D. and {Hryciuk}, A. and {Huang}, N. and {Hubmayr}, J. and {Irwin}, K.~D. and {James}, D.~J. and {K{\'e}ruzor{\'e}}, F. and {Khalife}, A.~R. and {Klein}, M. and {Knox}, L. and {Korman}, M. and {Kuehn}, K. and {Kuo}, C.-L. and {Lahav}, O. and {Lee}, A.~T. and {Lee}, S. and {Levy}, K. and {Li}, D. and {Lima}, M. and {Lowitz}, A.~E. and {Lowitz}, A. and {Lu}, C. and {Mahler}, Guillaume and {Maniyar}, A. and {Marshal}, J.~L. and {Marshall}, J.~L. and {Martsen}, E.~S. and {McDonald}, M. and {McMahon}, J.~J. and {Menanteau}, F. and {Millea}, M. and {Miquel}, R. and {Mohr}, J.~J. and {Montgomery}, J. and {Myles}, J. and {Nakato}, Y. and {Natoli}, T. and {Nibarger}, J.~P. and {Noble}, G.~I. and {Novosad}, V. and {Ogando}, R.~L.~C. and {Omori}, Y. and {Ouellette}, A. and {Padin}, S. and {Pan}, Z. and {Patil}, S. and {Pereira}, M.~E.~S. and {Phadke}, K.~A. and {Pieres}, A. and {Plazas Malag\&ooacute} and {n}, A.~A. and {Pollak}, A.~W. and {Prabhu}, K. and {Pryke}, C. and {Quan}, W. and {Rahimi}, M. and {Rahlin}, A. and {Reichardt}, C.~L. and {Rodr{\'\i}guez-Monroy}, M. and {Romer}, A.~K. and {Rouble}, M. and {Ruhl}, J.~E. and {Saliwanchik}, B.~R. and {Salvati}, L. and {Samuroff}, S. and {Sanchez}, E. and {Sarkar}, Arnab and {Saro}, A. and {Schaffer}, K.~K. and {Schiappucci}, E. and {Schrabback}, T. and {Sevilla-Noarbe}, I. and {Sievers}, C. and {Smecher}, G. and {Smith}, M. and {Sobrin}, J.~A. and {Somboonpanyakul}, T. and {Stalder}, B. and {Stark}, A.~A. and {Strazzullo}, V. and {Suchyta}, E. and {Swanson}, M.~E.~C. and {Tandoi}, C. and {Tarle}, G. and {Thorne}, B. and {To}, C. and {Trendafilova}, C. and {Tucker}, C. and {Umilta}, C. and {Veach}, T. and {Vieira}, J.~D. and {Vincenzi}, M. and {Vitrier}, A. and {Wan}, Y. and {Wang}, G. and {Weaverdyck}, N. and {Weller}, J. and {Whitehorn}, N. and {Wiseman}, P. and {Wu}, W.~L.~K. and {Yefremenko}, V. and {Young}, M.~R. and {Zebrowski}, J.~A. and {Zhang}, Y.},
        title = "{The SPT-deep Cluster Catalog: Sunyaev─Zel'dovich Selected Clusters from Combined SPT-3G and SPTpol Measurements over 100 Square Degrees}",
      journal = {\apj},
         year = 2026,
        month = jun,
       volume = {1004},
       number = {1},
          eid = {48},
        pages = {48},
          doi = {10.3847/1538-4357/ae5f7b},
archivePrefix = {arXiv},
       eprint = {2503.17271},
 primaryClass = {astro-ph.CO},
       adsurl = {https://ui.adsabs.harvard.edu/abs/2026ApJ..1004...48K}
}

@ARTICLE{carlstrom_2010,
       author = {{Carlstrom}, J.~E. and {Ade}, P.~A.~R. and {Aird}, K.~A. and {Benson}, B.~A. and {Bleem}, L.~E. and {Busetti}, S. and {Chang}, C.~L. and {Chauvin}, E. and {Cho}, H.-M. and {Crawford}, T.~M. and {Crites}, A.~T. and {Dobbs}, M.~A. and {Halverson}, N.~W. and {Heimsath}, S. and {Holzapfel}, W.~L. and {Hrubes}, J.~D. and {Joy}, M. and {Keisler}, R. and {Lanting}, T.~M. and {Lee}, A.~T. and {Leitch}, E.~M. and {Leong}, J. and {Lu}, W. and {Lueker}, M. and {Luong-Van}, D. and {McMahon}, J.~J. and {Mehl}, J. and {Meyer}, S.~S. and {Mohr}, J.~J. and {Montroy}, T.~E. and {Padin}, S. and {Plagge}, T. and {Pryke}, C. and {Ruhl}, J.~E. and {Schaffer}, K.~K. and {Schwan}, D. and {Shirokoff}, E. and {Spieler}, H.~G. and {Staniszewski}, Z. and {Stark}, A.~A. and {Tucker}, C. and {Vanderlinde}, K. and {Vieira}, J.~D. and {Williamson}, R.},
        title = "{The 10 Meter South Pole Telescope}",
      journal = {\pasp},
         year = 2011,
        month = may,
       volume = {123},
       number = {903},
        pages = {568},
          doi = {10.1086/659879},
archivePrefix = {arXiv},
       eprint = {0907.4445},
 primaryClass = {astro-ph.IM},
       adsurl = {https://ui.adsabs.harvard.edu/abs/2011PASP..123..568C}
}

@ARTICLE{Vanderlinde_2010,
       author = {{Vanderlinde}, K. and {Crawford}, T.~M. and {de Haan}, T. and {Dudley}, J.~P. and {Shaw}, L. and {Ade}, P.~A.~R. and {Aird}, K.~A. and {Benson}, B.~A. and {Bleem}, L.~E. and {Brodwin}, M. and {Carlstrom}, J.~E. and {Chang}, C.~L. and {Crites}, A.~T. and {Desai}, S. and {Dobbs}, M.~A. and {Foley}, R.~J. and {George}, E.~M. and {Gladders}, M.~D. and {Hall}, N.~R. and {Halverson}, N.~W. and {High}, F.~W. and {Holder}, G.~P. and {Holzapfel}, W.~L. and {Hrubes}, J.~D. and {Joy}, M. and {Keisler}, R. and {Knox}, L. and {Lee}, A.~T. and {Leitch}, E.~M. and {Loehr}, A. and {Lueker}, M. and {Marrone}, D.~P. and {McMahon}, J.~J. and {Mehl}, J. and {Meyer}, S.~S. and {Mohr}, J.~J. and {Montroy}, T.~E. and {Ngeow}, C.-C. and {Padin}, S. and {Plagge}, T. and {Pryke}, C. and {Reichardt}, C.~L. and {Rest}, A. and {Ruel}, J. and {Ruhl}, J.~E. and {Schaffer}, K.~K. and {Shirokoff}, E. and {Song}, J. and {Spieler}, H.~G. and {Stalder}, B. and {Staniszewski}, Z. and {Stark}, A.~A. and {Stubbs}, C.~W. and {van Engelen}, A. and {Vieira}, J.~D. and {Williamson}, R. and {Yang}, Y. and {Zahn}, O. and {Zenteno}, A.},
        title = "{Galaxy Clusters Selected with the Sunyaev-Zel'dovich Effect from 2008 South Pole Telescope Observations}",
      journal = {\apj},
         year = 2010,
        month = oct,
       volume = {722},
       number = {2},
        pages = {1180-1196},
          doi = {10.1088/0004-637X/722/2/1180},
archivePrefix = {arXiv},
       eprint = {1003.0003},
 primaryClass = {astro-ph.CO},
       adsurl = {https://ui.adsabs.harvard.edu/abs/2010ApJ...722.1180V}
}

@ARTICLE{Kacprzak_2023_COSMOGRID,
       author = {{Kacprzak}, Tomasz and {Fluri}, Janis and {Schneider}, Aurel and {Refregier}, Alexandre and {Stadel}, Joachim},
        title = "{CosmoGridV1: a simulated ��CDM theory prediction for map-level cosmological inference}",
      journal = {\jcap},
         year = 2023,
        month = feb,
       volume = {2023},
       number = {2},
          eid = {050},
        pages = {050},
          doi = {10.1088/1475-7516/2023/02/050},
archivePrefix = {arXiv},
       eprint = {2209.04662},
 primaryClass = {astro-ph.CO},
       adsurl = {https://ui.adsabs.harvard.edu/abs/2023JCAP...02..050K}
}

@article{Wang2018BOSSTomography,
    author = {Wang, Yuting and Zhao, Gong-Bo and Chuang, Chia-Hsun
              and Pellejero-Ibanez, Marcos and Zhao, Cheng
              and Kitaura, Francisco-Shu and Rodriguez-Torres, Sergio},
    title = {The clustering of galaxies in the completed SDSS-III
             Baryon Oscillation Spectroscopic Survey: a tomographic
             analysis of structure growth and expansion rate from
             anisotropic galaxy clustering},
    journal = {Monthly Notices of the Royal Astronomical Society},
    volume = {481},
    number = {3},
    pages = {3160--3166},
    year = {2018},
    doi = {10.1093/mnras/sty2449},
    eprint = {1709.05173},
    archivePrefix = {arXiv},
    primaryClass = {astro-ph.CO}
}

@article{Tanidis2023BOSSHarmonic,
    author = {Tanidis, Konstantinos and Camera, Stefano},
    title = {Model-independent Constraints on Clustering and Growth
             of Cosmic Structures from BOSS DR12 Galaxies in
             Harmonic Space},
    journal = {The Astrophysical Journal},
    volume = {948},
    number = {1},
    pages = {6},
    year = {2023},
    doi = {10.3847/1538-4357/acbfb2},
    eprint = {2107.00026},
    archivePrefix = {arXiv},
    primaryClass = {astro-ph.CO}
}

@article{Ruggeri2019eBOSS,
    author = {Ruggeri, Rossana and Percival, Will J. and
              Gil-Mar{\'i}n, H{\'e}ctor and Beutler, Florian and
              Mueller, Eva-Maria and Zhu, Fangzhou and
              Padmanabhan, Nikhil and Zhao, Gong-Bo and
              Zarrouk, Pauline and S{\'a}nchez, Ariel G. and
              Bautista, Julian and Brinkmann, Jonathan and
              Brownstein, Joel R. and Baumgarten, Falk and
              Chuang, Chia-Hsun and Dawson, Kyle and
              Seo, Hee-Jong and Tojeiro, Rita and Zhao, Cheng},
    title = {The clustering of the SDSS-IV extended Baryon
             Oscillation Spectroscopic Survey DR14 quasar sample:
             measuring the evolution of the growth rate using
             redshift-space distortions between redshift 0.8 and 2.2},
    journal = {Monthly Notices of the Royal Astronomical Society},
    volume = {483},
    number = {3},
    pages = {3878--3887},
    year = {2019},
    doi = {10.1093/mnras/sty3395},
    eprint = {1801.02891},
    archivePrefix = {arXiv},
    primaryClass = {astro-ph.CO}
}

@article{Kolmogorov1933,
    author = {Kolmogorov, A. N.},
    title = {Sulla determinazione empirica di una legge di distribuzione},
    journal = {Giornale dell'Istituto Italiano degli Attuari},
    volume = {4},
    pages = {83--91},
    year = {1933}
}

@article{Smirnov1948,
    author = {Smirnov, N.},
    title = {Table for Estimating the Goodness of Fit of Empirical Distributions},
    journal = {The Annals of Mathematical Statistics},
    volume = {19},
    number = {2},
    pages = {279--281},
    year = {1948},
    doi = {10.1214/aoms/1177730256}
}

@ARTICLE{Anbajagane:2026:Colliders,
       author = {{Anbajagane}, Dhayaa and {Lee}, Hayden},
        title = "{Primordial physics in the nonlinear Universe: Mapping cosmological collider models to weak-lensing observables}",
      journal = {\jcap},
         year = 2026,
        month = may,
       volume = {2026},
       number = {5},
          eid = {076},
        pages = {076},
          doi = {10.1088/1475-7516/2026/05/076},
archivePrefix = {arXiv},
       eprint = {2509.02693},
 primaryClass = {astro-ph.CO},
       adsurl = {https://ui.adsabs.harvard.edu/abs/2026JCAP...05..076A}
}

@ARTICLE{Anbajagane_2024_ULAGAM,
       author = {{Anbajagane}, Dhayaa and {Chang}, Chihway and {Lee}, Hayden and {Gatti}, Marco},
        title = "{Primordial non-Gaussianities with weak lensing: information on non-linear scales in the ULAGAM full-sky simulations}",
      journal = {\jcap},
         year = 2024,
        month = mar,
       volume = {2024},
       number = {3},
          eid = {062},
        pages = {062},
          doi = {10.1088/1475-7516/2024/03/062},
archivePrefix = {arXiv},
       eprint = {2310.02349},
 primaryClass = {astro-ph.CO},
       adsurl = {https://ui.adsabs.harvard.edu/abs/2024JCAP...03..062A}
}

@ARTICLE{Wright_2010_WISE,
       author = {{Wright}, Edward L. and {Eisenhardt}, Peter R.~M. and {Mainzer}, Amy K. and {Ressler}, Michael E. and {Cutri}, Roc M. and {Jarrett}, Thomas and {Kirkpatrick}, J. Davy and {Padgett}, Deborah and {McMillan}, Robert S. and {Skrutskie}, Michael and {Stanford}, S.~A. and {Cohen}, Martin and {Walker}, Russell G. and {Mather}, John C. and {Leisawitz}, David and {Gautier}, III, Thomas N. and {McLean}, Ian and {Benford}, Dominic and {Lonsdale}, Carol J. and {Blain}, Andrew and {Mendez}, Bryan and {Irace}, William R. and {Duval}, Valerie and {Liu}, Fengchuan and {Royer}, Don and {Heinrichsen}, Ingolf and {Howard}, Joan and {Shannon}, Mark and {Kendall}, Martha and {Walsh}, Amy L. and {Larsen}, Mark and {Cardon}, Joel G. and {Schick}, Scott and {Schwalm}, Mark and {Abid}, Mohamed and {Fabinsky}, Beth and {Naes}, Larry and {Tsai}, Chao-Wei},
        title = "{The Wide-field Infrared Survey Explorer (WISE): Mission Description and Initial On-orbit Performance}",
      journal = {\aj},
         year = 2010,
        month = dec,
       volume = {140},
       number = {6},
        pages = {1868-1881},
          doi = {10.1088/0004-6256/140/6/1868},
archivePrefix = {arXiv},
       eprint = {1008.0031},
 primaryClass = {astro-ph.IM},
       adsurl = {https://ui.adsabs.harvard.edu/abs/2010AJ....140.1868W}
}

@ARTICLE{Meisner_2022_WISE,
       author = {{Meisner}, Aaron M. and {Lang}, Dustin and {Schlafly}, Edward F. and {Schlegel}, David J.},
        title = "{9-yr Deep Sky unWISE Coadds}",
      journal = {Research Notes of the American Astronomical Society},
         year = 2022,
        month = sep,
       volume = {6},
       number = {9},
          eid = {188},
        pages = {188},
          doi = {10.3847/2515-5172/ac913e},
       adsurl = {https://ui.adsabs.harvard.edu/abs/2022RNAAS...6..188M}
}

@ARTICLE{Fazio_2004_IRAC,
       author = {{Fazio}, G.~G. and {Hora}, J.~L. and {Allen}, L.~E. and {Ashby}, M.~L.~N. and {Barmby}, P. and {Deutsch}, L.~K. and {Huang}, J.-S. and {Kleiner}, S. and {Marengo}, M. and {Megeath}, S.~T. and {Melnick}, G.~J. and {Pahre}, M.~A. and {Patten}, B.~M. and {Polizotti}, J. and {Smith}, H.~A. and {Taylor}, R.~S. and {Wang}, Z. and {Willner}, S.~P. and {Hoffmann}, W.~F. and {Pipher}, J.~L. and {Forrest}, W.~J. and {McMurty}, C.~W. and {McCreight}, C.~R. and {McKelvey}, M.~E. and {McMurray}, R.~E. and {Koch}, D.~G. and {Moseley}, S.~H. and {Arendt}, R.~G. and {Mentzell}, J.~E. and {Marx}, C.~T. and {Losch}, P. and {Mayman}, P. and {Eichhorn}, W. and {Krebs}, D. and {Jhabvala}, M. and {Gezari}, D.~Y. and {Fixsen}, D.~J. and {Flores}, J. and {Shakoorzadeh}, K. and {Jungo}, R. and {Hakun}, C. and {Workman}, L. and {Karpati}, G. and {Kichak}, R. and {Whitley}, R. and {Mann}, S. and {Tollestrup}, E.~V. and {Eisenhardt}, P. and {Stern}, D. and {Gorjian}, V. and {Bhattacharya}, B. and {Carey}, S. and {Nelson}, B.~O. and {Glaccum}, W.~J. and {Lacy}, M. and {Lowrance}, P.~J. and {Laine}, S. and {Reach}, W.~T. and {Stauffer}, J.~A. and {Surace}, J.~A. and {Wilson}, G. and {Wright}, E.~L. and {Hoffman}, A. and {Domingo}, G. and {Cohen}, M.},
        title = "{The Infrared Array Camera (IRAC) for the Spitzer Space Telescope}",
      journal = {\apjs},
         year = 2004,
        month = sep,
       volume = {154},
       number = {1},
        pages = {10-17},
          doi = {10.1086/422843},
archivePrefix = {arXiv},
       eprint = {astro-ph/0405616},
 primaryClass = {astro-ph},
       adsurl = {https://ui.adsabs.harvard.edu/abs/2004ApJS..154...10F}
}

@software{Lewis_Challinor_2011_CAMB,
       author = {{Lewis}, Antony and {Challinor}, Anthony},
        title = "{CAMB: Code for Anisotropies in the Microwave Background}",
 howpublished = {Astrophysics Source Code Library, record ascl:1102.026},
         year = 2011,
        month = feb,
          eid = {ascl:1102.026},
archivePrefix = {ascl},
       eprint = {1102.026},
       adsurl = {https://ui.adsabs.harvard.edu/abs/2011ascl.soft02026L}
}

@ARTICLE{Mead_2021,
       author = {{Mead}, A.~J. and {Brieden}, S. and {Tr{\"o}ster}, T. and {Heymans}, C.},
        title = "{HMCODE-2020: improved modelling of non-linear cosmological power spectra with baryonic feedback}",
      journal = {\mnras},
         year = 2021,
        month = mar,
       volume = {502},
       number = {1},
        pages = {1401-1422},
          doi = {10.1093/mnras/stab082},
archivePrefix = {arXiv},
       eprint = {2009.01858},
 primaryClass = {astro-ph.CO},
       adsurl = {https://ui.adsabs.harvard.edu/abs/2021MNRAS.502.1401M}
}

@ARTICLE{Zuntz_2015,
       author = {{Zuntz}, J. and {Paterno}, M. and {Jennings}, E. and {Rudd}, D. and {Manzotti}, A. and {Dodelson}, S. and {Bridle}, S. and {Sehrish}, S. and {Kowalkowski}, J.},
        title = "{CosmoSIS: Modular cosmological parameter estimation}",
      journal = {Astronomy and Computing},
         year = 2015,
        month = sep,
       volume = {12},
        pages = {45-59},
          doi = {10.1016/j.ascom.2015.05.005},
archivePrefix = {arXiv},
       eprint = {1409.3409},
 primaryClass = {astro-ph.CO},
       adsurl = {https://ui.adsabs.harvard.edu/abs/2015A&C....12...45Z}
}

@ARTICLE{Foreman-Mackey_2013,
       author = {{Foreman-Mackey}, Daniel and {Hogg}, David W. and {Lang}, Dustin and {Goodman}, Jonathan},
        title = "{emcee: The MCMC Hammer}",
      journal = {\pasp},
         year = 2013,
        month = mar,
       volume = {125},
       number = {925},
        pages = {306},
          doi = {10.1086/670067},
archivePrefix = {arXiv},
       eprint = {1202.3665},
 primaryClass = {astro-ph.IM},
       adsurl = {https://ui.adsabs.harvard.edu/abs/2013PASP..125..306F}
}

@ARTICLE{Wechsler_2006,
       author = {{Wechsler}, Risa H. and {Zentner}, Andrew R. and {Bullock}, James S. and {Kravtsov}, Andrey V. and {Allgood}, Brandon},
        title = "{The Dependence of Halo Clustering on Halo Formation History, Concentration, and Occupation}",
      journal = {\apj},
         year = 2006,
        month = nov,
       volume = {652},
       number = {1},
        pages = {71-84},
          doi = {10.1086/507120},
archivePrefix = {arXiv},
       eprint = {astro-ph/0512416},
 primaryClass = {astro-ph},
       adsurl = {https://ui.adsabs.harvard.edu/abs/2006ApJ...652...71W}
}

@ARTICLE{Klein_2024a,
       author = {{Klein}, M. and {Mohr}, J.~J. and {Davies}, C.~T.},
        title = "{The ACT-DR5 MCMF galaxy cluster catalog}",
      journal = {\aap},
         year = 2024,
        month = oct,
       volume = {690},
          eid = {A322},
        pages = {A322},
          doi = {10.1051/0004-6361/202451203},
archivePrefix = {arXiv},
       eprint = {2406.14754},
 primaryClass = {astro-ph.CO},
       adsurl = {https://ui.adsabs.harvard.edu/abs/2024A&A...690A.322K}
}

@ARTICLE{Klein_2018,
       author = {{Klein}, M. and {Mohr}, J.~J. and {Desai}, S. and {Israel}, H. and {Allam}, S. and {Benoit-L{\'e}vy}, A. and {Brooks}, D. and {Buckley-Geer}, E. and {Carnero Rosell}, A. and {Carrasco Kind}, M. and {Cunha}, C.~E. and {da Costa}, L.~N. and {Dietrich}, J.~P. and {Eifler}, T.~F. and {Evrard}, A.~E. and {Frieman}, J. and {Gruen}, D. and {Gruendl}, R.~A. and {Gutierrez}, G. and {Honscheid}, K. and {James}, D.~J. and {Kuehn}, K. and {Lima}, M. and {Maia}, M.~A.~G. and {March}, M. and {Melchior}, P. and {Menanteau}, F. and {Miquel}, R. and {Plazas}, A.~A. and {Reil}, K. and {Romer}, A.~K. and {Sanchez}, E. and {Santiago}, B. and {Scarpine}, V. and {Schubnell}, M. and {Sevilla-Noarbe}, I. and {Smith}, M. and {Soares-Santos}, M. and {Sobreira}, F. and {Suchyta}, E. and {Swanson}, M.~E.~C. and {Tarle}, G. and {DES Collaboration}},
        title = "{A multicomponent matched filter cluster confirmation tool for eROSITA: initial application to the RASS and DES-SV data sets}",
      journal = {\mnras},
         year = 2018,
        month = mar,
       volume = {474},
       number = {3},
        pages = {3324-3343},
          doi = {10.1093/mnras/stx2929},
archivePrefix = {arXiv},
       eprint = {1706.06577},
 primaryClass = {astro-ph.CO},
       adsurl = {https://ui.adsabs.harvard.edu/abs/2018MNRAS.474.3324K}
}

@ARTICLE{Klein_2019,
       author = {{Klein}, M. and {Grandis}, S. and {Mohr}, J.~J. and {Paulus}, M. and {Abbott}, T.~M.~C. and {Annis}, J. and {Avila}, S. and {Bertin}, E. and {Brooks}, D. and {Buckley-Geer}, E. and {Rosell}, A. Carnero and {Kind}, M. Carrasco and {Carretero}, J. and {Castander}, F.~J. and {Cunha}, C.~E. and {D'Andrea}, C.~B. and {da Costa}, L.~N. and {De Vicente}, J. and {Desai}, S. and {Diehl}, H.~T. and {Dietrich}, J.~P. and {Doel}, P. and {Evrard}, A.~E. and {Flaugher}, B. and {Fosalba}, P. and {Frieman}, J. and {Garc{\'\i}a-Bellido}, J. and {Gaztanaga}, E. and {Giles}, P.~A. and {Gruen}, D. and {Gruendl}, R.~A. and {Gschwend}, J. and {Gutierrez}, G. and {Hartley}, W.~G. and {Hollowood}, D.~L. and {Honscheid}, K. and {Hoyle}, B. and {James}, D.~J. and {Jeltema}, T. and {Kuehn}, K. and {Kuropatkin}, N. and {Lima}, M. and {Maia}, M.~A.~G. and {March}, M. and {Marshall}, J.~L. and {Menanteau}, F. and {Miquel}, R. and {Ogando}, R.~L.~C. and {Plazas}, A.~A. and {Romer}, A.~K. and {Roodman}, A. and {Sanchez}, E. and {Scarpine}, V. and {Schindler}, R. and {Serrano}, S. and {Sevilla-Noarbe}, I. and {Smith}, M. and {Smith}, R.~C. and {Soares-Santos}, M. and {Sobreira}, F. and {Suchyta}, E. and {Swanson}, M.~E.~C. and {Tarle}, G. and {Thomas}, D. and {Vikram}, V. and {DES Collaboration}},
        title = "{A new RASS galaxy cluster catalogue with low contamination extending to z {\ensuremath{\sim}} 1 in the DES overlap region}",
      journal = {\mnras},
         year = 2019,
        month = sep,
       volume = {488},
       number = {1},
        pages = {739-769},
          doi = {10.1093/mnras/stz1463},
archivePrefix = {arXiv},
       eprint = {1812.09956},
 primaryClass = {astro-ph.CO},
       adsurl = {https://ui.adsabs.harvard.edu/abs/2019MNRAS.488..739K}
}

@ARTICLE{Kaiser_1987_RSD,
       author = {{Kaiser}, Nick},
        title = "{Clustering in real space and in redshift space}",
      journal = {\mnras},
         year = 1987,
        month = jul,
       volume = {227},
        pages = {1-21},
          doi = {10.1093/mnras/227.1.1},
       adsurl = {https://ui.adsabs.harvard.edu/abs/1987MNRAS.227....1K}
}

@INPROCEEDINGS{Hamilton_1998_RSD_Review,
       author = {{Hamilton}, A.~J.~S.},
        title = "{Linear Redshift Distortions: a Review}",
    booktitle = {The Evolving Universe},
         year = 1998,
       editor = {{Hamilton}, Donald},
       series = {Astrophysics and Space Science Library},
       volume = {231},
        month = jan,
        pages = {185},
          doi = {10.1007/978-94-011-4960-0_17},
archivePrefix = {arXiv},
       eprint = {astro-ph/9708102},
 primaryClass = {astro-ph},
       adsurl = {https://ui.adsabs.harvard.edu/abs/1998ASSL..231..185H}
}

@ARTICLE{LSST_2018,
       author = {{The LSST Dark Energy Science Collaboration} and {Mandelbaum}, Rachel and {Eifler}, Tim and {Hlo{\v{z}}ek}, Ren{\'e}e and {Collett}, Thomas and {Gawiser}, Eric and {Scolnic}, Daniel and {Alonso}, David and {Awan}, Humna and {Biswas}, Rahul and {Blazek}, Jonathan and {Burchat}, Patricia and {Chisari}, Nora Elisa and {Dell'Antonio}, Ian and {Digel}, Seth and {Frieman}, Josh and {Goldstein}, Daniel A. and {Hook}, Isobel and {Ivezi{\'c}}, {\v{Z}}eljko and {Kahn}, Steven M. and {Kamath}, Sowmya and {Kirkby}, David and {Kitching}, Thomas and {Krause}, Elisabeth and {Leget}, Pierre-Fran{\c{c}}ois and {Marshall}, Philip J. and {Meyers}, Joshua and {Miyatake}, Hironao and {Newman}, Jeffrey A. and {Nichol}, Robert and {Rykoff}, Eli and {Sanchez}, F. Javier and {Slosar}, An{\v{z}}e and {Sullivan}, Mark and {Troxel}, M.~A.},
        title = "{The LSST Dark Energy Science Collaboration (DESC) Science Requirements Document}",
      journal = {arXiv e-prints},
         year = 2018,
        month = sep,
          eid = {arXiv:1809.01669},
        pages = {arXiv:1809.01669},
          doi = {10.48550/arXiv.1809.01669},
archivePrefix = {arXiv},
       eprint = {1809.01669},
 primaryClass = {astro-ph.CO},
       adsurl = {https://ui.adsabs.harvard.edu/abs/2018arXiv180901669T}
}

@ARTICLE{EUCLID_2025,
       author = {{Euclid Collaboration} and {Mellier}, Y. and {Abdurro'uf} and {Acevedo Barroso}, J.~A. and {Ach{\'u}carro}, A. and {Adamek}, J. and {Adam}, R. and {Addison}, G.~E. and {Aghanim}, N. and {Aguena}, M. and {Ajani}, V. and {Akrami}, Y. and {Al-Bahlawan}, A. and {Alavi}, A. and {Albuquerque}, I.~S. and {Alestas}, G. and {Alguero}, G. and {Allaoui}, A. and {Allen}, S.~W. and {Allevato}, V. and {Alonso-Tetilla}, A.~V. and {Altieri}, B. and {Alvarez-Candal}, A. and {Alvi}, S. and {Amara}, A. and {Amendola}, L. and {Amiaux}, J. and {Andika}, I.~T. and {Andreon}, S. and {Andrews}, A. and {Angora}, G. and {Angulo}, R.~E. and {Annibali}, F. and {Anselmi}, A. and {Anselmi}, S. and {Arcari}, S. and {Archidiacono}, M. and {Aric{\`o}}, G. and {Arnaud}, M. and {Arnouts}, S. and {Asgari}, M. and {Asorey}, J. and {Atayde}, L. and {Atek}, H. and {Atrio-Barandela}, F. and {Aubert}, M. and {Aubourg}, E. and {Auphan}, T. and {Auricchio}, N. and {Aussel}, B. and {Aussel}, H. and {Avelino}, P.~P. and {Avgoustidis}, A. and {Avila}, S. and {Awan}, S. and {Azzollini}, R. and {Baccigalupi}, C. and {Bachelet}, E. and {Bacon}, D. and {Baes}, M. and {Bagley}, M.~B. and {Bahr-Kalus}, B. and {Balaguera-Antolinez}, A. and {Balbinot}, E. and {Balcells}, M. and {Baldi}, M. and {Baldry}, I. and {Balestra}, A. and {Ballardini}, M. and {Ballester}, O. and {Balogh}, M. and {Ba{\~n}ados}, E. and {Barbier}, R. and {Bardelli}, S. and {Baron}, M. and {Barreiro}, T. and {Barrena}, R. and {Barriere}, J.-C. and {Barros}, B.~J. and {Barthelemy}, A. and {Bartolo}, N. and {Basset}, A. and {Battaglia}, P. and {Battisti}, A.~J. and {Baugh}, C.~M. and {Baumont}, L. and {Bazzanini}, L. and {Beaulieu}, J.-P. and {Beckmann}, V. and {Belikov}, A.~N. and {Bel}, J. and {Bellagamba}, F. and {Bella}, M. and {Bellini}, E. and {Benabed}, K. and {Bender}, R. and {Benevento}, G. and {Bennett}, C.~L. and {Benson}, K. and {Bergamini}, P. and {Bermejo-Climent}, J.~R. and {Bernardeau}, F. and {Bertacca}, D. and {Berthe}, M. and {Berthier}, J. and {Bethermin}, M. and {Beutler}, F. and {Bevillon}, C. and {Bhargava}, S. and {Bhatawdekar}, R. and {Bianchi}, D. and {Bisigello}, L. and {Biviano}, A. and {Blake}, R.~P. and {Blanchard}, A. and {Blazek}, J. and {Blot}, L. and {Bosco}, A. and {Bodendorf}, C. and {Boenke}, T. and {B{\"o}hringer}, H. and {Boldrini}, P. and {Bolzonella}, M. and {Bonchi}, A. and {Bonici}, M. and {Bonino}, D. and {Bonino}, L. and {Bonvin}, C. and {Bon}, W. and {Booth}, J.~T. and {Borgani}, S. and {Borlaff}, A.~S. and {Borsato}, E. and {Bose}, B. and {Botticella}, M.~T. and {Boucaud}, A. and {Bouche}, F. and {Boucher}, J.~S. and {Boutigny}, D. and {Bouvard}, T. and {Bouwens}, R. and {Bouy}, H. and {Bowler}, R.~A.~A. and {Bozza}, V. and {Bozzo}, E. and {Branchini}, E. and {Brando}, G. and {Brau-Nogue}, S. and {Brekke}, P. and {Bremer}, M.~N. and {Brescia}, M. and {Breton}, M.-A. and {Brinchmann}, J. and {Brinckmann}, T. and {Brockley-Blatt}, C. and {Brodwin}, M. and {Brouard}, L. and {Brown}, M.~L. and {Bruton}, S. and {Bucko}, J. and {Buddelmeijer}, H. and {Buenadicha}, G. and {Buitrago}, F. and {Burger}, P. and {Burigana}, C. and {Busillo}, V. and {Busonero}, D. and {Cabanac}, R. and {Cabayol-Garcia}, L. and {Cagliari}, M.~S. and {Caillat}, A. and {Caillat}, L. and {Calabrese}, M. and {Calabro}, A. and {Calderone}, G. and {Calura}, F. and {Camacho Quevedo}, B. and {Camera}, S. and {Campos}, L. and {Ca{\~n}as-Herrera}, G. and {Candini}, G.~P. and {Cantiello}, M. and {Capobianco}, V. and {Cappellaro}, E. and {Cappelluti}, N. and {Cappi}, A. and {Caputi}, K.~I. and {Cara}, C. and {Carbone}, C. and {Cardone}, V.~F. and {Carella}, E. and {Carlberg}, R.~G. and {Carle}, M. and {Carminati}, L. and {Caro}, F. and {Carrasco}, J.~M. and {Carretero}, J. and {Carrilho}, P. and {Carron Duque}, J. and {Carry}, B.},
        title = "{Euclid: I. Overview of the Euclid mission}",
      journal = {\aap},
         year = 2025,
        month = may,
       volume = {697},
          eid = {A1},
        pages = {A1},
          doi = {10.1051/0004-6361/202450810},
archivePrefix = {arXiv},
       eprint = {2405.13491},
 primaryClass = {astro-ph.CO},
       adsurl = {https://ui.adsabs.harvard.edu/abs/2025A&A...697A...1E}
}

@ARTICLE{Adame_2025,
       author = {{Adame}, A.~G. and {Aguilar}, J. and {Ahlen}, S. and {Alam}, S. and {Alexander}, D.~M. and {Allende Prieto}, C. and {Alvarez}, M. and {Alves}, O. and {Anand}, A. and {Andrade}, U. and {Armengaud}, E. and {Avila}, S. and {Aviles}, A. and {Awan}, H. and {Bahr-Kalus}, B. and {Bailey}, S. and {Baltay}, C. and {Bault}, A. and {Behera}, J. and {BenZvi}, S. and {Beutler}, F. and {Bianchi}, D. and {Blake}, C. and {Blum}, R. and {Bonici}, M. and {Brieden}, S. and {Brodzeller}, A. and {Brooks}, D. and {Buckley-Geer}, E. and {Burtin}, E. and {Calderon}, R. and {Canning}, R. and {Carnero Rosell}, A. and {Cereskaite}, R. and {Cervantes-Cota}, J.~L. and {Chabanier}, S. and {Chaussidon}, E. and {Chaves-Montero}, J. and {Chebat}, D. and {Chen}, S. and {Chen}, X. and {Claybaugh}, T. and {Cole}, S. and {Cuceu}, A. and {Davis}, T.~M. and {Dawson}, K. and {de la Macorra}, A. and {de Mattia}, A. and {Deiosso}, N. and {Dey}, A. and {Dey}, B. and {Ding}, Z. and {Doel}, P. and {Edelstein}, J. and {Eftekharzadeh}, S. and {Eisenstein}, D.~J. and {Elbers}, W. and {Elliott}, A. and {Fagrelius}, P. and {Fanning}, K. and {Ferraro}, S. and {Ereza}, J. and {Findlay}, N. and {Flaugher}, B. and {Font-Ribera}, A. and {Forero-S{\'a}nchez}, D. and {Forero-Romero}, J.~E. and {Frenk}, C.~S. and {Garcia-Quintero}, C. and {Garrison}, L.~H. and {Gazta{\~n}aga}, E. and {Gil-Mar{\'\i}n}, H. and {Gontcho}, S. Gontcho A. and {Gonzalez-Morales}, A.~X. and {Gonzalez-Perez}, V. and {Gordon}, C. and {Green}, D. and {Gruen}, D. and {Gsponer}, R. and {Gutierrez}, G. and {Guy}, J. and {Hadzhiyska}, B. and {Hahn}, C. and {Hanif}, M.~M.~S. and {Herrera-Alcantar}, H.~K. and {Honscheid}, K. and {Howlett}, C. and {Huterer}, D. and {Ir{\v{s}}i{\v{c}}}, V. and {Ishak}, M. and {Joyce}, R. and {Juneau}, S. and {Kara{\c{c}}ayl{\i}}, N.~G. and {Kehoe}, R. and {Kent}, S. and {Kirkby}, D. and {Kong}, H. and {Koposov}, S.~E. and {Kremin}, A. and {Krolewski}, A. and {Lahav}, O. and {Lai}, Y. and {Lan}, T.-W. and {Landriau}, M. and {Lang}, D. and {Lasker}, J. and {Le Goff}, J.~M. and {Le Guillou}, L. and {Leauthaud}, A. and {Levi}, M.~E. and {Li}, T.~S. and {Lodha}, K. and {Magneville}, C. and {Manera}, M. and {Margala}, D. and {Martini}, P. and {Matthewson}, W. and {Maus}, M. and {McDonald}, P. and {Medina-Varela}, L. and {Meisner}, A. and {Mena-Fern{\'a}ndez}, J. and {Miquel}, R. and {Moon}, J. and {Moore}, S. and {Moustakas}, J. and {Mudur}, N. and {Mueller}, E. and {Mu{\~n}oz-Guti{\'e}rrez}, A. and {Myers}, A.~D. and {Nadathur}, S. and {Napolitano}, L. and {Neveux}, R. and {Newman}, J.~A. and {Nguyen}, N.~M. and {Nie}, J. and {Niz}, G. and {Noriega}, H.~E. and {Padmanabhan}, N. and {Paillas}, E. and {Palanque-Delabrouille}, N. and {Pan}, J. and {Penmetsa}, S. and {Percival}, W.~J. and {Pieri}, M.~M. and {Pinon}, M. and {Poppett}, C. and {Porredon}, A. and {Prada}, F. and {P{\'e}rez-Fern{\'a}ndez}, A. and {P{\'e}rez-R{\`a}fols}, I. and {Rabinowitz}, D. and {Raichoor}, A. and {Ram{\'\i}rez-P{\'e}rez}, C. and {Ramirez-Solano}, S. and {Rashkovetskyi}, M. and {Ravoux}, C. and {Rezaie}, M. and {Rich}, J. and {Rocher}, A. and {Rockosi}, C. and {Roe}, N.~A. and {Rosado-Marin}, A. and {Ross}, A.~J. and {Rossi}, G. and {Ruggeri}, R. and {Ruhlmann-Kleider}, V. and {Samushia}, L. and {Sanchez}, E. and {Saulder}, C. and {Schlafly}, E.~F. and {Schlegel}, D. and {Schubnell}, M. and {Seo}, H. and {Shafieloo}, A. and {Sharples}, R. and {Silber}, J. and {Slosar}, A. and {Smith}, A. and {Sprayberry}, D. and {Tan}, T. and {Tarl{\'e}}, G. and {Taylor}, P. and {Trusov}, S. and {Vaisakh}, R. and {Valcin}, D. and {Valdes}, F. and {Valogiannis}, G. and {Vargas-Maga{\~n}a}, M. and {Verde}, L. and {Walther}, M. and {Wang}, B. and {Wang}, M.~S. and {Weaver}, B.~A. and {Weaverdyck}, N. and {Wechsler}, R.~H. and {Weinberg}, D.~H. and {White}, M. and {Wilson}, M.~J. and {Yi}, L.},
        title = "{DESI 2024 VII: cosmological constraints from the full-shape modeling of clustering measurements}",
      journal = {\jcap},
         year = 2025,
        month = jul,
       volume = {2025},
       number = {7},
          eid = {028},
        pages = {028},
          doi = {10.1088/1475-7516/2025/07/028},
archivePrefix = {arXiv},
       eprint = {2411.12022},
 primaryClass = {astro-ph.CO},
       adsurl = {https://ui.adsabs.harvard.edu/abs/2025JCAP...07..028A}
}

@ARTICLE{Euclid_2024_Bias,
       author = {{Euclid Collaboration} and {Castro}, T. and {Fumagalli}, A. and {Angulo}, R.~E. and {Bocquet}, S. and {Borgani}, S. and {Costanzi}, M. and {Dakin}, J. and {Dolag}, K. and {Monaco}, P. and {Saro}, A. and {Sefusatti}, E. and {Aghanim}, N. and {Amendola}, L. and {Andreon}, S. and {Baccigalupi}, C. and {Baldi}, M. and {Bodendorf}, C. and {Bonino}, D. and {Branchini}, E. and {Brescia}, M. and {Caillat}, A. and {Camera}, S. and {Capobianco}, V. and {Carbone}, C. and {Carretero}, J. and {Casas}, S. and {Castellano}, M. and {Castignani}, G. and {Cavuoti}, S. and {Cimatti}, A. and {Colodro-Conde}, C. and {Congedo}, G. and {Conselice}, C.~J. and {Conversi}, L. and {Copin}, Y. and {Costille}, A. and {Courbin}, F. and {Courtois}, H.~M. and {Da Silva}, A. and {Degaudenzi}, H. and {De Lucia}, G. and {Di Giorgio}, A.~M. and {Douspis}, M. and {Dupac}, X. and {Dusini}, S. and {Farina}, M. and {Farrens}, S. and {Ferriol}, S. and {Fosalba}, P. and {Frailis}, M. and {Franceschi}, E. and {Fumana}, M. and {Galeotta}, S. and {Gillis}, B. and {Giocoli}, C. and {G{\'o}mez-Alvarez}, P. and {Grazian}, A. and {Grupp}, F. and {Guzzo}, L. and {Haugan}, S.~V.~H. and {Holmes}, W. and {Hormuth}, F. and {Hornstrup}, A. and {Ili{\'c}}, S. and {Jahnke}, K. and {Jhabvala}, M. and {Joachimi}, B. and {Keih{\"a}nen}, E. and {Kermiche}, S. and {Kiessling}, A. and {Kilbinger}, M. and {Kubik}, B. and {Kunz}, M. and {Kurki-Suonio}, H. and {Lilje}, P.~B. and {Lindholm}, V. and {Lloro}, I. and {Maiorano}, E. and {Mansutti}, O. and {Marggraf}, O. and {Markovic}, K. and {Martinelli}, M. and {Martinet}, N. and {Marulli}, F. and {Massey}, R. and {Maurogordato}, S. and {Medinaceli}, E. and {Melchior}, M. and {Mellier}, Y. and {Meneghetti}, M. and {Merlin}, E. and {Meylan}, G. and {Moscardini}, L. and {Munari}, E. and {Niemi}, S.-M. and {Padilla}, C. and {Paltani}, S. and {Pasian}, F. and {Pedersen}, K. and {Percival}, W.~J. and {Pettorino}, V. and {Pires}, S. and {Polenta}, G. and {Poncet}, M. and {Popa}, L.~A. and {Pozzetti}, L. and {Raison}, F. and {Renzi}, A. and {Riccio}, G. and {Romelli}, E. and {Roncarelli}, M. and {Saglia}, R. and {Sakr}, Z. and {Salvignol}, J.-C. and {S{\'a}nchez}, A.~G. and {Sapone}, D. and {Sartoris}, B. and {Schirmer}, M. and {Secroun}, A. and {Serrano}, S. and {Sirignano}, C. and {Sirri}, G. and {Stanco}, L. and {Steinwagner}, J. and {Tallada-Cresp{\'\i}}, P. and {Taylor}, A.~N. and {Tereno}, I. and {Toledo-Moreo}, R. and {Torradeflot}, F. and {Tutusaus}, I. and {Valenziano}, L. and {Vassallo}, T. and {Verdoes Kleijn}, G. and {Wang}, Y. and {Weller}, J. and {Zacchei}, A. and {Zamorani}, G. and {Zucca}, E. and {Biviano}, A. and {Bolzonella}, M. and {Bozzo}, E. and {Burigana}, C. and {Calabrese}, M. and {Di Ferdinando}, D. and {Escartin Vigo}, J.~A. and {Finelli}, F. and {Gracia-Carpio}, J. and {Matthew}, S. and {Mauri}, N. and {Pezzotta}, A. and {P{\"o}ntinen}, M. and {Porciani}, C. and {Scottez}, V. and {Tenti}, M. and {Viel}, M. and {Wiesmann}, M. and {Akrami}, Y. and {Allevato}, V. and {Anselmi}, S. and {Archidiacono}, M. and {Atrio-Barandela}, F. and {Balaguera-Antolinez}, A. and {Ballardini}, M. and {Bertacca}, D. and {Bethermin}, M. and {Blanchard}, A. and {Blot}, L. and {B{\"o}hringer}, H. and {Bruton}, S. and {Cabanac}, R. and {Calabro}, A. and {Ca{\~n}as-Herrera}, G. and {Cappi}, A. and {Caro}, F. and {Carvalho}, C.~S. and {Chambers}, K.~C. and {Cooray}, A.~R. and {De Caro}, B. and {de la Torre}, S. and {Desprez}, G. and {D{\'\i}az-S{\'a}nchez}, A. and {Diaz}, J.~J. and {Di Domizio}, S. and {Dole}, H. and {Escoffier}, S. and {Ferrari}, A.~G. and {Ferreira}, P.~G. and {Ferrero}, I. and {Finoguenov}, A. and {Fontana}, A. and {Fornari}, F. and {Gabarra}, L. and {Ganga}, K. and {Garc{\'\i}a-Bellido}, J. and {Gasparetto}, T. and {Gautard}, V. and {Gaztanaga}, E. and {Giacomini}, F. and {Gianotti}, F.},
        title = "{Euclid preparation: L. Calibration of the halo linear bias in {\ensuremath{\Lambda}}(v)CDM cosmologies}",
      journal = {\aap},
         year = 2024,
        month = nov,
       volume = {691},
          eid = {A62},
        pages = {A62},
          doi = {10.1051/0004-6361/202451230},
archivePrefix = {arXiv},
       eprint = {2409.01877},
 primaryClass = {astro-ph.CO},
       adsurl = {https://ui.adsabs.harvard.edu/abs/2024A&A...691A..62E}
}

@ARTICLE{Gaia_2021,
       author = {{Gaia Collaboration} and {Brown}, A.~G.~A. and {Vallenari}, A. and {Prusti}, T. and {de Bruijne}, J.~H.~J. and {Babusiaux}, C. and {Biermann}, M. and {Creevey}, O.~L. and {Evans}, D.~W. and {Eyer}, L. and {Hutton}, A. and {Jansen}, F. and {Jordi}, C. and {Klioner}, S.~A. and {Lammers}, U. and {Lindegren}, L. and {Luri}, X. and {Mignard}, F. and {Panem}, C. and {Pourbaix}, D. and {Randich}, S. and {Sartoretti}, P. and {Soubiran}, C. and {Walton}, N.~A. and {Arenou}, F. and {Bailer-Jones}, C.~A.~L. and {Bastian}, U. and {Cropper}, M. and {Drimmel}, R. and {Katz}, D. and {Lattanzi}, M.~G. and {van Leeuwen}, F. and {Bakker}, J. and {Cacciari}, C. and {Casta{\~n}eda}, J. and {De Angeli}, F. and {Ducourant}, C. and {Fabricius}, C. and {Fouesneau}, M. and {Fr{\'e}mat}, Y. and {Guerra}, R. and {Guerrier}, A. and {Guiraud}, J. and {Jean-Antoine Piccolo}, A. and {Masana}, E. and {Messineo}, R. and {Mowlavi}, N. and {Nicolas}, C. and {Nienartowicz}, K. and {Pailler}, F. and {Panuzzo}, P. and {Riclet}, F. and {Roux}, W. and {Seabroke}, G.~M. and {Sordo}, R. and {Tanga}, P. and {Th{\'e}venin}, F. and {Gracia-Abril}, G. and {Portell}, J. and {Teyssier}, D. and {Altmann}, M. and {Andrae}, R. and {Bellas-Velidis}, I. and {Benson}, K. and {Berthier}, J. and {Blomme}, R. and {Brugaletta}, E. and {Burgess}, P.~W. and {Busso}, G. and {Carry}, B. and {Cellino}, A. and {Cheek}, N. and {Clementini}, G. and {Damerdji}, Y. and {Davidson}, M. and {Delchambre}, L. and {Dell'Oro}, A. and {Fern{\'a}ndez-Hern{\'a}ndez}, J. and {Galluccio}, L. and {Garc{\'\i}a-Lario}, P. and {Garcia-Reinaldos}, M. and {Gonz{\'a}lez-N{\'u}{\~n}ez}, J. and {Gosset}, E. and {Haigron}, R. and {Halbwachs}, J.-L. and {Hambly}, N.~C. and {Harrison}, D.~L. and {Hatzidimitriou}, D. and {Heiter}, U. and {Hern{\'a}ndez}, J. and {Hestroffer}, D. and {Hodgkin}, S.~T. and {Holl}, B. and {Jan{\ss}en}, K. and {Jevardat de Fombelle}, G. and {Jordan}, S. and {Krone-Martins}, A. and {Lanzafame}, A.~C. and {L{\"o}ffler}, W. and {Lorca}, A. and {Manteiga}, M. and {Marchal}, O. and {Marrese}, P.~M. and {Moitinho}, A. and {Mora}, A. and {Muinonen}, K. and {Osborne}, P. and {Pancino}, E. and {Pauwels}, T. and {Petit}, J.-M. and {Recio-Blanco}, A. and {Richards}, P.~J. and {Riello}, M. and {Rimoldini}, L. and {Robin}, A.~C. and {Roegiers}, T. and {Rybizki}, J. and {Sarro}, L.~M. and {Siopis}, C. and {Smith}, M. and {Sozzetti}, A. and {Ulla}, A. and {Utrilla}, E. and {van Leeuwen}, M. and {van Reeven}, W. and {Abbas}, U. and {Abreu Aramburu}, A. and {Accart}, S. and {Aerts}, C. and {Aguado}, J.~J. and {Ajaj}, M. and {Altavilla}, G. and {{\'A}lvarez}, M.~A. and {{\'A}lvarez Cid-Fuentes}, J. and {Alves}, J. and {Anderson}, R.~I. and {Anglada Varela}, E. and {Antoja}, T. and {Audard}, M. and {Baines}, D. and {Baker}, S.~G. and {Balaguer-N{\'u}{\~n}ez}, L. and {Balbinot}, E. and {Balog}, Z. and {Barache}, C. and {Barbato}, D. and {Barros}, M. and {Barstow}, M.~A. and {Bartolom{\'e}}, S. and {Bassilana}, J.-L. and {Bauchet}, N. and {Baudesson-Stella}, A. and {Becciani}, U. and {Bellazzini}, M. and {Bernet}, M. and {Bertone}, S. and {Bianchi}, L. and {Blanco-Cuaresma}, S. and {Boch}, T. and {Bombrun}, A. and {Bossini}, D. and {Bouquillon}, S. and {Bragaglia}, A. and {Bramante}, L. and {Breedt}, E. and {Bressan}, A. and {Brouillet}, N. and {Bucciarelli}, B. and {Burlacu}, A. and {Busonero}, D. and {Butkevich}, A.~G. and {Buzzi}, R. and {Caffau}, E. and {Cancelliere}, R. and {C{\'a}novas}, H. and {Cantat-Gaudin}, T. and {Carballo}, R. and {Carlucci}, T. and {Carnerero}, M.~I. and {Carrasco}, J.~M. and {Casamiquela}, L. and {Castellani}, M. and {Castro-Ginard}, A. and {Castro Sampol}, P. and {Chaoul}, L. and {Charlot}, P. and {Chemin}, L. and {Chiavassa}, A. and {Cioni}, M.-R.~L. and {Comoretto}, G. and {Cooper}, W.~J. and {Cornez}, T. and {Cowell}, S. and {Crifo}, F. and {Crosta}, M. and {Crowley}, C. and {Dafonte}, C. and {Dapergolas}, A. and {David}, M. and {David}, P.},
        title = "{Gaia Early Data Release 3. Summary of the contents and survey properties}",
      journal = {\aap},
         year = 2021,
        month = may,
       volume = {649},
          eid = {A1},
        pages = {A1},
          doi = {10.1051/0004-6361/202039657},
archivePrefix = {arXiv},
       eprint = {2012.01533},
 primaryClass = {astro-ph.GA},
       adsurl = {https://ui.adsabs.harvard.edu/abs/2021A&A...649A...1G}
}

@ARTICLE{Fumagalli_2024,
       author = {{Fumagalli}, A. and {Costanzi}, M. and {Saro}, A. and {Castro}, T. and {Borgani}, S.},
        title = "{Cosmological constraints from the abundance, weak lensing, and clustering of galaxy clusters: Application to the SDSS}",
      journal = {\aap},
         year = 2024,
        month = feb,
       volume = {682},
          eid = {A148},
        pages = {A148},
          doi = {10.1051/0004-6361/202348296},
archivePrefix = {arXiv},
       eprint = {2310.09146},
 primaryClass = {astro-ph.CO},
       adsurl = {https://ui.adsabs.harvard.edu/abs/2024A&A...682A.148F}
}

@ARTICLE{Taylor:2024:CombineHarvestor,
       author = {{Taylor}, Peter L. and {Cuceu}, Andrei and {To}, Chun-Hao and {Zaborowski}, Erik A.},
        title = "{CombineHarvesterFlow: Joint Probe Analysis Made Easy with Normalizing Flows}",
      journal = {The Open Journal of Astrophysics},
         year = 2024,
        month = oct,
       volume = {7},
          eid = {86},
        pages = {86},
          doi = {10.33232/001c.124495},
archivePrefix = {arXiv},
       eprint = {2406.06687},
 primaryClass = {astro-ph.CO},
       adsurl = {https://ui.adsabs.harvard.edu/abs/2024OJAp....7E..86T}
}

@ARTICLE{Anbajagane:2026:ColliderBias,
       author = {{Anbajagane}, Dhayaa and {Dalal}, Neal},
        title = "{Primordial Physics in the Nonlinear Universe: Revealing the oscillating halo bias from cosmological collider models}",
      journal = {arXiv e-prints},
         year = 2026,
        month = jul,
          eid = {arXiv:2607.24939},
        pages = {arXiv:2607.24939},
          doi = {10.48550/arXiv.2607.24939},
archivePrefix = {arXiv},
       eprint = {2607.24939},
 primaryClass = {astro-ph.CO},
       adsurl = {https://ui.adsabs.harvard.edu/abs/2026arXiv260724939A}
}

@ARTICLE{Zhan_2026,
       author = {{Zhan}, Junhao and {Reichardt}, Christian L.},
        title = "{Built-in precision: Improving cluster cosmology through the self-calibration of a galaxy cluster sample}",
      journal = {\pasa},
         year = 2026,
        month = jan,
       volume = {43},
          eid = {e003},
        pages = {e003},
          doi = {10.1017/pasa.2025.10132},
archivePrefix = {arXiv},
       eprint = {2510.14464},
 primaryClass = {astro-ph.CO},
       adsurl = {https://ui.adsabs.harvard.edu/abs/2026PASA...43....3Z}
}

@ARTICLE{Wu_Rozo_Wechsler_2008,
       author = {{Wu}, Hao-Yi and {Rozo}, Eduardo and {Wechsler}, Risa H.},
        title = "{The Effects of Halo Assembly Bias on Self-Calibration in Galaxy Cluster Surveys}",
      journal = {\apj},
         year = 2008,
        month = dec,
       volume = {688},
       number = {2},
        pages = {729-741},
          doi = {10.1086/591929},
archivePrefix = {arXiv},
       eprint = {0803.1491},
 primaryClass = {astro-ph},
       adsurl = {https://ui.adsabs.harvard.edu/abs/2008ApJ...688..729W}
}

@ARTICLE{London_2026,
       author = {{London}, Alexander Spencer and {Rogers}, Keir K. and {Lagu{\"e}}, Alex and {Hlo{\v{z}}ek}, Ren{\'e}e and {Zaman}, Zara},
        title = "{Cosmic voids as a probe of the nature of dark matter: simulations and galaxy survey forecasts}",
      journal = {arXiv e-prints},
         year = 2026,
        month = feb,
          eid = {arXiv:2602.22990},
        pages = {arXiv:2602.22990},
          doi = {10.48550/arXiv.2602.22990},
archivePrefix = {arXiv},
       eprint = {2602.22990},
 primaryClass = {astro-ph.CO},
       adsurl = {https://ui.adsabs.harvard.edu/abs/2026arXiv260222990L}
}

@ARTICLE{Schneider2019Baryonification,
       author = {{Schneider}, Aurel and {Teyssier}, Romain and {Stadel}, Joachim and {Chisari}, Nora Elisa and {Le Brun}, Amandine M.~C. and {Amara}, Adam and {Refregier}, Alexandre},
        title = "{Quantifying baryon effects on the matter power spectrum and the weak lensing shear correlation}",
      journal = {\jcap},
         year = 2019,
        month = mar,
       volume = {2019},
       number = {3},
          eid = {020},
        pages = {020},
          doi = {10.1088/1475-7516/2019/03/020},
archivePrefix = {arXiv},
       eprint = {1810.08629},
 primaryClass = {astro-ph.CO},
       adsurl = {https://ui.adsabs.harvard.edu/abs/2019JCAP...03..020S}
}

@ARTICLE{Arico:2021:Bacco,
       author = {{Aric{\`o}}, Giovanni and {Angulo}, Raul E. and {Contreras}, Sergio and {Ondaro-Mallea}, Lurdes and {Pellejero-Iba{\~n}ez}, Marcos and {Zennaro}, Matteo},
        title = "{The BACCO simulation project: a baryonification emulator with neural networks}",
      journal = {\mnras},
         year = 2021,
        month = sep,
       volume = {506},
       number = {3},
        pages = {4070-4082},
          doi = {10.1093/mnras/stab1911},
archivePrefix = {arXiv},
       eprint = {2011.15018},
 primaryClass = {astro-ph.CO},
       adsurl = {https://ui.adsabs.harvard.edu/abs/2021MNRAS.506.4070A}
}

@ARTICLE{Anbajagane:2024:Baryons,
       author = {{Anbajagane}, Dhayaa and {Pandey}, Shivam and {Chang}, Chihway},
        title = "{Map-level baryonification: Efficient modelling of higher-order correlations in the weak lensing and thermal Sunyaev-Zeldovich fields}",
      journal = {The Open Journal of Astrophysics},
         year = 2024,
        month = dec,
       volume = {7},
          eid = {108},
        pages = {108},
          doi = {10.33232/001c.126788},
archivePrefix = {arXiv},
       eprint = {2409.03822},
 primaryClass = {astro-ph.CO},
       adsurl = {https://ui.adsabs.harvard.edu/abs/2024OJAp....7E.108A}
}

\appendix

\section{Covariance Matrix Model}\label{appx:CovMatrix}

As discussed in Section \ref{sec:theory:cov}, our parameter inference pipeline uses an analytic covariance estimate. In Section \ref{appx:CovMatrix:Theory} we detail this model and present validations of it in Section \ref{appx:CovMatrix:Valid}.

\subsection{Covariance Matrix Estimates}\label{appx:CovMatrix:Theory}

Following the above model, we can also compute an analytic covariance matrix estimate for our statistic. This matrix comprises three separate terms --- (i) the Gaussian shot noise term, (ii) the Gaussian cosmic variance term, and (iii) the leading-order non-Gaussian term. We find the cosmic variance term is significantly subdominant for our setup, but generates some off-diagonal correlations and hence is relevant for our model. We now describe our model for the different terms.

\begin{widetext}
Given the transformation from $P(\kpar, \kperp)$ to $\wrp(\rp)$ is linear (Equation \eqref{eq:final theory}), the covariance can be first written in Fourier space and then converted into configuration space via linear transformations,
\begin{equation}\label{eq:E8}
    C[\wrp(\rp^i), \wrp(\rp^j)] 
    = \frac{1}{V_{\rm eff}}\int\int_0^\infty \frac{d\kperp^a \kperp^a}{2\pi} 
    \frac{d\kperp^b \kperp^b}{2\pi}
    J_0(\kperp^a \rp^i) J_0(\kperp^b \rp^j)\,
    C^T[P_\perp(\kperp^a, \kpar), P_\perp(\kperp^b, \kpar)]
\end{equation}
where the total covariance, $C^T$, is the sum of the Gaussian and non-Gaussian terms, $C^T = C^G + C^{NG}$. The Gaussian covariance matrix of the power spectra, using Wick's theorem, can be represented as 
\begin{equation}\label{eqn:covmat:CV}
    C^G[P_\perp(\kperp^a, \kpar), P_\perp(\kperp^b, \kpar)] = \bigg(X[P(\kperp^a, \kpar)] + \frac{1}{n_h}\bigg) \times \bigg(X[P(\kperp^b, \kpar)] + \frac{1}{n_h}\bigg).
\end{equation}
This matrix $C^G$ includes both shot noise and cosmic variance, which are Gaussian in nature. Here, the transformation $X$ encompasses the line-of-sight projections, redshift-averaging, etc. of the 3D power spectrum, while $n$ is the effective (surface) number density of the sample. The former transformation is defined using the inner integrals discussed above,
\begin{align}
    X[P(\kperp, \kpar)] = & \int_0^2 dz \,\, \frac{n(z)^2}{dV/dz}\,\, \int_{-\infty}^{\infty} \frac{d\kpar}{2\pi}  \bigg\{P(\kpar, \kperp, z)\, e^{-\kpar^2\sigma_\chi^2(z)} \frac{2\sin(\kpar \pimax)} {\kpar} \times \nonumber\\[10pt]
    & \bigg[\bigg(\frac{1}{\int_{0}^{200} d\xi\,\, {n(z, \xi)}}\int_{4}^{200}d\xi \,\, b(z, \xi) n(z, \xi)\bigg) + f(z)\frac{\kpar^2}{\kpar^2 + \kperp^2} \bigg]^2\bigg\}.
  \label{eq:E8}
\end{align}

The effective volume factor, $V_{\rm eff}$, in Equation \eqref{eq:E8} is calculated as
\begin{equation}
    V_{\rm eff} = \bigg[\Omega_s \int dz \frac{n(z)^2}{dV/(dzd\Omega)}\bigg]^{-1}
\end{equation}
and accounts for the distribution of objects pairs across redshift. If, for example, the sample spans a broad redshift $0 < z < 2$ but 99\% of objects are localized within $0.49 < z < 0.51$, then the above estimator explicitly accounts for this. Note that $n(z)$ is normalized so that $\int n(z) dz = 1$. The quantity $\Omega_s$ is the survey area in units of {steradians}.

Thus far we have discussed the Gaussian contributions to the covariance matrix. Given the low number density of the cluster sample, some non-Gaussian terms also become relevant \citep{Fumagalli:2024:CovEuclid}. The leading order non-Gaussian term is given by
\begin{equation}
    C[\wrp(\rp^i), \wrp(\rp^j)] 
    = \delta_{ij} \iint_0^\infty 
    \frac{d\kperp^a \kperp^a}{2\pi} 
    \frac{d\kperp^b \kperp^b}{2\pi}
    J_0(\kperp^a \rp^i) 
    J_0(\kperp^b \rp^j) 
    \bigg[\frac{2}{n_h^2} \times X[P(\kperp)]\bigg]
    \label{eq:cov:k2r}
\end{equation}
following the modeling choices of \citet{Fumagalli:2024:CovEuclid}. This term accounts for a four-point correlation between pairs of clusters AB and AC; that is, two correlations formed from three objects. This configuration causes a {shot noise} enhanced correlation between Fourier modes. Our empirical tests have shown this term is vital for accurately modeling the covariance measured in our simulations. Note the $\delta_{ij}$ function enforcing that this term only contributes to the diagonals of the matrix.

The inverse square of the number density is computed as 
\begin{align}
    \bigg\langle \frac{1}{n_h^2}\bigg\rangle = \frac{\int dz (dV/dz)}{\int dz n(z)^2 (dV / dz)}
\end{align}
where once again $\int n(z) dz = 1$. As was the case with $V_{\rm eff}$, this expression accounts for the full distribution of objects across redshift.

Finally, we also account for the effect of bin-averaging, i.e. the correlations at a given $r_p$ represent the average correlation between the edges, $r_{p, {\rm min}} < r_p < r_{p, {\rm max}}$. This can be accounted for using the integral
\begin{equation}
    C[\wrp(\hat{\rp}^i), \wrp(\hat{\rp}^j)] = 
    \int_{r_{p, {\rm min}}^i}^{r_{p, {\rm max}}^i} 
    \frac{2 r^i_p dr^i_p}{[(r_{p, {\rm max}}^i)^2 - (r_{p, {\rm min}}^i)^2]}  \int_{r_{p, {\rm min}}^j}^{r_{p, {\rm max}}^j} 
    \frac{2 r^j_p dr^j_p}{[(r_{p, {\rm max}}^j)^2 - (r_{p, {\rm min}}^j)^2]}
    dr^j_p\,\, C[\wrp(\rp^i), \wrp(\rp^j)]
\end{equation}
where $\hat{\rp}$ represent the binned measurements and the integrals average the covariance values within a given annulus. However, this integration can be done in Fourier space by replacing the kernel in Equation \eqref{eq:cov:k2r} with
\begin{equation}
    J_0(\kperp \rp^j) \rightarrow 
    \frac{2}{(r_{p, {\rm max}}^j)^2 - (r_{p, {\rm min}}^j)^2}
    \frac{r_{p, {\rm max}}^j J_1(\kperp r_{p, {\rm max}}^j) - r_{p, {\rm min}}^j J_1(\kperp r_{p, {\rm min}}^j)}{\kperp}.
\end{equation}
\end{widetext}

\subsection{Covariance Validations} \label{appx:CovMatrix:Valid}

\begin{figure}
    \centering
    \includegraphics[width=\linewidth]{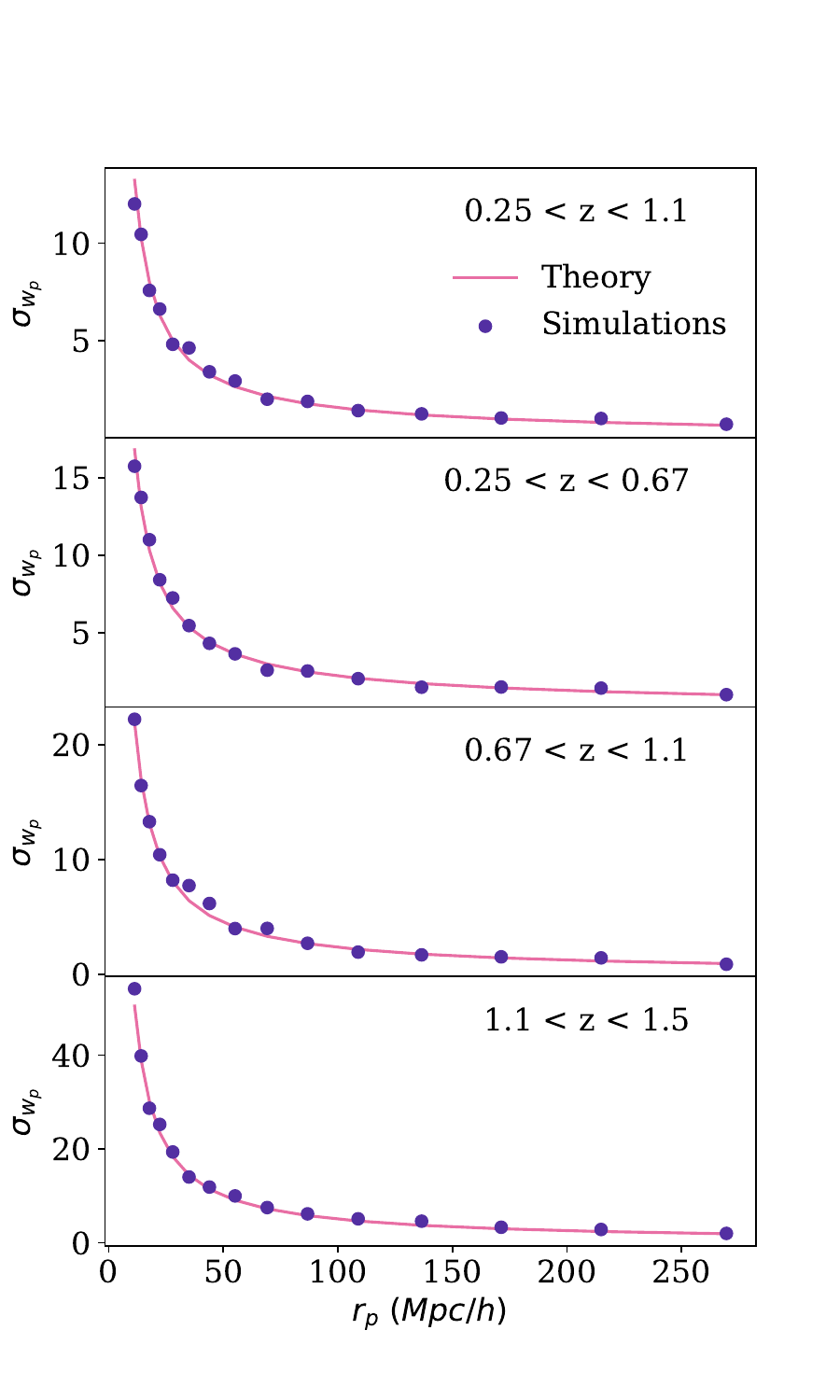}
    \caption{Comparison of standard deviations in $\wrp$ across separation between the theoretical model (pink line) and the simulation-based values (purple circles) for all redshift bins. An additional bin at $1.1 < z < 1.5$ has been included to show robustness out to the higher redshift end of the SPT-3G catalog that may be measurable in the future with improved photometric redshifts from \textit{Euclid}.}  
    \label{fig:STD_DEV_VALID}
\end{figure}

\begin{figure*}
    \centering
    \includegraphics[width=\linewidth]{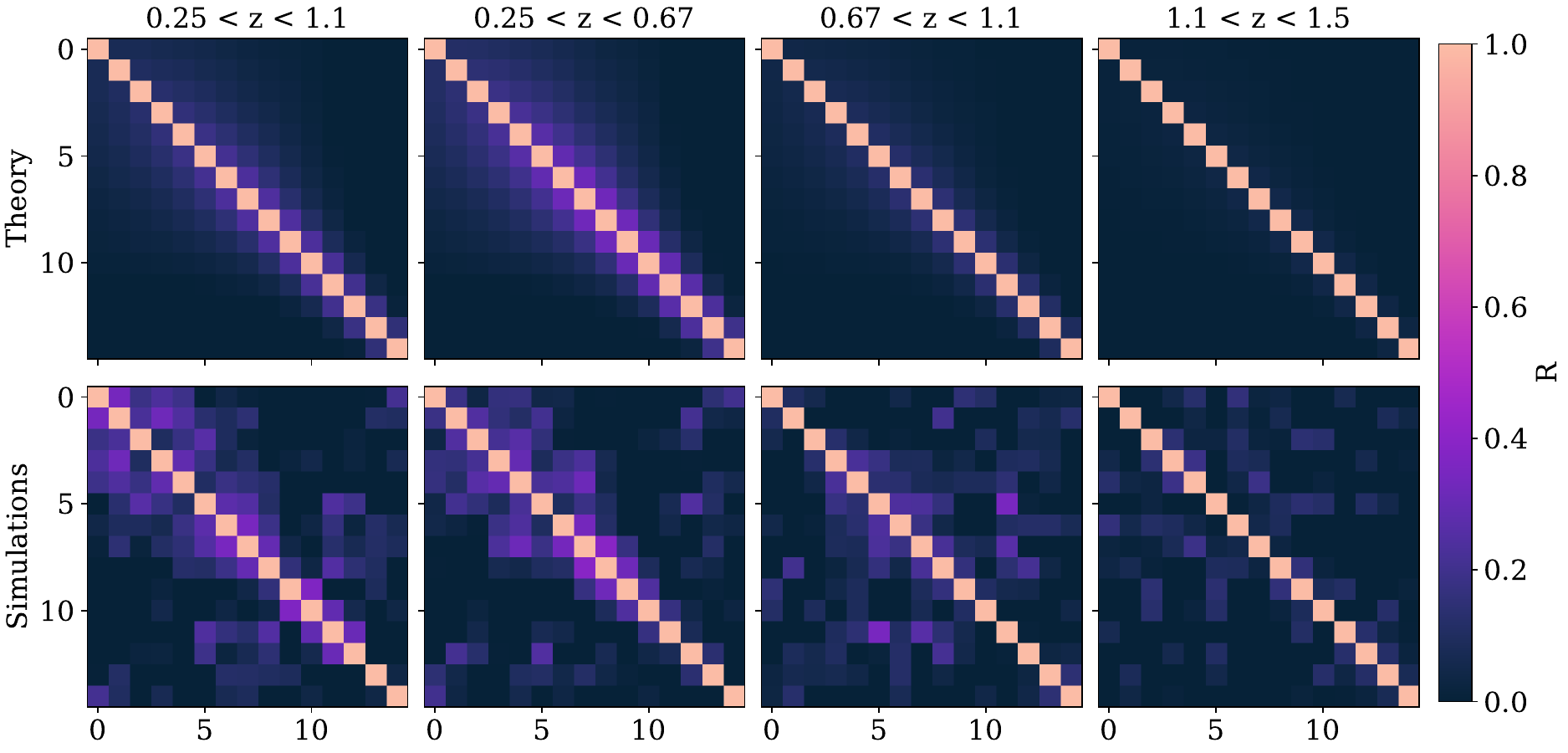}
    \caption{Comparison of correlation, R, in the theoretical (top) and simulation-based (bottom) covariance matrices. The indices correspond to $\rp$ bin index and are logarithmically spaced from $10 \mpc/h < \rp < 300 \mpc/h$. The matrices are largely uncorrelated with some minor correlations at low-redshift. This is found consistently in both simulations and our modeling.}
    \label{fig:COV_VAL}
\end{figure*}

To validate the analytic covariance model described in {Appendix} \ref{appx:CovMatrix}, we tested its performance against simulations. This is done for all three redshift-bin choices used in this analysis, as well as for an additional bin at very high redshifts that is not used in this analysis but further showcases the model accuracy. For robustness, we test the model in a number of cases:
\begin{enumerate}
    \item Fiducial configuration.
    \item {Shot noise} only (no cosmological clustering).
    \item Varying $z_{\rm min}$ and $z_{\rm max}$ values in the span of $ 0.1 < z < 1.5$.
    \item Varying cluster number densities between $0.1$ to $10$ times the fiducial {value}.
    \item Varying $\pimax$ in the span of $20 \mpc/h < \pimax < 200\mpc/h $.
    \item Varying the number of $\rp$ points, N$_{\rp}$, in the span of $5 < \textrm{N}_{\rp} < 20$.
\end{enumerate}
In each case, we generate simulations for a given configuration and then make predictions with our model. The comparisons of the two, in all cases, are better than 10\% {when comparing their diagonal elements}. We are limited from making more precise characterizations of the model as we have limited simulation realizations to use for this check. Nonetheless, $\lesssim 10\%$ shifts in the covariance are negligible in our analysis. Figure \ref{fig:STD_DEV_VALID} shows the comparison between our model and simulated covariances. For brevity, we only show the comparison for our fiducial setup and for a few different redshift bin choices. Figure \ref{fig:STD_DEV_VALID} also extends to $\rp = 300 \mpc/h$, while our main analysis is limited to $\rp < 100 \mpc/h$. We show the former to highlight the accuracy of our model in extended regimes, outside just our particular configurations.

So far, we have only shown and discussed the amplitude of the diagonals in the covariance matrix. We now characterize the off-diagonals using the correlation matrix {given as}
\begin{equation}
    R_{ij} = \frac{C_{ij}}{\sigma_i \sigma_j},
\end{equation}
where $\sigma_i = \sqrt{C_{ii}}$ as usual. Figure \ref{fig:COV_VAL} shows the correlation matrices for both the theoretical and simulation-based estimates across all redshift bins shown previously. Both matrices show only weak correlations between adjacent $\rp$ bins, as expected for a {shot noise} dominated measurement. However, there are some mild off-diagonal correlations, especially at lower-redshift where {shot noise} is less dominant, and we see the theory-based and simulation-based estimates show such correlation. We once again cannot make a precise comparison due to noise in the simulation-based estimate due to the number of realizations available to us. 

Note that while our model is heavily based on that given {in} \citet{Fumagalli:2024:CovEuclid}, we do not follow their semi-analytic fit to simulations. Our validations against simulations found such a fit was not necessary for the datavector setup in this work. Such a shift is not surprising as our work uses a different cluster sample (with different {shot noise} properties) and considers a projected correlation function rather than the 3D one. Future tSZ cluster clustering analyses may find semi-analytic modeling to be necessary given their precision. However, the analysis in this work can be precisely carried out using the analytic modeling alone. 

\section{Correlation with Observing Conditions}\label{appx:syscorr}

This work focuses on the spatial correlations in the number density of tSZ clusters. To ensure our measurements represent only the cosmological clustering signal, we must quantify the amplitude of any artificial, non-cosmological clustering in our measurements. The latter can arise if the survey footprint is generally inhomogeneous in the properties that affect cluster detection. In general, the most important property for cluster detection (regardless of the wavelength) is the map depth or noise level. The tSZ regime could also have secondary impacts from the cosmic infrared background (CIB) and galactic dust extinction; however, \citet{Kornoelje_2025} shows that the SPT-3G cluster detection pipeline is robust to the effects of the CIB, and the galactic dust is adequately handled {by a combination of area selection and} highpass filtering of the maps (\citetalias{Bleem:2026}).

Additionally, tSZ clusters detected with SPT also rely on optical follow-up, both for external confirmation and to compute redshifts for these clusters. {The optical data are prone to typical effects affecting uniformity, such as depth, point-spread function, and stellar density variations}. These latter quantities do not impact the tSZ detection, but could have a second-order impact on our catalog due to our reliance on optical data for redshift estimates. In the case where these effects are sufficiently large, one can account for them either by defining ``systematic weights'' which re-weight the number density of the cluster sample so that it is uncorrelated with a set of systematic template maps (such as depth and noise). One could alternatively also account for these effects by generating a random catalog that explicitly considers the observational selection effects; therefore, the randoms will trace all of these non-cosmological correlations, which in turn will allow us to remove their effects accurately. Contrastingly, if the effect of observational conditions is sufficiently small, the only necessary consideration to forming the random catalog is that the catalog occupies the same survey mask and redshift distribution as the data.

We examine the significance of six observing conditions that are most likely to impact the selection of SPT-3G clusters, namely: the SPT weight (synonymous with detector and atmosphere noise levels), stellar density estimated from \citet{Gaia_2021}, Galactic dust extinction \citep[the E(B-V) maps from][]{Schlegel:1998:DustMaps}, and the optical $g$-, $r$-, and $z$-band depths from the DESI
Legacy Surveys DR10 \citep[LS-DR10;][]{dey19}. 

To quantify the correlation of our cluster spatial variations with these observing conditions, we create pixelized maps of the SPT-3G cluster counts, as well as of each observing condition within the 3G mask. The maps are made with pixel sizes of $0.46$ deg or $\texttt{NSIDE} = 256$. The results we present below are similar for larger pixel sizes as well. Note also that the cluster count map explicitly accounts for what fraction of a healpix pixel contains observations from SPT-3G (after masking),
\begin{equation}
    n_{c}(p_i) = \frac{N_{c,i}}{\Omega_{\rm pix} f_{\rm obs}(i)}, 
\end{equation}
where $f_{\rm obs}$ is the observed fraction, $N_{c, i}$ is the counts of clusters in the pixel, and $\Omega_{\rm pix}$ is the pixel area. To check the correlations with some observing condition, $\alpha$, we bin the pixels into 10 bins spanning the range of $\alpha$ values and compute the average cluster number density within those pixels. We show this relation for the six observation conditions in Figure \ref{fig:num_dense}. 

\begin{figure*}
    \centering
    \includegraphics[width=\linewidth]{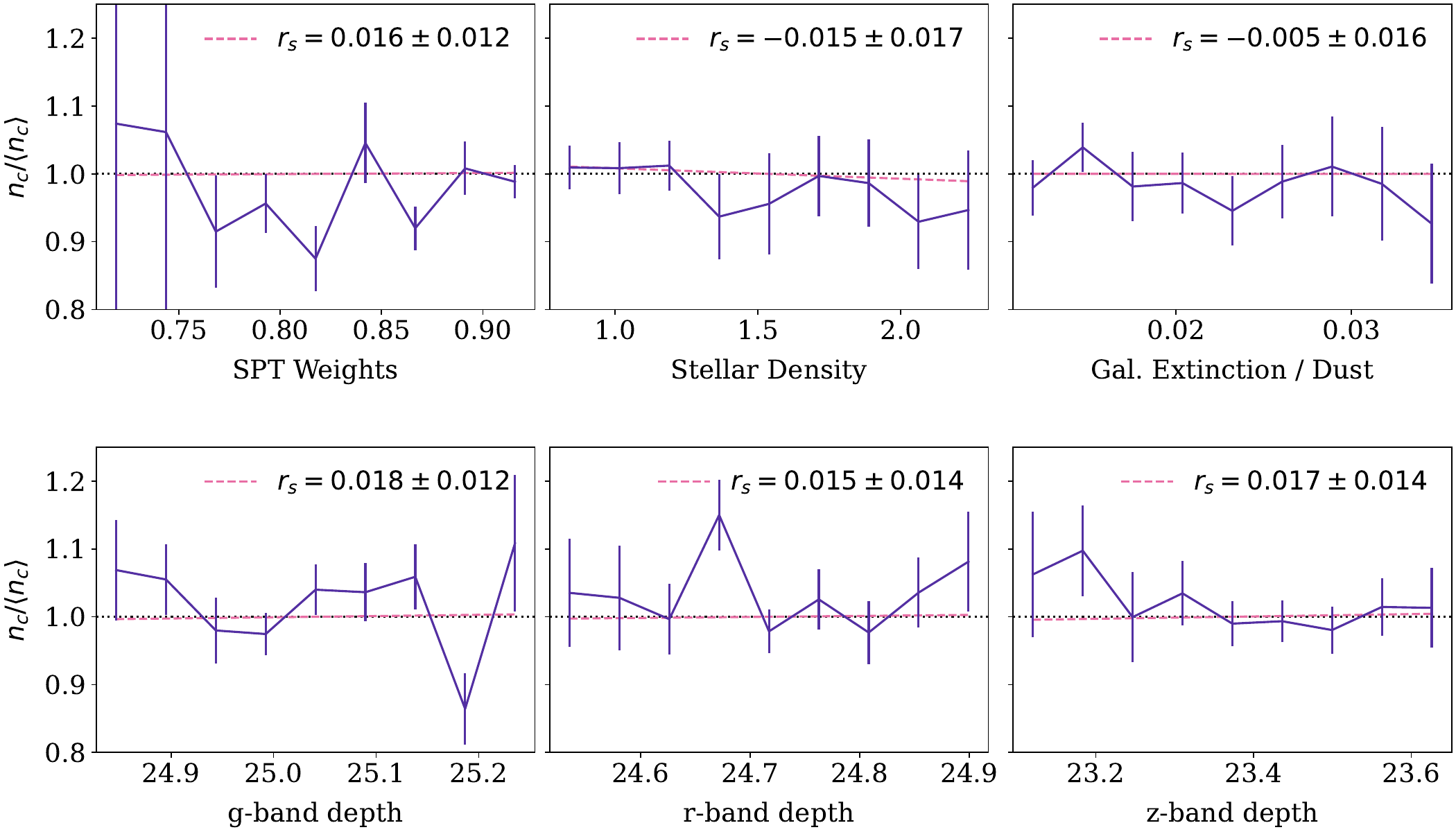}
    \caption{Comparison between the cluster number density in pixels and the corresponding systematic map weight (solid purple lines). Correlation is shown in terms of the linear best-fit as the dashed pink lines.}
    \label{fig:num_dense}
\end{figure*}

We additional quantify the correlation between $n_c$ and $\alpha$ using a Spearman's rank correlation coefficient, $r_s$, which is given as
\begin{equation}
    r_s = 1 -\frac{6\sum d_{i}^{2}}{N_o(N_o^{2}-1)}
\end{equation}
where $d$ is difference in ranks of a pixel when ordered by $n_c$ compared to when it is ordered by $\alpha$ and $N_o$ is the number of observations or data points.

Figure \ref{fig:num_dense} shows the correlation coefficient alongside jackknife uncertainties estimated from splitting the sky into 50 equal-area patches. Of the six observing conditions, all six are consistent with null correlation at $\leq 1.5\sigma$. In principle, it is possible a different observing map can exhibit a statistically significant correlation with the data. However, the above six maps represent the key properties expected to impact our measurements. The spatial correlations presented in this work are therefore dominated by the cosmological signal. This is a key advantage of tSZ cluster clustering from SPT-3G compared to those from optical or X-ray datasets, where the latter must contend with significant spatially varying observing conditions that directly affect the cluster detection at zeroth-order (e.g., exposure time or neutral hydrogen absorption).

\end{document}